\documentclass[prb,twocolumn,amsmath,amssymb,superscriptaddress,longbibliography]{revtex4-1}
\usepackage{graphicx}
\usepackage{dcolumn}
\usepackage{bm}
\usepackage{color}
\usepackage{ulem}
\begin{document}
\title{Construction and operation of a tabletop system for nanoscale magnetometry with single nitrogen-vacancy centers in diamond}
\author{Daiki Misonou}
\affiliation{School of Fundamental Science and Technology, Keio University, 3-14-1 Hiyoshi, Kohoku-ku, Yokohama 223-8522, Japan}
\author{Kento Sasaki}
\affiliation{School of Fundamental Science and Technology, Keio University, 3-14-1 Hiyoshi, Kohoku-ku, Yokohama 223-8522, Japan}
\author{Shuntaro Ishizu}
\affiliation{School of Fundamental Science and Technology, Keio University, 3-14-1 Hiyoshi, Kohoku-ku, Yokohama 223-8522, Japan}
\author{Yasuaki Monnai}
\affiliation{School of Fundamental Science and Technology, Keio University, 3-14-1 Hiyoshi, Kohoku-ku, Yokohama 223-8522, Japan}
\author{Kohei M. Itoh}
\email{kitoh@appi.keio.ac.jp}
\affiliation{School of Fundamental Science and Technology, Keio University, 3-14-1 Hiyoshi, Kohoku-ku, Yokohama 223-8522, Japan}
\affiliation{Center for Spintronics Research Network, Keio University, 3-14-1 Hiyoshi, Kohoku-ku, Yokohama 223-8522, Japan}
\author{Eisuke Abe}
\email{eisuke.abe@riken.jp}
\affiliation{School of Fundamental Science and Technology, Keio University, 3-14-1 Hiyoshi, Kohoku-ku, Yokohama 223-8522, Japan}
\affiliation{RIKEN Center for Emergent Matter Science, Wako, Saitama 351-0198, Japan}
\date{\today}
\begin{abstract}
A single nitrogen-vacancy (NV) center in diamond is a prime candidate for a solid-state quantum magnetometer capable of detecting single nuclear spins
with prospective application to nuclear magnetic resonance (NMR) at the nanoscale.
Nonetheless, an NV magnetometer is still less accessible to many chemists and biologists, as its experimental setup and operational principle are starkly different from those of conventional NMR.
Here, we design, construct, and operate a compact tabletop-sized system for quantum sensing with a single NV center,
built primarily from commercially available optical components and electronics.
We show that our setup can implement state-of-the-art quantum sensing protocols that enable the detection of single $^{13}$C nuclear spins in diamond and the characterization of their interaction parameters,
as well as the detection of a small ensemble of proton nuclear spins on the diamond surface.
This article providing extensive discussions on the details of the setup and the experimental procedures, our system will be reproducible by those who have not worked on the NV centers previously.
\end{abstract}

\maketitle
\section{Introduction}
Quantum sensing aims to measure physical quantities such as magnetic field (magnetometry), electric field (electrometry), temperature (thermometry) 
with high sensitivity and precision by exploiting a controlled quantum system as a sensor.~\cite{DRC17}
Its rapid growth as a subfield of quantum technology owes in no small part to the negatively-charged nitrogen-vacancy (NV) center in diamond,
which is among the most promising platforms for a solid-state quantum sensor,~\cite{MSH+08,BCK+08,RTH+14,SCLD14,CvdSY18,AS18}
with a wide range of applications including nanoscale magnetic resonance imaging (nanoMRI).~\cite{MKS+13,SSP+13,MKC+14,HSR+15,DPL+15,KSD+15,SRP+15,PHHH16,KJM+17}
In a diamond substrate with sufficiently low NV defect density, a single isolated NV center can be resolved optically,
and its electronic spin can be initialized by optical pumping, be coherently manipulated by a series of microwave pulses.
The use of the single-NV quantum sensor significantly reduces the required volume of analyte for nuclear magnetic resonance (NMR), ultimately down to the single molecular level.
Recent demonstrations toward this ambitious goal include detection of single protons,~\cite{SLC+14} spectroscopy of single proteins,~\cite{LSU+16}
sub-hertz spectral resolution,~\cite{SGS+17,BCZD17,APN+17,GBL+18,SDK+19}
spectroscopy and tracking of single nuclear spins via weak measurements,~\cite{PWS+19,CBH+19}
determination of the positions of single nuclear spins,~\cite{ZHS+12,ZCS+18,ZHCD18,SIA18,ARB+19} and so forth.
Remarkably, many of them have been realized at room temperature.

Given such rapid progress, there is a growing need for making the NV magnetometry more accessible to wider research communities
such as chemists and biologists, who are the most frequent users of conventional NMR and potentially benefit from this new avenue.
From technological point of view, standard knowledge on optics and microwave engineering are sufficient to construct an NV magnetometry system in the laboratory.
Even so, combining the knowledge from the two areas and designing, building, and testing a working system are not trivial, and present a number of challenges to nonspecialists (those who have never worked on the NV centers).
A few platforms targeted for NV magnetometry have become commercially available,~\cite{attocube,qnami,qutools,qzabre} but industrial-level turn-key systems are still yet to come.
We therefore feel that at this stage it is important and helpful to provide a comprehensive and detailed description about the construction and operation of a simple single-NV magnetometer,
especially the one which we believe is reproducible by nonspecialists without sacrificing immense amount of their time and effort.
A circumstantial evidence of this belief is that the setup we describe below was constructed from scratch and tested by then-undergraduate students with no prior hands-on experience on either optics or microwave engineering.

We note that there are recent reports describing experimental setups for NV centers in detail.~\cite{ZBL+18,BAB+19}
A major distinction between those and ours is whether a system is designed for ensemble or single NV centers.
The setup for the former is more robust against misalignment than that for the latter,
because (i) it does not require high-precision scanning stages and (ii) the photon counts are significantly enhanced.
It is even possible to miniaturize the entire system including microwave and electrical components.
By replacing bulky electronics with integrated circuits, portable NV magnetometers usable outside of the laboratory environment may be realized.~\cite{SBK+19,WCT+19,KIF+19}
On the other hand, the use of scanning stages to resolve single NV centers means additional system complexity and size,
even though we made an effort to keep the system compact by utilizing commercially available, off-the-shelf fiber optic components and optical cage system.
Still limited to the laboratory use, our system is capable of working with single NV centers, and is fully fledged, in the sense that in principle many of state-of-the-art NMR experiments with single NV centers can be performed.
Therefore, the two approaches based on ensemble and single NV centers are complementary.

This article is organized as follows.
We describe the design of our system in Sec.~\ref{sec_design}.
After giving an overview of the physics of the NV center and discussing the key parameters (Sec.~\ref{sec_nv_center}),
various aspects of the design, optics (Sec.~\ref{sec_optics}), microwave antenna (Sec.~\ref{sec_antenna}), DC magnetic field (Sec.~\ref{sec_magnet}), and electronics (Sec.~\ref{sec_electronics}) are explained in more detail.
Section~\ref{sec_electronics} also discusses multipulse sensing protocols used for nanoscale magnetometry.
In Sec.~\ref{sec_operation}, we present experimental data obtained from our system.
Diamond samples used are summarized in Sec.~\ref{sec_sample}.
We first test the scan range and calibrate the magnetic field strength using ensemble NV centers (Sec.~\ref{sec_ensemble}).
We then show the basic results such as photoluminescence (PL) imaging, continuous-wave optically detected magnetic resonance (CW ODMR),
Rabi oscillation, pulsed ODMR, Ramsey interferometry, and Hahn echo, measured with a single NV center (Sec.~\ref{sec_single}).
Quantum sensing of internal and external nuclear spins is demonstrated in Sec.~\ref{sec_13c} ($^{13}$C nuclei in diamond) and Sec.~\ref{sec_proton} (protons in oil), respectively.
In Sec.~\ref{sec_photon}, extendibility of our setup to measure PL spectra and photon statistics are discussed briefly.
The conclusion is given in Sec.~\ref{sec_conclusion}, followed by Appendices providing the lists of items and the design details.
Throughout this article, we have nonspecialists in mind and aim to provide comprehensive and pedagogical descriptions.

\section{Design\label{sec_design}}
\subsection{NV center\label{sec_nv_center}}
\begin{figure*}
\begin{center}
\includegraphics{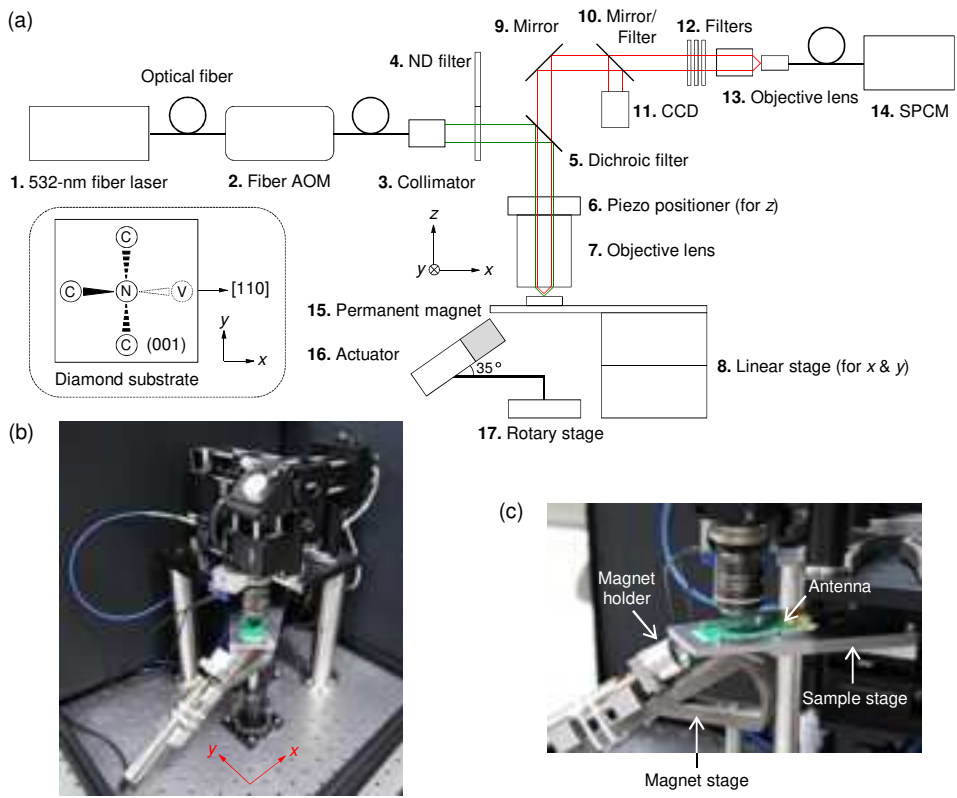}
\caption{(a) Schematic setup showing major optical components, scanning and DC magnetic field control systems.
See Tables~\ref{tab_a1}--\ref{tab_a4} in Appendix~\ref{sec_list} for details.
The inset shows a schematic of an NV center in a (001)-faced, square-shaped diamond substrate with $\langle$110$\rangle$ edges.
The N--C bonds and the NV symmetry axis point along the $\langle$111$\rangle$ directions, off by $\pm$35$^{\circ}$ from the (001) plane.
AOM: acousto-optic modulator. ND: neutral density. CCD: charge coupled device. SPCM: single photon counting module.
(b) Photo of the setup.
The optical table on which the system is mounted has M6 taps on 25 mm centers.
Not captured in this photo are the laser (\#1) and the AOM (\#2), which are connected to the blue fiber cable (to the left), and the SPCM (\#14), which is connected to the fiber cable covered with aluminum foil (to the right).
The foil is to prevent stray light.
(c) Close-up of the setup around the sample.
The sample is visible at the center, placed on the microwave antenna and fixed with black double sided tape.
See Fig.~\ref{fig13} (Fig.~\ref{fig14}) in Appendix~\ref{sec_cad} for drawings of the sample stage (the magnet holder/stage).
}
\label{fig1}
\end{center}
\end{figure*}

The (negatively-charged) NV center in diamond is an optically-active, $S$ = 1 paramagnetic defect in diamond.
The defect structure projected onto the (001) plane of a diamond crystal is shown in the inset of Fig.~\ref{fig1}(a). 
The electronic structure and optical and spin properties of the NV center have been discussed in a large amount of literature.~\cite{JW06,DMD+11,MGT+11,DMD+13}
Here, we explain them briefly, just enough to identify the physical parameters (wavelength, frequency, and magnetic field strength) that directly influence our design of the setup.
The orbital ground state of the NV center is spin-triplet, labeled as $^{3}A_{2}$.
The Hamiltonian $H_{\mathrm{e}}$ of this triplet electronic spin of the NV center, when the static magnetic field $B_0$ is applied along the NV symmetry axis (one of the $\langle$111$\rangle$ axes), are given by
\begin{equation}
H_{\mathrm{e}}/h = D S_z^2 + \gamma_{\mathrm{e}} B_0 S_z,
\end{equation}
where $h$ is the Planck constant, $S_z$ is the $S = 1$ spin operator, $D$ = 2.87~GHz is the zero-field splitting, and $\gamma_{\mathrm{e}}$ = 28.0~MHz/mT is the gyromagnetic ratio of the NV spin.
$H_{\mathrm{e}}$ is diagonal, and its eigenenergies $E$ are straightforwardly given by
\begin{equation}
E/h = D (m_S)^2 + \gamma_{\mathrm{e}} B_0 m_S,
\end{equation}
where $m_S = 0, \pm1$ are the spin quantum numbers.
In our case, we solely work with the $m_S = 0 \leftrightarrow -1$ magnetic dipole transition at $f_{\mathrm{NV}} = D - \gamma_{\mathrm{e}} B_0$
and use the $m_S$ = 0 and $-$1 states as the ``sensor qubit'' $|0\rangle$ and $|1\rangle$ states, respectively.
We apply $B_0$ of up to 30~mT, which is most conveniently produced by a permanent magnet (Sec.~\ref{sec_magnet}).

There is an orbital excited state labeled as $^{3}E$, higher in energy by 1.95~eV from the $^{3}A_{2}$ state.
When the defect is illuminated by a green laser, it emits photons at the wavelength ranging from 637~nm (zero-phonon line) to 800~nm (the long wavelength end of the phonon sideband).
Importantly, this fluorescence is spin-state dependent.
The photon counts are higher (lower) if the spin state is in the $m_S = 0$ ($-$1) state.
This is because the $m_S = -1$ state prefers the non-radiative decay process (intersystem crossing) 
via intermediate $^{1}A_{1}$--$^{1}E$ states that involves a spin-flip ($m_S = -1 \rightarrow 0$).
Therefore, the laser excitation at 532~nm (optical pumping) and the counting of photons from the phonon sideband (650--800~nm) provide a simple means to initialize and determine the spin state.
We note that with a 532-nm laser the second-order Raman spectrum of diamond appears in the 600--625~nm range (Sec.~\ref{sec_photon}), and therefore the photons in this range must be filtered out.
Combined with spin control by microwaves, ODMR of a single spin is realized at room temperature.

\subsection{Optics\label{sec_optics}}
Our aim is to construct a compact, easy-to-handle, yet single-NV-resolving nanoscale magnetometry system.
For optical components, we rely on commercially available fiber optics and optical cage system.
The use of these components provides self-aligned free space optics, and enhances the accessibility and reproducibility by nonspecialists.
Figure~\ref{fig1} shows a schematic of our setup.
Fiber optics is used from the laser (\#1) to the collimator (\#3), and at the single photon counting module (SPCM, \#14).
Between the input and output parts is the cage system, constituting a confocal scanning microscope.

The optical diffraction limit $\delta$ is given by $\delta = \lambda/(2\mathrm{NA})$
with $\lambda$ the excitation wavelength and $\mathrm{NA} \approx$ 1 the numerical aperture (0.8 for the air objective lens and 1.42 for the oil objective lens listed in Table~\ref{tab_a1} in Appendix~\ref{sec_list}).
With a 532-nm laser the focal spot size is about a quarter of a micron, and a single NV center can be regarded as a point light source.
Lateral scanning is usually achieved by the use of mirror galvanometers or translation stages (often piezoelectric).
We adopt a dual-axis linear feedback stage from Sigma Tech (\#8), which provides the minimum resolution of 10~nm ($\ll \delta$) and the maximum scan range of 20 $\times$ 20~mm$^2$.
Diamond substrates used for quantum sensing experiments are typically a few millimeters wide.
Therefore, the scan range of 20 $\times$ 20~mm$^2$ enables the search of the whole surface of a diamond substrate or even multiple samples at the same time.
This is particularly convenient to characterize NV-containing diamond samples prepared in the laboratory
by using chemical vapor deposition (CVD) with N-doping near the surface or ion-implantation of N$^{+}$.~\cite{ORW+13,MBD+05}
Since the presence of nitrogen is only a prerequisite to create and stabilize NV centers,
additional steps such as annealing, electron irradiation, ion-implantation of C$^{+}$ or He$^{+}$, and chemical treatment are often required.~\cite{OPC+12,OHB+12,OHdlC+14,HLS+13,WJGM13} 
The ability to quickly examine the sample conditions in between different steps is highly desirable.

In standard confocal microscopy, the depth resolution is improved by the use of a pinhole to spatially filter the out-of-focus background light.
In our case, the emitted photons are focused (via an objective lens of \#13) onto a single-mode fiber connected to the SPCM.
This effectively works as a spatial filter, and provides the depth resolution of about a micron.
In addition to saving the space, the absence of mirror galvanometers and a pinhole in our setup facilitates the task of optical alignment.
The depth scan is carried out by a piezo positioner (\#6).
Taken together, the system achieves the confocal volume much smaller than 1~$\mu$m$^3$,
and therefore single NV centers can be resolved optically when the NV defect density is on the order of or less than 10$^{12}$~cm$^{-3}$ (= 1~NV center/$\mu$m$^3$),
which is satisfied in high-quality diamond substrates or low-dose N$^+$ ion implantation (Sec.~\ref{sec_sample}).
Furthermore, super-resolution imaging techniques, such as stimulated emission depletion (STED) microscopy and stochastic optical reconstruction microscopy (STORM),
have been applied to NV centers, successfully resolving multiple NV centers separated by less than the optical diffraction limit.~\cite{RHI+09,MMS+10,PAW+14,JBA+17}

For clarity, we make passing reference to the roles of respective optical components shown in Fig.~\ref{fig1}, in the order of light propagation.
The 532-nm light from the laser (\#1) is passed through or blocked by the fiber acousto-optic modulator (AOM, \#2).
Pulsed ODMR requires an AOM to have high extinction ratio, because a leakage of green light during the spin manipulation leads to undesired initialization of the spin state, degrading the fidelity of spin control.
With a free-space AOM, one can employ a double-pass configuration, which effectively enhances the extinction ratio with a single device.
This is not possible with a fiber AOM.
The fiber AOM we use (Gooch \& Housego Fiber-Q) provides the extinction ratio of 50~dB at 532~nm, satisfying the requirement.

The collimator (\#3) prepares a collimated beam that enters into the objective lens (\#7) after being attenuated to a desired power by the neutral density (ND) filter (\#4) and reflected at the dichroic filter (\#5).
The diameter of the collimated beam should match with the pupil diameter of the objective lens.
The NV center emits red photons, and a portion of them are collected by the same objective lens.
Although this is a widespread configuration for detecting NV centers, the collection efficiency is intrinsically low due to the large mismatch of the refractive indices between air ($n$ = 1) and diamond ($n$ = 2.4).
The total internal reflection at the air--diamond interface limits the number of photons coming out of the surface.~\cite{HRCS93,LPB+12}
The dipole emission pattern of the NV center also contributes to the low collection efficiency.~\cite{EMKA05,ZLC+13}
Typically, less than 10\% of the photons emitted from the NV center enters into the objective lens.
Various approaches to enhancing the collection efficiency, such as the use of a solid immersion lens or a photonic waveguide, have been demonstrated.~\cite{AEV+18,AHWZ18,HSB18}

The photons entering the objective lens pass through the dichroic filter and are directed to the charge coupled device (CCD, \#11) or the second objective lens (\#13) by the mirrors (\#9 and \#10).
The CCD is used to find the diamond surface prior to the scan.
The second mirror (\#10) is removable, and is replaced by the longpass filter (cut-on at 650~nm) when using the SPCM.
The filter set (\#12) is composed of notch (to reject residual laser light), longpass (cut-on at 600~nm), and shortpass (cut-off at 800~nm) filters.
The reason for the 650-nm longpass filter being made removable is that we wanted to keep the ability to perform optical spectroscopy below 650~nm (Sec.~\ref{sec_photon}).
If such spectroscopy is not needed, the 600-nm longpass filter in the filter set can be replaced with the 650-nm one.

The signal red photons are imaged onto the single-mode fiber and collected by the SPCM (\#14).
Choosing an SPCM with high photon detection efficiency, low dark count and high timing resolution is essential.
The detection efficiency, as well as the coupling efficiency to the fiber port, affects the total collection efficiency at the detector,
which typically ends up in a few \%.
We use Excelitas Technologies SPCM-AQRH-16-FC, which is specified to have detection efficiency $>$ 70\% at 700~nm, dark count rate $<$ 25~cps (counts/second), and timing resolution $<$ 500~ps.
This model is a popular choice in the community, but comparable performance can also be achieved by products from other manufactures (e.g., Laser Components COUNT-10C).

\subsection{Microwave antenna\label{sec_antenna}}
\begin{figure}
\begin{center}
\includegraphics{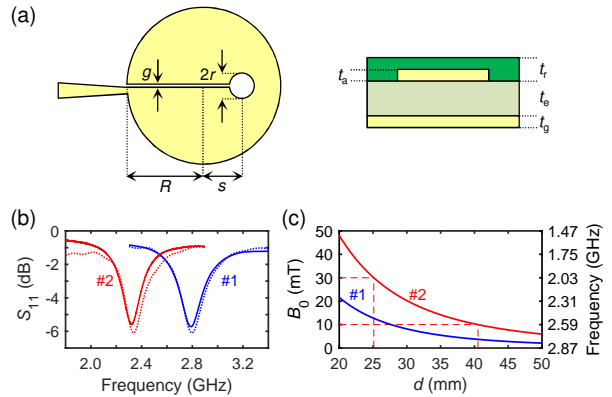}
\caption{(a) Schematic of a microwave planar ring antenna.~\cite{SMS+16}
A printed circuit board consists of, from top to bottom, resist, antenna pattern (copper), epoxy glass (FR4), and ground plane (copper).
(b) $S_{11}$ of Antennas \#1 and \#2.
The solid (dotted) lines are experimental (simulated) data.
The data for Antenna \#1 is reproduced from Ref.~\onlinecite{SMS+16}.
(c) $B_0$ as functions of $d$ for Magnets \#1 and \#2.
The axis on the right hand side is calculated by $f_{\mathrm{NV}} = D-\gamma_{\mathrm{e}} B_0$.
}
\label{fig2}
\end{center}
\end{figure}

To take advantage of the wide scan range available in our system, microwaves should also be applied across the wide area of the sample surface.
In Ref.~\onlinecite{SMS+16}, we designed a broadband, spatially-uniform microwave planer ring antenna for ODMR of NV centers.
The antenna has the resonance frequency $f_{\mathrm{res}}$ at around 2.87 GHz, the bandwidth of 400 MHz,
and uniform microwave magnetic fields oscillating perpendicular to the antenna plane within a 1-mm-diameter hole of the antenna [Fig.~\ref{fig2}(a)].
The identical design can be adopted here for the use in $B_0$ = 0--10~mT.
The design parameters are listed in Table~\ref{tab_parameter} as Antenna \#1.
\begin{table}
\caption{Parameters of microwave antennas and permanent magnets.
See Fig.~\ref{fig2}(a) for the definitions of the parameters.
$B_{\mathrm{r}}$: remanence field.
$u$: radius of the magnet.
$h$: height of the magnet.
The length is given in millimeters (mm).}
\label{tab_parameter}
\begin{tabular}{ccc|ccc}
\hline
Antenna & \#1 & \#2 & Magnet & \#1 & \#2\\
\hline
$f_{\mathrm{res}}$ (GHz) & 2.790 & 2.325 & $B_0$ (mT) & 0--10 & 10--30 \\
\hline
$R$ & 7.0 & 8.0 & $B_{\mathrm{r}}$ (mT) & \multicolumn{2}{c}{1270$\pm$20}\\
$s$ & 3.9 & 5.4 & $u$ & \multicolumn{2}{c}{10} \\
$r$ & \multicolumn{2}{d|}{0.5} & $h$ & 5 & 20 \\
$g$ & \multicolumn{2}{d|}{0.1} & & \\
$t_{\mathrm{r}}$ & \multicolumn{2}{d|}{0.03} & & \\
$t_{\mathrm{a}}$, $t_{\mathrm{g}}$ & \multicolumn{2}{d|}{0.018} & & \\
$t_{\mathrm{e}}$ & \multicolumn{2}{d|}{1.6} & & \\
\hline
\end{tabular}
\end{table}
Its simulated and measured $S_{11}$ properties are shown in Fig.~\ref{fig2}(b).

To work at $B_0 \approx$ 20~mT, the design needs to be modified in order to bring the resonance frequency down to 2.3~GHz.
In Ref.~\onlinecite{SMS+16}, we made a qualitative argument to view the antenna as a series LC circuit,
where the circuit inductance $L$ is proportional to the hole radius $r$, and the capacitance $C$ is inversely proportional to the gap $g$ and proportional to the gap width $R+s-r$.
Then, the circuit has $f_{\mathrm{res}}$ that is proportional to $(LC)^{-1/2} \propto (g/r(R+s-r))^{-1/2}$.
This suggests that a reasonable strategy to reduce $f_{\mathrm{res}}$ is to increase $R$ and $s$.
We performed three-dimensional electromagnetic simulations using CST MICROWAVE STUDIO$^{\textregistered}$ software,
and identified the values of $R$ and $s$ that provide a desirable $f_{\mathrm{res}}$ (Antenna \#2 of Table~\ref{tab_parameter}).
The fabricated antenna shows the $S_{11}$ property in good agreement with the simulation.
We note that the data in Fig.~\ref{fig2}(b) were simulated or measured without placing a diamond sample on top of the antenna.
It has been confirmed that the presence of the diamond barely affects the $S_{11}$ properties of these antennas.~\cite{SMS+16}

\subsection{DC magnetic field\label{sec_magnet}}
We apply $B_0$ by using a cylindrically shaped permanent magnet and aligning the axis of the cylinder parallel to the NV axis, which is one of the $\langle$111$\rangle$ directions.
To realize this efficiently, we use the sample stage and the magnet stage shown in Fig.~\ref{fig1}(c).
In designing them, the following factors were considered.
First, when picking up a single NV center, we do not know in advance which one of the $\langle$111$\rangle$ directions this particular NV center is pointing in.
Hence, it is desired that we are able to apply $B_0$ along any one of four possible $\langle$111$\rangle$ directions.
Second, in our configuration of scanning confocal microscopy, the laser focal spot is fixed in the laboratory frame upon horizontal image scan, and the vertical movement of the objective lens is typically restricted to a few microns.
Third, diamond substrates we use are most often (001)-faced single crystals.
In this case, the $\langle$111$\rangle$ directions are off by $\pm$35$^{\circ}$ from the (001) plane.
Hence, if we mount the sample so that its orthogonal [110] and [1$\bar{1}$0] directions may be parallel to the lateral scan directions,
which are defined as the $x$ and $y$ axes in the laboratory frame [the inset of Fig.~\ref{fig1}(a)], possible NV axes lie in the $xz$ and $yz$ planes.
(Parenthetically, the $\langle$110$\rangle$ directions of a given sample is usually specified in its specification sheet or can be known by inspecting facets of the crystal.)

The sample stage itself is fixed on the translation stage, and the microwave antenna with the diamond sample on top is fixed on it by four screws.
Its cantilever shape creates a free space beneath the antenna to allow for the access of the magnet.
Importantly, the photon counts in this configuration are confirmed to be stable over long time (i.e., negligible vibrations).
The magnet stage has a fixed tilt angle of 35$^{\circ}$ and is combined with the linear actuator (\#16) and the rotary stage (\#17). 

Upon installation, the height of the magnet is manually adjusted so that the axis of the cylinder hits the focal spot.
This needs to be done only once, as long as the vertical displacement of the focal spot is much smaller than the radius of the cylindrical magnet (chosen to be $u$ = 10~mm in our case),
which is placed in the magnet holder.
The direction of $B_0$ is roughly aligned along the desired NV axis by rotating the magnet stage near the $xz$ or $yz$ planes, but the exact rotation angle is determined by examining CW ODMR spectra (Sec.~\ref{sec_ensemble}).
Once $B_0$ is aligned with the symmetry axis of a chosen NV center, one will find that other NV centers pointing in the same direction are brighter than those not, due to the magnetic-field-dependent fluorescence of the NV centers.~\cite{TRS+12}
Hence one does not have to align the rotation angle of the magnet stage for each NV center as long as one stays in the same $B_0$ direction.

The strength of $B_0$ is controlled by changing the distance $d$ between the focal spot and the magnet surface.
The relation between $B_0$ and $d$ can be calculated analytically.
Namely, the magnetic field $B_0$ that a cylindrically shaped permanent magnet with the remanence field $B_{\mathrm{r}}$, the radius $u$, and the height $h$
produces at the distance $d$ from the center of the top surface along the symmetry axis is calculated as
\begin{equation}
B_0 = \frac{ B_{\mathrm{r}} }{2} \left( \frac{h+d}{ \sqrt{u^2 + (h+d)^2} } - \frac{d}{ \sqrt{u^2 + d^2} } \right).
\label{eq_b0}
\end{equation}
By geometry, $d$ cannot be made arbitrarily small.
The magnet can approach the sample until a periphery of the magnet touches the back side of the sample stage, i.e., $d_{\mathrm{min}} = u/\tan 35^{\circ} + t_{\mathrm{h}}/\sin 35^{\circ}$ with $t_{\mathrm{h}}$ = 5~mm the thickness of the sample stage.
On the other hand, the maximum $d$ is limited by the travel range of the actuator ($d_{\mathrm{tr}}$ = 25~mm in our case), but it is not necessary to move the magnet in the full range as long as the desired field range is covered.

Figure~\ref{fig2}(c) shows two curves of $B_0$ as functions of $d$ with the parameter values listed in Table~\ref{tab_parameter}.
We select the parameters so that Magnet \#1 (\#2) is compatible with Antenna \#1 (\#2) within the acceptable range of $d$ [the horizontal axis on the right hand side of Fig.~\ref{fig2}(c)].
With $u$ = 10~mm, we obtain $d_{\mathrm{min}} \approx$ 23~mm.
We observe that Magnet \#2 can generate $B_0$ = 10--30~mT by setting $d \approx$ 40--25~mm ($d_{\mathrm{min}} <$ 25~mm and $d_{\mathrm{min}}+d_{\mathrm{tr}} >$ 40~mm), as desired.
The lower magnetic field is realized by Magnet \#1, except for $B_0$ = 0~mT, which is obtained by simply removing the magnet.
Magnet \#1 has $h$ that is 15~mm shorter than Magnet \#2, with other parameters being equal between the two.
If we replace Magnet \#2 with Magnet \#1, $d_{\mathrm{min}}$ will be about 40~mm, for which $B_0$ is expected to be 2~mT.
With a spacer, Magnet \#1 can be brought closer to the substrate and the desired field up to 10~mT is readily achieved. 

Our method described here is not without a few drawbacks.
The main concern is that a diamond substrate can have a miscut angle of a few degrees, meaning that the surface is not exactly the (001) plane.
In addition, it is possible that the top and back surfaces of the substrate are not perfectly parallel and/or manual sample mounting causes an unintentional tilt.
These errors can be cumulative or may cancel each other (accidentally), but in the end we are likely to have an angle error of a few degrees that is uncorrectable by the present method.
So far, we find this level of error not to be detrimental to our experiments.
A straightforward remedy will be to add a goniometer to the magnet stage, but we have not adopted it in order to keep our system as simple as possible.

Another concern is that the remanence field $B_{\mathrm{r}}$ of a permanent magnet is temperature-dependent.
We used neodymium (NdFeB) magnets, but it is known that samarium-cobalt (SmCo) magnets are more temperature-insensitive and thus are a good option when the temperature control of a laboratory space is not stable.
The spatial homogeneity achieved by a single permanent magnet is not problematic in dealing with single NV centers,
but the use of a pair of permanent magnets or electromagnets may be considered when dealing with ensemble NV centers and a larger spot size.~\cite{BAB+19}

Before concluding this section, we comment on the use of (111)-faced substrates.
In this case, the NV centers of interest are almost always those pointing perpendicular to the surface.
The application of $B_0$ becomes simpler, realized by setting a permanent magnet on the back of the substrate and moving it up and down. 
However, this means that the microwave antennas discussed in Sec.~\ref{sec_antenna}, which generate the microwave magnetic fields perpendicular to the antenna plane, are unsuitable for ODMR.
In this case, alternative designs of uniform microwave antennas compatible with (111)-faced substrates, such as discussed in Refs.~\onlinecite{HAS+16,MMO+18}, can be utilized.

\subsection{Electronics\label{sec_electronics}}
\begin{figure*}
\begin{center}
\includegraphics{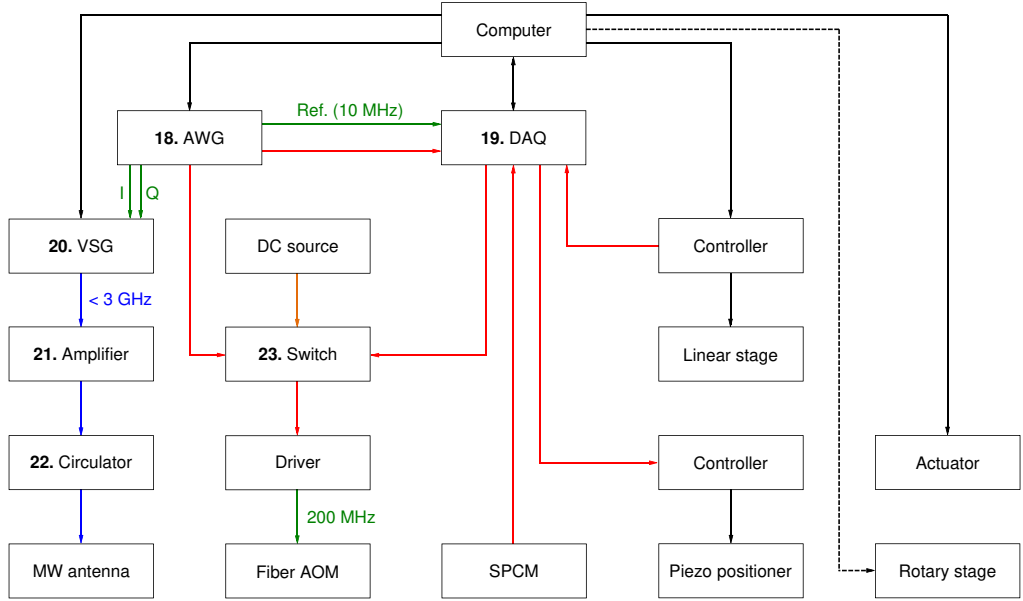}
\caption{Diagram of electronic devices.
See Table~\ref{tab_a5} in Appendix~\ref{sec_list} for the model numbers.
The black arrows indicate the flow of the control command (via GPIB or USB or other cables).
The blue arrows indicate the microwave signals (via SMA cables).
The green arrows indicate the RF signals for the synchronization, the IQ modulation, and the AOM driving (via BNC or SMA cables).
The orange arrow indicates the DC signal.
The red arrows indicate the TTL/trigger signals.
The reported setup uses a manual rotary stage, but we later replaced it with a motorized stage (Zaber Technologies X-RSW60A-E03) controllable with a computer (the dashed arrow).
The amplifier (\#21), the switch (\#23), the AOM driver, and the SPCM require dedicated DC power supplies, which are not shown.
AWG: arbitrary waveform generator. DAQ: data acquisition. VSG: vector signal generator.
}
\label{fig3}
\end{center}
\end{figure*}

Figure~\ref{fig3} shows a diagram of electronic devices used in our setup.
The devices without numbers have already appeared in Fig.~\ref{fig1}.
Some of them are controlled directly from the computer or via the data acquisition device (DAQ, \#19),
and even when dedicated controllers are required (the linear stage and the piezo positioner) they are relatively small in size.
The most space-occupying devices are rack-mount instruments, namely
the arbitrary waveform generator (AWG, \#18, Tektronix AWG7102) and the vector signal generator (VSG, \#20, Anritsu MG3700A).
While the development of dedicated fast electronics with small footprint are not the scope of the present work,
designing compact high-frequency devices will become essential in the near future, especially in the field of quantum computing,
where the increase in the number of qubits means the increase in the number of control devices and the use of bulky electronic devices is no longer intolerable.
The use of field-programmable gate array (FPGA) is becoming increasingly important to realize fast, compact, flexibly controllable circuitry.
In the below, we discuss a method of spin control based on our specific choice of devices, but many of the considerations there are sufficiently general and applicable to other configurations and device models.

It is important to recognize that, whatever we do, in the end all the information about the NV centers is obtained from photons emitted from them.
In this sense, the performance of the SPCM is crucial, as discussed in Sec.~\ref{sec_optics}.
The SPCM converts the photon counting events into TTL electrical signals, which are collected by the DAQ and processed in the computer.
The control of electronics and signal processing are all carried out via integrated software custom-written in MATLAB.~\cite{software}
In our setup, the timing and duration of the data acquisition are not triggered, but all the photon counting events are collected and time-tagged.
Based on the types of experiments executed, the software program judges which photon counting events contain meaningful information.
For instance, a PL image is obtained by moving the sample stage with the linear stage (for $x, y$) and the piezo positioner (for $z$) while continuously illuminating the laser light (with the AOM open).
The map of the photon counts at respective sample positions gives the PL image.
When the microwave frequency of the (V)SG is swept under continuous laser illumination and at a fixed sample position,
the photon counts as a function of the microwave frequency give a CW ODMR spectrum.
As will be shown in Figs.~\ref{fig5}(b) and \ref{fig6}(b), the vertical axis of a CW ODMR spectrum is usually given as {\it contrast},
which is defined as the ratio of the photon count under microwave irradiation to that without.
By turning on and off the microwave with an AWG-generated square wave at 2~kHz, the contrast is immune to the fluctuation of the photon counts
arising the temporal drift of the position or laser power that occurs slower than 2~kHz. 

Contrary to these continuous measurements, a time-domain experiment involves dynamical control of a sensor electronic spin.
To our knowledge, the most widespread approach to timing control in NV magnetometry setups is to employ SpinCore Technologies PulseBlaster as a multichannel pulse pattern generator (PPG).
On the other hand, advanced quantum sensing protocols make essential use of shaped microwave pulses, for which an AWG is required.
By using the trigger signals of the AWG for timing control, and thus letting the AWG play dual roles of pulse shaping and timing control, we have dispensed with a PPG in our setup.
The AWG is readily synchronized with the DAQ, and therefore the control sequences and subsequent detection events are correlated unambiguously.

Microwave pulse sequences for dynamical spin control and quantum sensing require that the microwave pulses be not only time- and shape-controlled but also phase-controlled.
The phase control here does {\it not} mean to control the absolute phase of a gigahertz oscillation or to realize the spin rotations around different axes in the laboratory frame.
The spin rotation is realized in the rotating frame, and only the relative phases among different pulses in a single run of pulse sequence are relevant.~\cite{VC04}
Moreover, in the scheme so called single-sideband suppressed carrier modulation, the relative phase is defined by the in-phase (I) and quadrature (Q) signals, over which the operator (user) has full control.
Suppose we prepare the I and Q signals given by
\begin{eqnarray}
I(t) &=& A(t) \cos(2 \pi f_{\mathrm{IF}} t + \theta_{\mathrm{IF}}) \\
Q(t) &=& A(t) \sin(2 \pi f_{\mathrm{IF}} t + \theta_{\mathrm{IF}})
\end{eqnarray}
using the AWG.
Here, $A(t)$ defines the pulse shape and length, $f_{\mathrm{IF}}$ is the modulation frequency (IF stands for the intermediate frequency), and $\theta_{\mathrm{IF}}$ is the phase offset.
They are fed into the I and Q ports of the VSG.
The VSG internally splits microwave at $f_{\mathrm{LO}}$ (= 2--3~GHz, LO stands for the local oscillator) into two signals that are phase shifted by 90$^{\circ}$, and mixes the respective signals with the I and Q signals.
They are subsequently summed to give the output signal
\begin{eqnarray}
& & I(t) \cos (2 \pi f_{\mathrm{LO}} t + \theta_{\mathrm{LO}} ) - Q(t) \sin(2 \pi f_{\mathrm{LO}} t + \theta_{\mathrm{LO}}) \notag \\
&=& A(t) \cos [2 \pi (f_{\mathrm{LO}} + f_{\mathrm{IF}}) t + \theta_{\mathrm{LO}} + \theta_{\mathrm{IF}}].
\end{eqnarray}
We observe that $\theta_{\mathrm{IF}}$ determines the relative phase across different pulses and the LO phase $\theta_{\mathrm{LO}}$ does not have to be known.
The lack of the knowledge on $\theta_{\mathrm{LO}}$ does not cause a problem in defining the rotation axes, as long as $\theta_{\mathrm{LO}}$ is constant during each run of the sequence ($<$ 1~ms).
We define the rotation about the $x$ ($y$) axis of the rotating frame by setting $\theta_{\mathrm{IF}}$ = 0$^{\circ}$ (90$^{\circ}$).

In the remainder of this section, we introduce microwave pulse sequences to be used in Sec.~\ref{sec_operation}.
These sequences consist of a combination of $\pi$/2 and $\pi$ pulses, and their pulse lengths have to be determined in advance.
This is done by observing the Rabi oscillation, in which a microwave pulse of variable length $T_{\mathrm{p}}$ is sandwiched by 532-nm optical pulses with typically length on the order of 1~$\mu$s.
The first optical pulse initializes the NV spin into the $m_S$ = 0 state, and the second optical pulse reads out the final spin state.
As increasing $T_{\mathrm{p}}$, the final spin state oscillates between the $m_S$ = 0 and $-$1 states (equivalently the sensor qubit's $|0 \rangle$ and $|1 \rangle$ states).
$T_{\mathrm{p}}$ at which the state is driven from $|0 \rangle$ to $|1 \rangle$ for the first time gives a $\pi$ pulse (a $\pi$ rotation in the Bloch sphere).
Setting a $\pi$/2 pulse length as a half of the $\pi$ pulse length,
we create a superposition of $|0 \rangle$ and $|1 \rangle$ with a $\pi$/2 pulse.

\begin{figure}
\begin{center}
\includegraphics{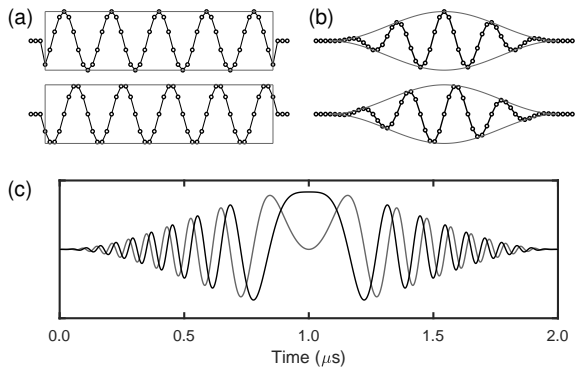}
\caption{(a) The waveforms of I (top) and Q (bottom) signals for a square pulse with $T_{\mathrm{p}}$ = 48~ns.
$f_{\mathrm{IF}}$ = 100~MHz and the sampling rate is 1~GS/s.
The gray lines indicate the pulse envelope.
(b) Same as (a) for a cosine-square pulse.
(c) The waveforms of I (black) and Q (gray) signals for a WURST pulse with $T_{\mathrm{w}}$ = 2~$\mu$s and $w$ = 2.
The frequency is chirped from $-$10~MHz to 10~MHz.
}
\label{fig4}
\end{center}
\end{figure}

As discussed above, the pulse shape and length are defined by $A(t)$ of the IQ signals, and we use either a square envelope or a cosine-square envelope.
We set $f_{\mathrm{IF}}$ = 100~MHz, and use 1~GS/s, giving 10 points in one cycle of the oscillation.
Exemplary IQ waveforms are shown in Figs.~\ref{fig4}(a) and (b).
For both pulse shapes, we define $T_{\mathrm{p}}$ as the time between the rising and falling edges of the pulses, i.e., the envelope of the cosine-square pulse is given as $\cos^2(2\pi t/T_{\mathrm{p}})$.
We use the square pulses for $\pi$/2 pulses and the cosine-square pulses for $\pi$ pulses.
The reason for these choices will be explained at the end of this section.
Concrete examples of the Rabi oscillations will be presented in Sec.~\ref{sec_single}.

Having determined the pulses lengths, we start with two most basic pulse sequences: Ramsey interferometry and Hahn echo.~\cite{R50,E50}
The pulse sequence for the Ramsey interferometry is given by 
\begin{equation}
X/2\mathrm{-}\tau\mathrm{-}X/2,
\end{equation}
and that for the Hanh echo as
\begin{equation}
X/2\mathrm{-}\tau\mathrm{-}X\mathrm{-}\tau\mathrm{-}X/2.
\label{eq_hahn_sq}
\end{equation}
Here, $X$/2 and $X$ denote $\pi$/2 and $\pi$ pulses about the $x$ axis of the rotating frame, respectively.
In both cases (and the rest of the sequences given below),
the 532-nm optical pulses are applied before and after the microwave pulse sequences.
In the Ramsey interferometry, the first $X$/2 pulse creates spin coherence along the $y$ axis of the rotating frame.
The NV spin then evolves freely during the time $\tau$, and is brought back to the $z$ axis of the rotating frame by the second $X$/2 pulse.
Here, the microwave frequency ($f_{\mathrm{LO}} + f_{\mathrm{IF}}$) does not have to be tuned exactly at the resonance frequency of the NV spin, but can be detuned by a few MHz.
The signal is seen to oscillate at the detuned frequency, and the resulting oscillation is called the Ramsey fringe.
In practice, the detuning is inevitable due to the presence of the hyperfine interaction with the nitrogen nuclear spin inherent to the NV center.
The two stable isotopes of nitrogen, $^{14}$N (99.6\%) and $^{15}$N (0.04\%), have nuclear spins $I$ = 1 and $\frac{1}{2}$, respectively.
The hyperfine interaction between the NV electronic spin and the $^{14}$N ($^{15}$N) nitrogen nuclear spin results in the triplet (doublet) spectrum separated by 2 (3)~MHz, which are simultaneously excited by fast, broadband pulses.

The Ramsey interferometry is subject to the temporal fluctuation of the magnetic environment of the NV spin, resulting in a fast decay of the Ramsey fringe upon time averaging.
This decay time scale is called the dephasing time and is denoted as $T_2^*$.
In the case of natural abundance diamond, where $^{13}$C isotopes with $I = \frac{1}{2}$ occupy 1.1\% of the lattice sites (the rest is occupied by $^{12}$C isotopes with $I$ = 0),
$T_2^*$ is typically on the order of a few $\mu$s.
If the fluctuation is slow compared with $\tau$, and is regarded as quasistatic, there is a way to counter the effect of the fluctuation.
The Hahn echo realizes this by adding the $X$ pulse in the middle and thereby refocuses the evolution during the first half of $\tau$.
The Hahn echo often improves the spin coherence by a few orders of magnitude, and its decay time scale is called the coherence time $T_2$.
Concrete examples of the Ramsey interferometry and the Hahn echo will be presented in Sec.~\ref{sec_single}.
There, we will observe that a longer time scale allowed by the Hahn echo reveals complex dynamics due to the interaction between the NV electronic spin and the environmental $^{13}$C nuclear spins.

The refocusing $X$ pulse can be repeated many times.
In fact, such multipulse sequences realize more general control of qubit--environment dynamics, called dynamical decoupling, and are used to extend qubit coherence times.~\cite{dLWR+10,BPB+12,BPJ+13,ACB+18}
Many of them share a common form
\begin{equation}
X/2\mathrm{-}(\cdots)_{N/k}\mathrm{-}X/2,
\end{equation}
where $(\cdots)_{N/k}$ denotes the repetition of $N/k$ times with $k$ the number of $\pi$ pulses in the repetition block and $N$ the total number of $\pi$ pulses. 
For instance, the Carr-Purcell (CP) and Carr-Purcell-Meiboom-Gill (CPMG) sequences repeat the following sequences $N$ times (with $k$ = 1):
\begin{eqnarray}
\mathrm{CP} &=& \tau/2\mathrm{-}X\mathrm{-}\tau/2 \\
\mathrm{CPMG} &=& \tau/2\mathrm{-}Y\mathrm{-}\tau/2,
\end{eqnarray}
where $Y$ is the $\pi$ pulse about the $y$ axis of the rotating frame.~\cite{CP54,MG58}

More complex multipulse sequences are a family of XY sequences.~\cite{GBC90}
The repetition blocks for XY$k$, with $k$ = 4, 8, 16, are given by
\begin{eqnarray}
\mathrm{XY4} &=& \tau/2\mathrm{-}X\mathrm{-}\tau\mathrm{-}Y\mathrm{-}\tau\mathrm{-}X\mathrm{-}\tau\mathrm{-}Y\mathrm{-}\tau/2 \\
\mathrm{XY8} &=& \mathrm{XY4}\mathrm{-}\mathrm{YX4} \\
\mathrm{XY16} &=& \mathrm{XY8}\mathrm{-}\mathrm{\bar{X}\bar{Y}8},
\end{eqnarray}
where $\bar{X}$ and $\bar{Y}$ are the $\pi$ pulses about $-x$ and $-y$ axes of the rotating frame, respectively, and the definitions of YX4 and $\mathrm{\bar{X}\bar{Y}8}$ follow logically.
When XY$k$ is repeated $N/k$ times in the sequence, we denote XY$k$-$N$.

From a different perspective, these periodic pulse sequences can be viewed as a filter function primarily sensitive to the frequency component at $(2\tau)^{-1}$.
For AC magnetometry, we repeat a sequence as incrementing $\tau$.
When $\tau$ matches with the half period of an oscillating field, dynamical decoupling fails and the loss of coherence signals the presence of a magnetic field oscillating at $(2\tau)^{-1}$ .
The spectral resolution of AC magnetometry (i.e., how long we can sample an oscillation) is limited by $T_2$,
but multipulse sequences for AC magnetometry simultaneously achieve dynamical decoupling, extending $T_2$.

We will make a frequent use of correlation spectroscopy, the sequence of which is given by
\begin{equation}
X/2\mathrm{-}(\mathrm{\cdots})_{N/k}\mathrm{-}Y/2\mathrm{-}t_{\mathrm{corr}}\mathrm{-}Y/2\mathrm{-}(\mathrm{\cdots})_{N/k}\mathrm{-}X/2.
\end{equation}
$\tau$ in the $(\cdots)_{N/k}$ sequence is fixed at the value where the signal is observed in multipulse sequences, and $t_{\mathrm{corr}}$ is swept.
The first block of the pulse sequence start with the $X$/2 pulse, creating coherence along the $y$ axis of the rotating frame, and ends with the $Y$/2 pulse.
This means that the $Y$/2 pulse projects the angle between the $y$ axis and the Bloch vector of the sensor, $\varphi \approx \sin \varphi$, onto the $z$ axis.
The information on $\varphi$ is stored along the $z$ axis during $t_{\mathrm{corr}}$, and is retrieved back to the $xy$ plane by the $Y$/2 pulse for further sensing.
In this way, $t_{\mathrm{corr}}$ can be extended up to the limit of spin relaxation time (denoted as $T_1$), which is usually much longer than $T_2$ in this system.
Correlation spectroscopy thus achieves higher resolution compared with standard $T_2$-limited AC magnetometry.~\cite{LDB+13,KSD+15,SRP+15,BCA+16}
A variant of correlation spectroscopy is to apply $\pi$ pulses $M$ times instead of simply waiting $t_{\mathrm{corr}}$:
\begin{equation}
X/2\mathrm{-}(\mathrm{\cdots})_{N/k}\mathrm{-}Y/2\mathrm{-}(\cdots)_{M}\mathrm{-}Y/2\mathrm{-}(\mathrm{\cdots})_{N/k}\mathrm{-}X/2.
\end{equation}
Correlation spectroscopy turns out to be a powerful tool to analyze the hyperfine parameters of a single nuclear spin (Sec.~\ref{sec_13c}).

Finally, we comment on some of the further details of our approach to pulsed experiments.
First, the waveform length of our AWG goes up to 32.4 $\times$ 10$^6$ points, and with 2 channels of 1~GS/s sampling, it covers 16.2~ms,
meaning that the AWG can generate not only the individual pulses, but the entire sequence.
For a long sequence, a majority of them could be zeros (free evolutions), but by generating the entire sequence as a single waveform, we do not have to worry about the timing control during the sequence.

Second, for the same $T_{\mathrm{p}}$, the pulse area is larger for the square pulse, and thus a $\pi$ pulse is achieved in a shorter pulse duration.
We use square pulses for $\pi$/2 pulses to create and retrieve coherence, for which short (ideally instantaneous) pulses are desired.
On the other hand, the cosine-square (or other shaped) pulses can overcome the timing resolution limited by the sampling rate of the AWG,~\cite{ZSC+17}
and are advantageous for $\pi$ pulses constituting multipulse sequences.
As mentioned above, for AC magnetometry the condition $f_{\mathrm{ac}} \approx (2\tau)^{-1}$ must be satisfied.
At high $f_{\mathrm{ac}}$ and with an increased pulse number, the total time spent for $\pi$ pulses become non-negligible.
It is thus convenient to define $N\tau$ as the time from the spin coherence is created till it is retrieved,
i.e., $\tau$ is defined as the time between the center positions of the adjacent $\pi$ pulses or the falling edge of the $\pi$/2 pulse to the center of the $\pi$ pulse.
Although $\tau$ is commonly referred to as ``free evolution time'' or ``interpulse delay'', this is not strictly correct in our definition.

Third, pulsed experiments often take a few hours or even longer than a day for signal accumulation, during which the laser light must be kept focused onto the target NV center.
We use a {\it tracking} technique whereby a narrow-range PL image is taken at regular time interval of $t_{\mathrm{track}}$ (typically set to be 5~min) and the focal position is readjusted to the brightest pixel.
Our AWG cannot handle this transition between repeated pulse sequences and PL imaging, as $t_{\mathrm{track}}$ is much longer than the maximum waveform length.
Instead, the 2-in-1-out switch (\#23) controlled by the DAQ (\#19) is utilized (Fig.~\ref{fig3}).
During repeated pulse sequences, the AWG sends TTL signals to open and close the AOM.
For PL imaging, this control signal line is switched to the DC source, making the AOM always open,
and a small volume (typically 0.5 $\times$ 0.5 $\times$ 3 $\mu$m$^3$) is scanned.
Note that microwave pulses continue to be generated and applied during the tracking procedure,
as the switching happens every $t_{\mathrm{track}}$ that is not synchronized with the timing of pulse sequences.
This does not affect the imaging quality, because of the continuous illumination of the laser light and the small duty cycle of the microwave pulses.
As mentioned earlier in this section, the photon counting events are time-tagged.
The photon counts recorded during PL imaging are correlated with the positions of the linear stage and are distinguished from the signals from pulsed experiments, which otherwise have been obtained.

\section{Operation\label{sec_operation}}
\subsection{Sample\label{sec_sample}}
Three diamond samples were prepared from CVD-grown, (001)-faced, type-IIa substrates (Element Six) with the dimension of 2 $\times$ 2 $\times$ 0.5~mm$^3$.
These substrates are specified to contain less than 0.03~ppb NV centers, corresponding to $<$ 5 $\times$ 10$^{12}$~cm$^{-3}$ or $<$ 5 NV centers/$\mu$m$^3$
(the atomic density of diamond is 1.77 $\times$ 10$^{23}$~cm$^{-3}$).
The actual NV densities are often smaller than the specified values, to the level where single NV centers are resolvable optically. 
Sample \#1 is as delivered and single NV centers contained in bulk are to be measured.

In Sample \#2, the entire surface was implanted by $^{14}$N$^+$ ions with the energy of 10~keV and the dose of 10$^{11}$~cm$^2$, producing ensemble NV centers.
Sample \#3 was ion-implanted together with Sample \#2, but with SiO$_2$ films deposited on the surface prior to the implantation.
This method generates single NV centers close to the diamond surface, which we use to detect proton nuclear spins placed on the diamond surface. 
The detail of the ion implantation with SiO$_2$ films is described in Ref.~\onlinecite{ISS+17}.

\subsection{PL imaging, CW ODMR, and calibration of $B_0$\label{sec_ensemble}}
We begin with ensemble NV centers.
Sample \#2 is mounted on Antenna \#2.
The purpose here is to check the basic performance of our setup.
In particular, we have to make sure that our designs of microwave antennas and magnets are working well, prior to examining single NV centers.

\begin{figure}
\begin{center}
\includegraphics{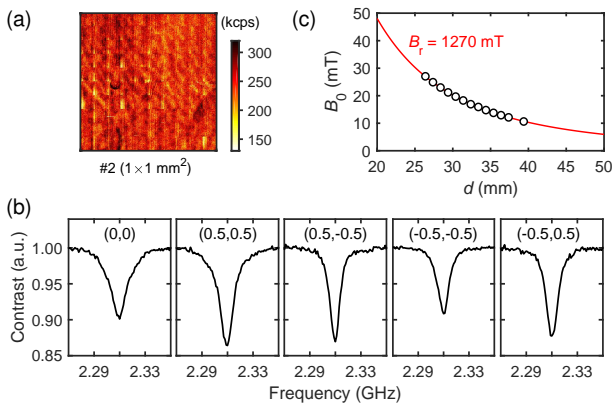}
\caption{(a) PL image of the surface of Sample \#2 with ensemble NV centers.
(b) ODMR spectra at different positions.
The origin $(0,0)$ is the center of the image in (a).
The length unit is in mm.
(c) Experimentally determined $B_0$ as a function of $d$.
The solid line is a fit by Eq.~(\ref{eq_b0}).
}
\label{fig5}
\end{center}
\end{figure}

One of preferable features of our setup is its wide lateral ($xy$) scan range.
This is demonstrated in Fig.~\ref{fig5}(a), showing a 1 $\times$ 1~mm$^2$ area of the diamond surface (at $B_0$ = 0~mT).
The presence of ensemble NV centers is confirmed by high photon counts (150--300~kcps) in the entire surface.
To obtain this image, the $z$ scan is also performed at each pixel to account for the tilt of the sample with respect to the laboratory frame.
We find that the $z$ position changes by $-$9.039~$\mu$m ($-$4.001~$\mu$m) along the $x$ ($y$) direction, corresponding to the tilt angle of $-$0.52$^{\circ}$ ($-$0.23$^{\circ}$).

Figure~\ref{fig5}(b) shows CW ODMR spectra taken at five different positions in Fig.~\ref{fig5}(a), one at the center $(0, 0)$ and four at the corners $(\pm0.5, \pm0.5)$.
We note that the position $(0, 0)$ coincides with the center of the 1-mm-diameter hole of the antenna.
As mentioned in Sec.~\ref{sec_nv_center}, the photon counts are spin-state-dependent, and the photon counts are reduced when the microwave frequency matches with the $m_S = 0 \leftrightarrow -1$ transition. 
$B_0$ is aligned along one of the NV axes with the strength of 20~mT (by Magnet \#2).
To achieve this, the following procedure was taken.
We first set the actuator position, which sets the field strength, and obtain an ODMR spectrum.
We then rotate the magnet using the rotary stage and take another ODMR spectrum.
We repeat the process until the resonance dip becomes the lowest frequency.
After alignment is done, we change the actuator position and take a series of ODMR spectra to calibrate the relation between $B_0$ and $d$.

The result for the Magnet \#2--Antenna \#2 pair is shown in Fig.~\ref{fig5}(c).
The solid line is a fit by Eq.~(\ref{eq_b0}) with $B_{\mathrm{r}}$ as the only fitting parameter.
We obtain $B_{\mathrm{r}}$ = 1270~mT, fully consistent with our design in Sec.~\ref{sec_magnet}.
The same calibration is conducted for the Magnet \#1--Antenna \#1 pair.
In this case, lower magnetic fields allow for the observation of the two resonances corresponding to the $m_S = 0 \leftrightarrow \pm1$ transitions simultaneously [Fig.~\ref{fig6}(b)].
The use of the information on the $m_S = 0 \leftrightarrow 1$ transition complements the smaller change of the magnetic fields as moving the positions of the actuator or rotary stage.

The successful observation of ODMR spectra from the positions separated by 1~mm demonstrates that the microwave fields from Antenna \#2 is distributed in the large area.
The variations in the width and contrast of the resonance lines can be attributed to two effects.
One is that the quality and distribution of the NV centers are not uniform across the sample, as also evidenced by the spatially inhomogeneous photon counts in Fig.~\ref{fig5}(a).
The other effect arises from the property of the antenna itself.
It is seen that the ODMR spectrum at the center is broader compared with the rest of the spectra.
This suggests that, under the fixed microwave input power, the microwave field is strongest at center, causing power broadening, while other positions experience comparatively weaker fields, yet enough to induce the spin resonance.
This is consistent with the fact that the four corners are slightly outside of the hole of the antenna.

\subsection{Basic test with a single NV center\label{sec_single}}
\begin{figure}
\begin{center}
\includegraphics{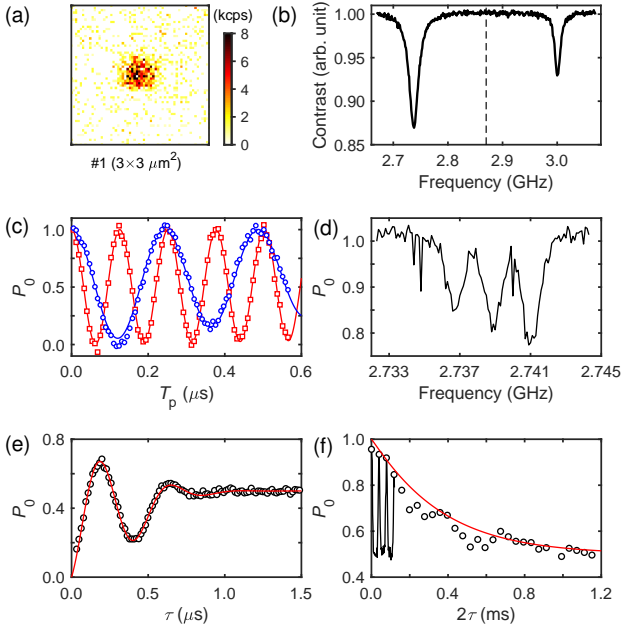}
\caption{(a) PL image of a single NV center.
(b) CW ODMR at $B_0$ = 4.6~mT.
The dashed vertical line indicates $D$ = 2.87~GHz.
(c) Rabi oscillations with the square ($\square$) and cosine-square ($\bigcirc$) pulses.
The solid lines are the fits by cosine oscillations superposed with the $e^{-T_{\mathrm{p}}/C}$ factor on the bottom side.
(d) Pulsed ODMR.
(e) Ramsey fringe.
The solid line is the fit given by Eq.~(\ref{ramsey_fit}).
(f) Hahn echo decay.
The solid line is the fit given by Eq.~(\ref{echo_fit}).
}
\label{fig6}
\end{center}
\end{figure}

We now examine the basic operations of our setup by carrying out standard characterization of single NV centers in diamond.
Figure~\ref{fig6} summarizes the results on Sample \#1 mounted on Antenna \#1.
An exemplary photoluminescence (PL) image from scanning confocal microscopy is shown in Fig.~\ref{fig6}(a).
It is the lateral scan of a 3 $\times$ 3 $\mu$m$^2$ area at a depth of 40~$\mu$m beneath the diamond surface.
The bright spot at the center of the image indicates the presence of the NV center.
It should however be noted that the size of the NV center itself is by far smaller than the fluorescent spot, and it is possible that multiple NV centers exist in a single spot.
A convincing way to confirm that a given fluorescent spot contains only one single-photon emitter is to take photon statistics and observe an antibunching behavior.
This can be done by slightly modifying our setup (Sec.~\ref{sec_photon}), but requires an additional SPCM, a fiber beam splitter and a timing analyzer.
Diffraction-unlimited super-resolution imaging, as mentioned in Sec.~\ref{sec_optics}, is another approach but is technically more demanding.
Empirically, we know typical photon counts we get from a single NV center under given conditions (laser power, depth, and so on), and if there are multiple, $k$ = 2, 3,$\cdots$, emitters in a single spot, the photon counts increases roughly $k$ times.
In addition, unless all the NV centers in a spot share the same NV axes (with probability 0.25$^{k-1}$), multiple resonances due to nondegenerate $m_S = 0 \leftrightarrow -1$ transitions will be observed in the ODMR spectrum.

Figure~\ref{fig6}(b) shows the CW ODMR spectrum on this spot.
The resonance frequencies of the two dips, 2.738~GHz and 3.001~GHz, are $-$132~MHz and 131~MHz away from $D$ = 2.870~GHz,
corresponding to the $m_S = 0 \leftrightarrow -1$ and $0 \leftrightarrow 1$ transitions, respectively.
The equal separation from $D$ ensures that the magnetic field direction (of Magnet \#1) is correctly aligned with the NV axis under the field strength of 4.7~mT.
It is observed that the $m_S = 0 \leftrightarrow -1$ resonance exhibits a larger contrast and a broader linewidth compared with the $m_S = 0 \leftrightarrow 1$ counterpart.
This is due to the $S_{11}$ property of Antenna \#1 [Fig.~\ref{fig2}(b)];
under the same microwave input power, a more power is absorbed at 2.738~GHz, causing a power broadening of the line but giving higher contrast.

Figure~\ref{fig6}(c) shows Rabi oscillations with square ($\square$) and cosine-square ($\bigcirc$) pulses, wherein the microwave frequency is set at the $m_S = 0 \leftrightarrow -1$ resonance of 2.7375~GHz.
The vertical axis is given as $P_0$, the probability of the final state being $m_S$ = 0.
It should be noted that when calculating $P_0$ the information of the photon counts from CW measurements cannot be used.
This is because in the CW mode the NV spin is continuously repumped while in the pulsed mode the optical pulse and the microwave control are temporally separated.
The photon counts from the $m_S$ = 0 state is straightforward to determine (the photon counts right after the optical pulse).
On the other hand, the photon counts from the $m_S$ = $-$1 state must be determined independently of the Rabi measurement, as there is no guarantee a priori that the applied microwave pulses can accurately rotate the NV spin.
To robustly flip the NV spin from laser-initialized $m_S$ = 0 to $m_S$ = $-$1, we use a chirped microwave pulse known as WURST (wideband, uniform rate, smooth truncation).~\cite{KF95}
The amplitude modulation of WURST is given by
\begin{equation}
1- \left| \sin \left( 2 \pi \frac{t}{T_{\mathrm{w}}} \right) \right|^{w},
\end{equation}
where we set $T_{\mathrm{w}}$ = 2 $\mu$s and $w$ = 2.
The frequency is chirped linearly from $-$10 to 10~MHz around the resonance frequency during the pulse.
The IQ signals are shown in Fig.~\ref{fig4}(c).
By slowly and widely sweeping the microwave frequency across the resonance, the WURST pulse adiabatically flips the target spin.
Prior to each pulse sequence, we record the reference photon counts with and without a chirped pulse and use them to calculate $P_0$.

Figure~\ref{fig6}(c) demonstrates clear oscillations for the two pulse shapes, but their frequencies (Rabi frequencies) are markedly different.
This difference is not essential and is due to the definition of $T_{\mathrm{p}}$ (Sec.~\ref{sec_electronics}).
Under the normalized microwave power, the pulse area of the square pulse is twice larger than that of the cosine-square pulse, accounting for the difference in the Rabi frequencies.
More importantly, it is observed that, as increasing $T_{\mathrm{p}}$, the oscillation minima (rotations by odd multiple of $\pi$) gradually deviate from $P_0$ = 0.
This is because the pulse bandwidth becomes narrower for longer $T_{\mathrm{p}}$, and the narrowband pulse can no longer excite the whole resonance line.
As the bandwidth of the cosine-square pulse is narrower for the same $T_{\mathrm{p}}$, this effect is seen to be more pronounced.
On the other hand, when full rotations (even multiple of $\pi$) are realized, the electronic spin always ends up in the $m_S$ = 0 state.
The oscillation maxima are thus not affected by the pulse bandwidth.
From Fig.~\ref{fig6}(c), we determine $T_{\mathrm{p}}$ for $\pi$/2 ($\pi$) with the square (cosine-square) pulse to 31.5~ns (122.1~ns).
These values will be used in multipulse experiments.

While fast, broadband microwave pulses are building blocks for time-domain experiments, we can use a slow, narrowband microwave pulse to perform pulsed ODMR.
In Fig.~\ref{fig6}(d), the length of the $\pi$ pulse is set at 1292~ns and the microwave frequency is swept across the resonance.
The observed three dips correspond to the $^{14}$N nuclear spin being in the $m_I$ = $-$1, 0, and 1 states in ascending order.
The observed linewidths of about 1~MHz are narrower than the pulse bandwidth,
and the adjacent $m_I$ = $-$1 and 0 (0 and 1) resonances are separated by 2.18~MHz (2.14~MHz), consistent with the known hyperfine parameter of the $^{14}$N nuclear spin.
These fine structures were masked in the CW ODMR spectrum of Fig.~\ref{fig6}(b), in which the resonance width of the $m_S = 0 \leftrightarrow -1$ transition is as broad as 20~MHz,
due to the microwave power broadening.

Figure~\ref{fig6}(e) shows a Ramsey fringe with the microwave frequency set at the $m_I$ = 0 resonance.
Consecutive $X$/2 pulses are equivalent to an $X$ pulse, so the fringe starts at $P_0$ = 0 ($m_S$ = $-$1).
For longer $\tau$, the spin state is completely randomized and the fringe converges to $P_0$ = 0.5.
The curve is fitted by
\begin{equation}
P_0(\tau) = \frac{1}{2} - e^{-(\tau/T_2^*)^p} \left[ \alpha_0 + \alpha_1 \cos (2 \pi \alpha_2 \tau + \alpha_3) \right],
\label{ramsey_fit}
\end{equation}
for which we obtain $T_2^*$ = 0.50~$\mu$s with the stretched exponent $p$ of 2.01 (i.e., Gaussian decay).
$\alpha_0$ ($\alpha_1$) is the portion of the signal arising from the $m_I$ = 0 ($\pm$1) state(s),
and indeed the fitted value of $\alpha_2$ = 2.1~MHz agrees with the $^{14}$N hyperfine splitting observed in pulsed ODMR.

To obtain Fig.~\ref{fig6}(e), we also ran a sequence in which the readout $X/2$ is replaced by $\bar{X}/2$.
The sequence with $\bar{X}/2$ should produce an inverted echo decay curve, and the both decay curves should converge to $P_0$ = 0.5.
The two data are subtracted and averaged.
This procedure, called phase cycling, serves to ensure that all the experimental parameters are properly set.
Large errors in the pulse rotation angles, short initialization time, and incorrect setting of the readout duration are among typical causes for the two decay curves to become asymmetric.
The phase cycling is used in all the multipulse experiments presented in this article.

We go on to perform a Hahn echo experiment [Fig.~\ref{fig6}(f)].
It is observed that the echo signal decays to $P_0$ = 0.5 within 20~$\mu$s, and grows back to $P_0 \approx$ 1 in the next 20~$\mu$s.
This collapse and revival behavior repeats many times.
The first three recurrences are fully shown, and after that, only the peak positions are acquired ($\bigcirc$).
The recurrence is due to the interaction with the $^{13}$C nuclear spin bath in diamond.~\cite{CDT+06,MTL08}
At $B_0$ = 4.7~mT, $^{13}$C nuclear spins precess at $f_{\mathrm{c}}$ = $\gamma_{\mathrm{c}} B_0$ = 50.3~kHz,
where $\gamma_{\mathrm{c}}$ = 10.705~kHz/mT is the gyromagnetic ratio of the $^{13}$C nuclear spin.
It is found that the revival period matches with twice the $^{13}$C precession period (20.9~$\mu$s).
During the free evolution, the NV electronic spin is in a superposition of the $m_S$ = 0 ($|0 \rangle$) and $-$1 ($|1 \rangle$) states, which couple differently to $^{13}$C nuclear spins.
The $m_S$ = 0 state is nonmagnetic and the $^{13}$C nuclear spins precess at their bare Larmor frequency,
whereas the hyperfine interactions with the $m_S$ = $-$1 state result in nuclear precession frequencies being slightly different for individual nuclear spins.
The refocusing $\pi$ pulse exchanges $|0 \rangle$ and $|1 \rangle$, and different evolutions of the nuclear spins before and after the $\pi$ pulse cause the decoherence of the NV electronic spin.
When the nuclear spins make integer multiple of the bare Larmor precession during $\tau$, the evolution of the nuclear spin bath before and after the $\pi$ pulse cancels exactly, leading to the echo revivals.

Other effects, such as dipolar interactions among nuclei, make this cancellation incomplete and the peak positions decay over time.
We fit this envelope decay with stretched exponential of the form
\begin{equation}
P_0(\tau) = \frac{1}{2} \left[ 1 + e^{-(2\tau/T_2^*)^p} \right],
\label{echo_fit}
\end{equation}
and obtain $T_2$ = 364~$\mu$s ($p$ = 1.06), much longer than $T_2^*$ obtained from the Ramsey fringe.
There are additional modulations superposed to the decay envelope, with broad dips at around 0.3 and 0.5~ms.
These features are likely due to the interaction with clusters of $^{13}$C nuclei.~\cite{MTL08,ZHH+11}

The results presented in this section clearly and sufficiently demonstrate that our compact setup fully functions as an ODMR spectrometer for single NV centers.
We thus move on to demonstrate the ability of detecting single and ensemble nuclear spins using a single NV electronic spin as a quantum sensor.

\subsection{Quantum sensing of $^{13}$C nuclei in diamond\label{sec_13c}}
Using the single NV center found and characterized in Sec.~\ref{sec_single}, we carry out quantum sensing of $^{13}$C nuclei in diamond.
As mentioned above, uncoupled, bare $^{13}$C nuclear spins precess at $f_{\mathrm{c}}$ = 50.3~kHz at $B_0$ = 4.7~mT.
On the other hand, when $^{13}$C nuclei are located relatively close to the NV center,
the hyperfine interaction acts as an effective magnetic field added to $B_0$ and their precession frequencies are spectrally separated from the bath nuclei.

\begin{figure}
\begin{center}
\includegraphics{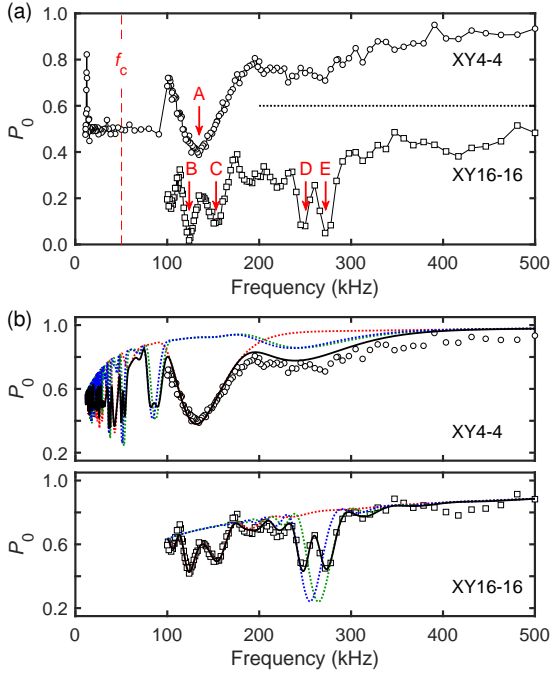}
\caption{(a) NMR spectra of $^{13}$C nuclei in diamond at $B_0$ = 4.7~mT taken by XY4-4 ($\bigcirc$) and XY16-16 ($\square$, shifted downward by 0.4).
(b) Comparison of experimental and theoretical NMR spectra.
The dotted lines are the spectra of the individual nuclei (red for A--C, blue for D, green for E).
}
\label{fig7}
\end{center}
\end{figure}

The NMR spectra obtained by XY4-4 are shown as circle points ($\bigcirc$) in Fig.~\ref{fig7}(a).
The sequence is repeated as incrementing $\tau$, and $P_0$ at respective $\tau$ are plotted as a function of $(2\tau)^{-1}$.
In the spectrum by XY4-4, the signals around $f_{\mathrm{c}}$ saturate at $P_0 \approx$ 0.5.
This is characteristic of AC signals with random amplitude and random phase,~\cite{DRC17}
in this case manifesting weakly-coupled nuclear spin bath.
On the other hand, we observe a dip at 134.41~kHz (marked as A) that goes below 0.5, indicative of an isolated single nuclear spin, which is coherently coupled with the NV center.
We further examine the region $\geq$100~kHz by applying XY16-16 ($\square$).
The spectrum changes dramatically, and we now observe four dips marked as B (123.76~kHz), C (152.43~kHz), D (250.0~kHz), and E (271.74~kHz), with additional small dips around 200~kHz.
In the below, we analyze these signals in detail, and determine the hyperfine parameters of the $^{13}$C nuclear spin responsible for the spectra.
Importantly, we show that the seemingly different spectra obtained by XY4-4 and XY16-16 in fact originate from the same set of nuclear spins and are explained consistently.
Central to this analysis is correlation spectroscopy.

\begin{figure}
\begin{center}
\includegraphics{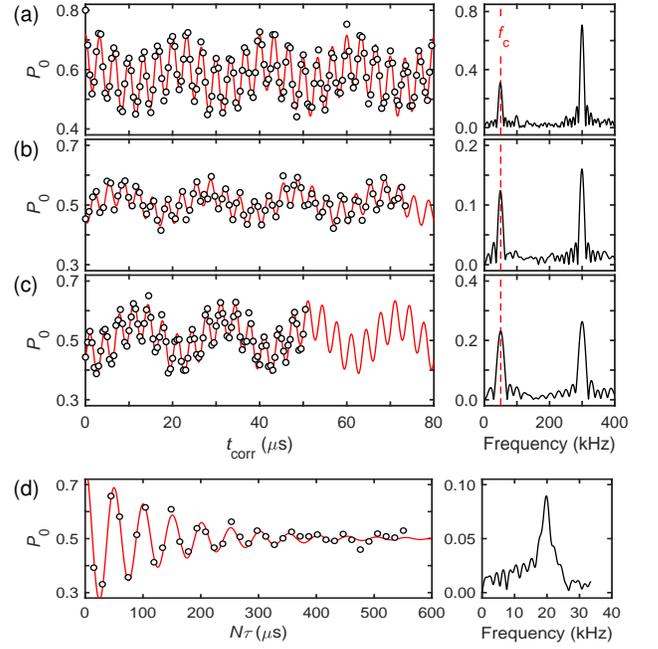}
\caption{(a) Correlation spectroscopy at A by XY4-32 with $\tau$ = 3.72~$\mu$s = (2 $\times$ 134.41~kHz)$^{-1}$.
FFT shows peaks at $f_0^{(\mathrm{A})}$ = 50~kHz and $f_1^{(\mathrm{A})}$ = 300~kHz.
The solid line in the left figure is a fit by two sinusoidal oscillations with amplitudes and phases being the fitting parameters.
(b) Same as (a), except that the data were taken at B by XY16-16 with $\tau$ = 4.04~$\mu$s = (2 $\times$ 123.76~kHz)$^{-1}$.
FFT shows peaks at $f_0^{(\mathrm{B})}$ = 49~kHz and $f_1^{(\mathrm{B})}$ = 299~kHz.
(c) Same as (a), except that the data were taken at C by XY16-16 with $\tau$ = 3.28~$\mu$s = (2 $\times$ 152.43~kHz)$^{-1}$.
FFT shows peaks at $f_0^{(\mathrm{C})}$ = 50~kHz and $f_1^{(\mathrm{C})}$ = 300~kHz.
(d) $P_0$ at A as incrementing $N$ with $\tau$ fixed as 3.72~$\mu$s.
FFT shows a peak at $f_{\mathrm{r}}^{(\mathrm{A})}$ = 19.80~kHz.
The solid line in the left figure is a fit by a damped cosine oscillation with amplitude and decay constant as the fitting parameters.
}
\label{fig8}
\end{center}
\end{figure}

We first examine A.
Figure~\ref{fig8}(a) shows the result of correlation spectroscopy as a function of $t_{\mathrm{corr}}$ (left) and its fast Fourier transform (FFT) spectrum (right).
FFT reveals two peaks, one at $f_0^{(\mathrm{A})}$ = 50~kHz and the other at and $f_1^{(\mathrm{A})}$ = 300~kHz.
Similar to the echo revival, the origin of this two-component oscillation is rooted in the interaction with $^{13}$C nuclei, but now with individual nuclear spins.
Again, the nonmagnetic $m_S$ = 0 state does not affect the $^{13}$C nuclear spin precession, whereas the $m_S$ = $-$1 state modifies it.
We observe that $f_0^{(\mathrm{A})}$ matches with $f_{\mathrm{c}}$, suggesting that it arises from the coupling to the $m_S$ = 0 state.
It follows that the $m_S$ = $-$1 state is responsible for $f_1^{(\mathrm{B})}$.
Theoretically, $f_{0,1}$ correspond to the eigenvalues of the NV--$^{13}$C-coupled Hamiltonian:
\begin{equation}
H_{\mathrm{e}\otimes\mathrm{n}}/h = H_{\mathrm{{e}}} - \gamma_{\mathrm{c}} B_0 I_z + S_z (a_{\perp} I_x + a_{\parallel} I_z).
\label{eq_h_en}
\end{equation}
Here, $I_{x,z}$ are the $I = \frac{1}{2}$ spin operators, and $a_{\parallel,\perp}$ are the components of the hyperfine parameters parallel and perpendicular to $B_0$, respectively.
Solving Eq.~(\ref{eq_h_en}), we find that $f_{0,1}$ are given as $f_{\mathrm{c}}$ and $\sqrt{(f_{\mathrm{c}} + a_{\parallel})^2 + a_{\perp}^2}$, respectively.

The fact that the information on $a_{\parallel,\perp}$ is contained only in $f_1$ suggests that an additional measurement is necessary to determine both.
This is realized by measuring $P_0$ by incrementing $N$ (the number of the equally spaced $\pi$ pulses) in multipulse sequences.
For a nuclear spin coherently coupled with the NV electronic spin, $P_0$ as a function $N$ exhibits an oscillation that can be interpreted as a nuclear Rabi rotation around the $a_{\perp}$ axis.
The result is shown in Fig.~\ref{fig8}(d), with the oscillation frequency $f_{\mathrm{r}}^{(\mathrm{A})}$ = 19.80~kHz.
From the measured $f_{0,1}$ and $f_{\mathrm{r}}$, $a_{\parallel,\perp}$ are calculated as~\cite{TWS+12,KUBL12,BCA+16,SIA18}
\begin{eqnarray}
a_{\parallel} &=& \frac{\cos \phi_0 \cos \phi_1 - \cos \phi_{\mathrm{r}}}{\sin \phi_0 \sin \phi_1} f_1 - f_0
\label{eq_A_para} \\
a_{\perp} &=& \sqrt{f_1^2 - (f_0 + a_{\parallel})^2},
\label{eq_A_perp}
\end{eqnarray}
with $\phi_{0,1} = \pi f_{0,1} \tau$ and $\phi_{\mathrm{r}} = \pi - 2\pi f_{\mathrm{r}} \tau$.
Substituting the values of $f_0^{(\mathrm{A})}$, $f_1^{(\mathrm{A})}$, $f_{\mathrm{r}}^{(\mathrm{A})}$, and $\tau$ into Eqs.~(\ref{eq_A_para}) and (\ref{eq_A_perp}),
we obtain $a_{\parallel}$ = $-$226.2~kHz and $a_{\perp}$ = 242.8~kHz.

Although we have analyzed A in the XY4-4 spectrum successfully, it disappears in the XY16-16 spectrum and instead B and C appear on the lower and higher frequency sides, respectively.
It may be guessed that B and C have the same origin as A.
To confirm this experimentally, we repeat correlation spectroscopy on B and C.
The results are shown in Figs.~\ref{fig8}(b) and (c).
As expected, $f_{0,1}$ of B and C all appear at frequencies identical to those for A.
We conclude that A, B, and C originate from the same single $^{13}$C nuclear spin.

\begin{figure}
\begin{center}
\includegraphics{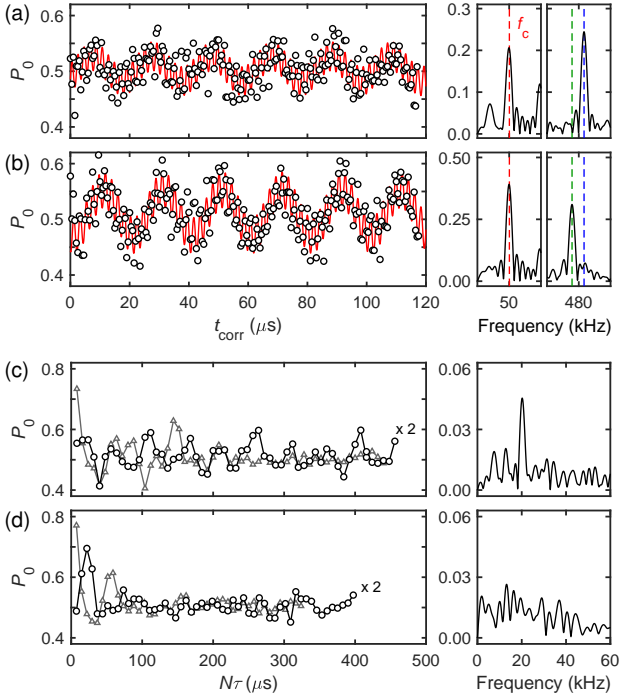}
\caption{(a) Correlation spectroscopy at D by XY16-16 with $\tau$ = 2.00~$\mu$s = (2 $\times$ 250.00~kHz)$^{-1}$.
FFT shows peaks at $f_0^{(\mathrm{D})}$ = 50~kHz and $f_1^{(\mathrm{D})}$ = 488~kHz.
(b) Same as (b), except that the data were taken at E with XY16-16 with $\tau$ = 1.84$\mu$s = (2 $\times$ 271.74~kHz)$^{-1}$.
FFT shows a peak at $f_0^{(\mathrm{E})}$ = 50~kHz and $f_1^{(\mathrm{E})}$ = 469~kHz.
(c) The points in $\triangle$ (gray) are obtained by the $N$ $\pi$-pulse sequence at D with $\tau$ fixed as 2.00~$\mu$s.
The points in $\bigcirc$ (black) are obtained by the $N$ $\pi$-sequence combined with correlation spectroscopy.
For clarity, the signal amplitude is multiplied by 2 with the baseline set at $P_0$ = 0.5.
The FFT is from the latter data.
FFT shows a peak at $f_{\mathrm{r}}^{(\mathrm{D})}$ = 20.3~kHz.
(d) Same as (c), except that the data were taken at E with $\tau$ fixed as 1.84~$\mu$s.
FFT shows no peak.
}
\label{fig9}
\end{center}
\end{figure}

Next, we analyze D and E.
Unlike B and C, the XY4-4 spectrum does not show a clear dip around 250--270~kHz,
providing no clue whether D and E originate from the same nuclear spin or have different origins.
We perform correlation spectroscopy on D and E for $t_{\mathrm{corr}}$ longer than 100~$\mu$s, so that the spectral resolution better than 10~kHz are achieved.
Figures~\ref{fig9} (a) and (b) show the results on D and E, respectively.
From the Fourier spectra, it is confirmed that the $f_1$ signals for D and E both appear at 50~kHz, exactly matching with the nuclear Larmor frequency as in the cases of A--C.
On the other hand, the $f_1$ signals for D and E appear at 488~kHz and 469~kHz, respectively, a difference larger than the spectral resolution.
We thus infer that D and E originate from different nuclear spins.
We further run a sequence as incrementing $N$ [$\triangle$ in Figs.~\ref{fig9} (c) and (d)].
In both cases, we observe only feeble oscillations.
This does not improve well even using correlation spectroscopy ($\bigcirc$).
The FFT spectrum of D shows a weak signal at 20.3~kHz, but no peak is observed in the FFT spectrum of E.
Relying on the FFT data of D, we obtain $a_{\parallel}$ = 357.0~kHz and $a_{\perp}$ = 270.2~kHz.

To determine the hyperfine parameters of E, we simulate the NMR spectra of Fig.~\ref{fig7}(a),
using the missing experimental input $f_{\mathrm{r}}^{(\mathrm{E})}$ as a fitting parameter.
The NMR signals from the individual nuclei are simulated as~\cite{TWS+12,KUBL12,BCA+16,SIA18}
\begin{equation}
P_0 = 1 - \left( \dfrac{ a_{\parallel} \sin \dfrac{\phi_0}{2} \sin \dfrac{\phi_1}{2} \sin \dfrac{N \phi_{\mathrm{r}}}{2} }{ f_1 \cos \dfrac{ \phi_{\mathrm{r}} }{2} } \right)^2,
\label{eq_p0_1}
\end{equation}
with
\begin{equation}
\cos \phi_{\mathrm{r}} = \cos \phi_0 \sin \phi_1 - \frac{ a_{\parallel} + f_0 }{ f_1 } \sin \phi_0 \sin \phi_1.
\end{equation}
The full spectrum is then calculated as 
\begin{equation}
P_0 = \frac{1}{2} + e^{-N\tau/T_2} \prod_i \left( P_0^{(i)} - \frac{1}{2} \right),
\label{eq_p0_2}
\end{equation}
where $P_0^{(i)}$ is $P_0$ simulated for $i$ = A, D, and E using Eq.~(\ref{eq_p0_1}),
and $T_2$ here is a fitting parameter that accounts for the effect of decoherence of the sensor spin.
We search the value of $f_{\mathrm{r}}^{(\mathrm{E})}$ that reproduces the XY16-16 spectrum in 100--500~kHz.
Once all the hyperfine parameters are determined, we simulate the XY4-4 spectrum in 0--500~kHz to check the consistency.
The results are shown in Fig.~\ref{fig7}(b), where the solid lines are the simulated full spectra [Eq.~(\ref{eq_p0_2})] and the dotted lines are the contributions from the respective nuclei [Eq.~(\ref{eq_p0_1})].
With $a_{\parallel}$ = 348.2~kHz and $a_{\perp}$ = 248.7~kHz for E, which are very different from those for D,
the spectra of both XY16-16 and XY4-4 are reproduced satisfactorily.
Notably, we can identify the origin of the small dips around 200~kHz of the XY16-16 spectrum as tails of B--E.
They arise due to the filter function that the periodic $\pi$ pulses form.~\cite{DRC17}
On the other hand, there is a small discrepancy between the simulation and the experiment at around 420~kHz.
Correlation spectroscopy by XY16-16 with $\tau$ = 1.20~$\mu$s = (2 $\times$ 416.67~kHz)$^{-1}$ did not show any signals, however.

From Fig.~\ref{fig7}(b), we can make a few instructive observations.
First and foremost, {\it appearances are deceiving}.~\cite{AS18}
Different multipulse sequences can give different spectra, even though they are applied to the same nuclear environment.
In addition, finite-length pulses can produce more complex structures.~\cite{LBR+15}
Therefore, correlation spectroscopy should be conducted to correctly interpret them.
We note that our simulation assumed infinitesimally short pulses.
This worked well in the present case, where the low frequency (long $\tau$) regime is probed.
Second, in XY16-16, we observe that the spectral overlap between D and E causes the signals to become weaker, due to Eq.~(\ref{eq_p0_2}).
The same mechanism is also responsible for feeble oscillations observed in Fig.~\ref{fig9}(b).
Third, the simulation of XY4-4 shows dense signals below 100~kHz.
These are fractional harmonic signals appearing at $1/3$ and $1/5$ of the original frequencies.
They have significant contributions to the spectrum below 100~kHz, where the bath nuclei with weak hyperfine parameters also exist.
Interestingly, we identify that the signals at 1/3 of frequencies of B and C are responsible for the sharp change observed at 100~kHz, a transition from the continuous spectrum to the discrete one (A).
Fourth, decoherence of the sensor spin causes $P_0$ not to return to 1.0 even in the absence of the signal.
This is more pronounced when $\tau$ becomes longer, i.e., at low frequencies and/or for XY16-16.

Finally, we have also performed a fitting assuming that D and E share the same hyperfine parameters.
Although a large parameter space was searched, the NMR spectra had never been reproduced as well as Fig.~\ref{fig7}(b).
We conclude confidently that the origin of D and E are different.

\subsection{Proton NMR\label{sec_proton}}
Quantum sensing of single nuclear spins, as demonstrated in Sec.~\ref{sec_13c}, holds great promise for structure analysis at the single molecular level.
Although outside of the scope of this article, the current setup is fully compatible with protocols to achieve sub-hertz spectral resolution.~\cite{SGS+17,BCZD17,GBL+18}
In addition, with a single-NV quantum sensor, we can utilize not only the spectral information but the spatial information to determine a molecular structure.
By localizing the positions of the individual nuclear spins in a molecule,~\cite{ZCS+18,ZHCD18,SIA18,ARB+19} the molecular structure emerges itself.
At present, experimental demonstrations are primarily performed on single $^{13}$C nuclear spins in diamond as a testbed.
There are still only a handful of reports on the detection of single or a few {\it external} nuclear spins.~\cite{MKC+14,SLC+14}
This is, however, not due to the limitation of protocols but to the difficulty of having a single NV center very close to the surface and with high coherence.
The continued efforts and advancement have been made on this subject.~\cite{WJGM13,OHB+12,ORW+13,SBAP14,dOAW+17,SDS+19}
Therefore, the next important task of our nanoscale magnetometer is to detect a small ensemble of external nuclear spins.
The simplest approach is to replace the air objective lens with oil immersion one (Table~\ref{tab_a1} of Appendix~\ref{sec_list}) and detect proton spins in the oil.
For this purpose, we use Sample \#3, Antenna \#2, and Magnet \#2 with $B_0$ set to 23.5~mT. 
The corresponding $^1$H Larmor frequency is $f_{\mathrm{h}}$ = $\gamma_{\mathrm{h}} B_0$ = 1~MHz, where $\gamma_{\mathrm{h}}$ = 42.577~kHz/mT is the gyromagnetic ratio of the $^{1}$H nuclear spin.

\begin{figure}
\begin{center}
\includegraphics{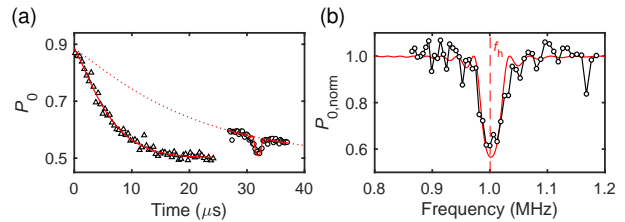}
\caption{
(a) Hahn echo decay ($\triangle$) and proton NMR with XY16-64 ($\bigcirc$).
(b) The normalized NMR spectrum.
}
\label{fig10}
\end{center}
\end{figure}

The top surface of Sample \#3 has a few-micron-thick CVD-grown, undoped $^{12}$C layer, in which single NV centers generated by $^{14}$N$^{+}$ ion implantation exist.~\cite{ISS+17}
The triangle points ($\triangle$) in Fig.~\ref{fig10}(a) are the Hahn echo decay curve.
The horizontal axis for the Hahn echo decay is given as 2$\tau$ with $\tau$ as defined in Eq.~(\ref{eq_hahn_sq}).
The fit with a stretched exponential decay gives $T_2$ = 6.2~$\mu$s and $p$ = 1.26.
The echo revival as observed in Fig.~\ref{fig6}(f) could appear at every 4.0~$\mu$s at this field strength, but is absent here owing to the use of $^{12}$C isotopes.
This short $T_2$ is typical of near surface NV centers, as mentioned above.
Nonetheless, multipulse sequences can extend the coherence.
We apply the XY16-64 sequence here.
For the proton precession frequency of 1~MHz, we expect the signal at around $N\tau$ = 32~$\mu$s [$N$ = 64, $(2\tau)^{-1}$ = 64/(2 $\times$ 32~$\mu$s) = 1~MHz],
and indeed observe a dip there ($\bigcirc$). 
To analyze the data, we define a decay envelope given as a stretched exponential decay as shown in the dashed line ($T_2$ = 16.1~$\mu$s and $p$ = 0.97).
The observed proton NMR signal $C(\tau)$ is expressed as~\cite{PDC+16}
\begin{eqnarray}
& & C(\tau) \approx \nonumber \\
& & \exp \left[ -8
(\gamma_{\mathrm{e}} B_{\mathrm{rms}} N\tau)^2
\mathrm{sinc}^2 \left\{\pi N \tau \left( f_{\mathrm{h}} - \frac{1}{2\tau} \right) \right\}
\right].
\label{eq_C}
\end{eqnarray}
Here, $B_{\mathrm{rms}}$ is the root-mean-square amplitude of AC magnetic field produced by the nuclear spins, given as
\begin{equation}
B_{\mathrm{rms}} = \frac{\mu_0}{4 \pi} h \gamma_{\mathrm{h}} \sqrt{ \frac{ 5 \pi \rho}{ 96 \, d_{\mathrm{NV}}^3 } }
\label{eq_b_rms}
\end{equation}
with $\rho$ the nuclear spin density and $d_{\mathrm{NV}}$ is the depth of the NV center from the surface.
For the oil we use (Olympus IMMOIL-F30CC), $\rho$ is known to be 6 $\times$ 10$^{28}$~m$^{-3}$.~\cite{LPMD14}
To derive Eq.~(\ref{eq_C}), the dephasing time of nuclear spins $T_{ \mathrm{2n} }^*$ is assumed to infinitely long (see Ref.~\onlinecite{PDC+16} for deviation and the expression for finite $T_{ \mathrm{2n} }^*$). 
Figure~\ref{fig10}(b) shows the {\it normalized} signal in the frequency unit after subtracting the decay envelope.
The fit (solid line) gives $d_{\mathrm{NV}}$ = 6.26~nm.
The depth information is crucial in studying the properties of shallow NV centers, and here proton NMR plays a vital role.

As demonstrated here, the principle of nanoscale NMR spectroscopy with a single NV center is quite different from that of conventional NMR.
The detection capability of the latter is governed by small thermal polarizations of nuclei and the sensitivity of inductive coils.
Various approaches are being pursued to enhance the nuclear polarizations statically (high magnetic fields $>$ 10~T and low temperatures) and dynamically (hyperpolarization)
and to reduce the analyte volume down to nanoliters.~\cite{LOC14,ABC+15}
In contrast, the present experiments were performed at low magnetic fields and room temperatures.
In addition, the detection volume of the sensor is on the order of $(d_{^\mathrm{NV}})^3 \approx$ 0.25~zL (zeptoliters).
The number of protons contained in this volume is $\rho(d_{^\mathrm{NV}})^3 \approx$ 1500.
The thermal nuclear polarization at room temperature and at low magnetic fields is on the order of 10$^{-7}$, and the number of polarized nuclear spins in this volume is much less than one.
In fact, what we relied on here is the statistical polarization, which is on the order of $\sqrt{1500} \approx$ 39 and exceeds the thermal polarization.
From Eq.~(\ref{eq_b_rms}), we estimate $B_{\mathrm{rms}} \approx$ 560~nT.
Still, in both conventional and nanoscale NMR, the magnetization signals from ensemble nuclei are detected.
To detect single nuclear spins, the coupling of the NV center to the target nuclear spin must be stronger than that to the ensemble spins, as in the case of $^{13}$C nuclei in diamond (Sec.~\ref{sec_13c}).~\cite{MKC+14}
To this end, creation of high-coherence shallow NV centers is crucial, as mentioned above.

\subsection{PL spectroscopy and photon correlation\label{sec_photon}}
In this section, we provide optical characterizations of NV centers obtained by interchangeably modifying the original setup.
As demonstrated so far, nanoscale magnetometry is fully possible without these measurements.
Nonetheless, the performance of the magnetometer critically depends on the optical properties of the NV centers and
the ability to characterize them within the same setup is highly advantageous.

\begin{figure}
\begin{center}
\includegraphics{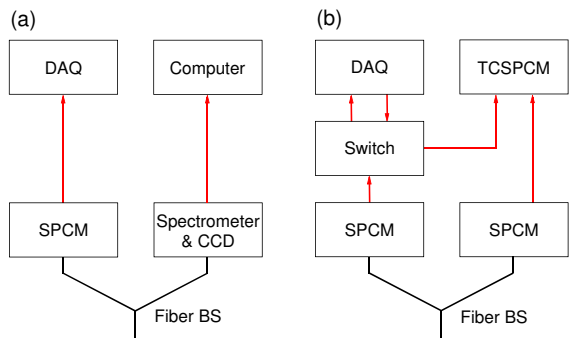}
\caption{Diagrams of the detection configurations modified for (a) PL spectroscopy and (b) photon correlation measurements.
See Table~\ref{tab_a6} of Appendix~\ref{sec_list} for the model numbers.
BS: beam splitter.
TCSPCM: time correlated single photon counting module.
}
\label{fig11}
\end{center}
\end{figure}

We use a double-grating spectrometer and a liquid-nitrogen-cooled CCD for spectroscopy.
The former is equipped with a fiber-coupled input port, so in principle all we have to do is to disconnect the output fiber from the SPCM, and connect it to the input port of the spectrometer.
However, this makes spectroscopy data susceptible to drift, as there is no mechanism to track the position of an NV center.
We utilize a fiber beam splitter (BS) to conduct both spectroscopy and tracking, as schematically shown in Fig.~\ref{fig11}(a).
Although the spectrometer and the tracking system (Sec.~\ref{sec_electronics}) are not synchronized, spectroscopy data are not affected by tracking owing to short times required for it.

\begin{figure}
\begin{center}
\includegraphics{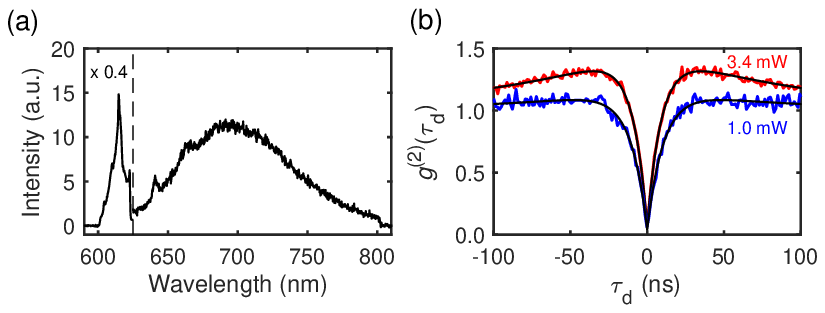}
\caption{(a) PL spectrum of a single NV center (625--800~nm), and the second-order Raman spectrum of diamond (600--625~nm).
The intensity of the latter is multiplied by 0.4.
(b) $g^{(2)}(\tau_{\mathrm{d}})$ of a single NV center with $P_{\mathrm{L}}$ = 1.0 and 3.4~mW.
The solid lines are the fits given by Eq.~(\ref{eq_g2}).
}
\label{fig12}
\end{center}
\end{figure}

Figure~\ref{fig12}(a) shows spectroscopy on the single NV center identical to the one we characterized in Sec.~\ref{sec_single}.
The weak zero-phonon line at around 640~nm and the broad phonon sideband extending up to 800~nm are observed.
In this range of wavelengths, the emission from neutral vacancies (V$^{0}$) can appear around 741~nm (called the GR1 line).~\cite{DLC+92}
The structure between 600~nm and 621~nm is the second-order Raman spectrum of diamond.
Considering the excitation wavelength of 532~nm, the corresponding Raman shift ranges from 2130~cm$^{-1}$ to 2690~cm$^{-1}$, consistent with the literature values.~\cite{SR70}
While the Raman photons do not spectrally overlap with the PL photons, and can be filtered out by use of a longpass filter at 650~nm (which is removed in this measurement, see Sec~\ref{sec_optics}),
they do overlap with the phonon sideband of the neutral NV centers.
This can be avoided by using a green laser operating at a shorter wavelength.
For instance, with excitation at 514.5~nm, the Raman spectrum can be brought to appear in the range shorter than 600~nm (though of course the selection of optical components must be modified accordingly).

The photon correlation measurement is also possible with a fiber-based setup.
Now connecting one end of the fiber BS to another SPCM instead of spectrometer and also using a time correlated single photon counting module (TCSPCM) [Fig.~\ref{fig11}(b)],
we can measure the second order correlation function $g^{(2)}(\tau_{\mathrm{d}})$.
Again, it is required to keep tracking an NV center and a switch is added to realize this.
The measurement mode is also switched by software, so that the photon correlation data are not affected by tracking.
$g^{(2)}(\tau_{\mathrm{d}})$ of the same single NV center is shown in Fig.~\ref{fig12}(b).
Here, the delay time $\tau_{\mathrm{d}}$ is defined as the temporal difference between the time at which one SPCM records a photon arrival and the time at which the other does.
The difference in the electrical lengths is calibrated by exchanging the roles of the two SPCMs.
For the two laser powers ($P_{\mathrm{L}}$) used, $g^{(2)}(0)$ goes very close to zero, 0.040 (0.064) for $P_{\mathrm{L}}$ = 1.0~mW (3.4~mW), 
demonstrating photon antibunching, a hallmark for a quantum emitter that emits one photon at a time.
The data are fitted by~\cite{KMZW00,BBPG00,BMD+15}
\begin{equation}
g^{(2)}(\tau_{\mathrm{d}}) = 1 - \zeta \eta e^{- \gamma_1 \tau_{\mathrm{d}}} + \zeta (\eta-1) e^{-\gamma_2 \tau_{\mathrm{d}}}.
\label{eq_g2}
\end{equation}
Here, $\zeta$ accounts for the effect of incoherent background photons: $g^{(2)}(0) = 1 - \zeta$.
$\gamma_1$ is related to the excitation and emission rates between the $^{3}A_{2}$ and $^{3}E$ states under off-resonant laser excitation.
The transition to and from the nonradiative $^{1}A_{1}$--$^{1}E$ states gives $\gamma_2 > 0$ and $\eta > 1$
(the absence of the intersystem crossing would give $\gamma_2$ = 0 and $\eta$ = 1).
When $P_{\mathrm{L}}$, and thus the excitation rate from $^{3}A_{2}$ to $^{3}E$, is increased,
the excitation--decay cycle becomes unbalanced, causing photon bunching $g^{(2)}(\tau_{\mathrm{d}}) > 1$ as observed in the $P_{\mathrm{L}}$ = 3.4~mW case.
From the fits, we obtain $\eta$ = 1.18 (1.53), $\gamma_1$ = 0.094~ns$^{-1}$ (0.108~ns$^{-1}$), $\gamma_2$ = 0.012~ns$^{-1}$ (0.010~ns$^{-1}$) for $P_{\mathrm{L}}$ = 1.0~mW (3.4~mW).
$\eta$, $\gamma_{1,2}$ provide the information on the photophysics of the NV center involving multiple energy levels.~\cite{KMZW00,BBPG00,BMD+15}
The understanding of the photophysics of the NV center is a key ingredient to achieve better collection efficiency and spin-initialization efficiency,~\cite{HSB18} which then has a far-reaching consequence not only for quantum sensing, but for quantum network~\cite{AEV+18,AHWZ18,WEH18} and fluorescent biomarkers~\cite{H07,MSHG12,WJPW16} utilizing NV centers.

\section{Conclusion\label{sec_conclusion}}
To conclude, we have designed, constructed, and operated a compact tabletop-sized system for quantum sensing with a single NV center.
Despite its compactness, our setup realizes the state-of-the-art quantum sensing protocols that enable the detection of single nuclear spins and the characterization of their interaction parameters.
The data also show that our setup possesses the temporal stability enough to keep tracking the same NV center for long time to conduct detailed nuclear spin sensing experiments with it.
We have also provided a wealth of practical and hands-on information so that nonspecialists can reproduce the setup, conduct experiments, and analyze the data.

\section*{Acknowledgments}
We thank T. Teraji for CVD growth of a $^{12}$C diamond layer of Sample \#3.
K.S. is supported by JSPS Grant-in-Aid for Research Fellowship for Young Scientists (DC1) No.~JP17J05890.
Y.M. is supported by JSPS Grant-in-Aid for Young Scientists (A) No.~17H04928.
K.M.I. is supported by JSPS Grant-in-Aid for Scientific Research (S) No.~26220602 and (B) No.~19H02547,
JST Development of Systems and Technologies for Advanced Measurement and Analysis (SENTAN), and Spintronics Research Network of Japan (Spin-RNJ).

\appendix
\section{Lists of items\label{sec_list}}
Tables~\ref{tab_a1}--\ref{tab_a6} provide full lists of optical and electronic components of the reported nanoscale magnetometer.
An important caveat is that, although carefully chosen, the listed items are by no means the only options and there may be a set of items which perform equally well or better than the listed ones.

For stable operation of the system, it is recommended that the system is fixed on an optical table or a breadboard and is housed in an optical enclosure to block external light.
We note that our design is metric and the optical table/breadboard on which the system is to be fixed should have M6 taps on 25 mm centers.
The use of imperial parts will require a modification of the design to fit one's own optical table/breadboard.

Some items which we assume are readily available in optics laboratories, for instance, laser safety glasses, an optical power meter, and electrical and communication cables (BNC, SMA, GPIB, LAN etc.), are omitted.

\begin{table*}
\caption{Optical components shown in Fig.~\ref{fig1}.
Note that the collimator (\#3) is selected based on the pupil diameter of the objective lens (\#7).
The current lineup of TC18APC-type collimators from Thorlabs do not cover 532 nm, but the alignment wavelength can be customized upon request.
When a different objective lens is used, a different collimator may be considered.
}
\label{tab_a1}
\begin{tabular}{dlll}
\hline
\multicolumn{1}{c}{\#} & \multicolumn{1}{c}{Item} & \multicolumn{1}{c}{Model \#} & \multicolumn{1}{c}{Manufacturer} \\
\hline
1 & 532-nm fiber laser & J030GS-1G-12-23-12 & SOC (Showa Optronics) \\
2 & Fiber AOM & S-M-200-0.4C2C-3-F2P \& 97-03307-26 (driver) & Gooch \& Housego \\
3 & Collimator & TC18APC-CUSTOM (customized at 532 nm) & Thorlabs \\
4 & ND filter & NDM2/M & Thorlabs \\
5 & Dichroic filter & FF552-Di02-25x36 & Semrock \\
6 & Piezo positioner & MIPOS100PL-SG \& 30V300CLE (controller) & piezosystem jena \\
7.1 & Objective lens & LMPLFLN100xBD (air) & Olympus \\
7.2 & Objective lens & PLAPON60XO \& IMMOIL-F30CC (oil) & Olympus \\
8 & Linear stage & FS-1020PXY (for two axes) \& FC-511 (controller) & Sigma Tech \\
9 & Mirror & BB1-E02 & Thorlabs \\
10.1 & Mirror &BB1-E02 & Thorlabs \\
10.2 & Longpass filter & FELH0650 & Thorlabs \\
11 & CCD & Basler ace & Basler \\
12.1 & Notch filter & NF533-17 & Thorlabs \\
12.2 & Longpass filter & FELH0600 & Thorlabs \\
12.3 & Shortpass filter & FESH0800 & Thorlabs \\
13 & Objective lens & M-10X & Newport \\
14 & SPCM & SPCM-AQRH-16-FC & Excelitas Technologies \\
\hline
\end{tabular}
\end{table*}

\begin{table*}
\caption{Optical components for the optical cage system.
The numbers in the Location column refer to those in Fig.~\ref{fig1} and Table~\ref{tab_a1}.
Several items (with identical model \#) appear multiple times as they are used in different locations.
In each entry, the quantity is 1 unless otherwise noted in the Model \# column.
}
\label{tab_a2}
\begin{tabular}{clll}
\hline
\multicolumn{1}{c}{Location} & \multicolumn{1}{c}{Item} & \multicolumn{1}{c}{Model \#} & \multicolumn{1}{c}{Manufacturer} \\
\hline
3 & Adapter & AD15F & Thorlabs \\
& Cage plate & CP02T/M & Thorlabs \\
& Optical post & TR30/M-JP & Thorlabs \\
& Pedestal post holder & PH20E/M & Thorlabs \\
& Clamping fork & CF125C/M & Thorlabs \\
\hline
Between 3 \& 4 & Cage assembly rod & ER1.5-P4 & Thorlabs \\
\hline
Between 4 \& 5 & Cage assembly rod & ER05-P4 & Thorlabs \\
\hline
5 & Cage cube with dichroic filter mount & CM1-DCH/M & Thorlabs \\
& Blank plate & CP01/M & Thorlabs \\
\hline
Between 5 \& 6 & Adaptor & SM1A34 & Thorlabs \\
\hline
Between 6 \& 7 & Adaptor for 7.1 & M32SM1S \& SM1A28 & Thorlabs \\
& Adaptor for 7.2 & 32RMSS & Thorlabs \\
\hline
8 & Vertical translation stage & MVS010/M & Thorlabs \\
& Dual threaded adapter & AE4M6M ($\times$4) & Thorlabs \\
\hline
Between 5 \& 9 & Cage assembly rod & ER1-P4 & Thorlabs \\
\hline
9 & Right-angle kinematic mirror mount & KCB1C/M & Thorlabs \\
\hline
10 & Pivoting optic mount & CP360R/M & Thorlabs \\
& Cage cube & C6W & Thorlabs \\
& Cage assembly rod & ER4-P4 & Thorlabs \\
& Cage cube platform & B3C/M & Thorlabs \\
& Optical post & TR75/M-JP & Thorlabs \\
& Pedestal post holder & PH75E/M & Thorlabs \\
& Clamping fork & CF125C/M & Thorlabs \\
\hline
11 & Adaptor & SM1A39 & Thorlabs \\
& Achromatic lens & AC254-100-A & Thorlabs \\
& Cage plate & CP02T/M & Thorlabs \\
& Cage plate & CP02/M & Thorlabs \\
& Optical post & TR75/M-JP & Thorlabs \\
& Pedestal post holder & PH75E/M & Thorlabs \\
& Clamping fork & CF125C/M & Thorlabs \\
\hline
12 & Cage plate & CP02/M & Thorlabs \\
\hline
13 & Adaptor & SM1A3 & Thorlabs \\
& XY translator & ST1XY-A/M & Thorlabs \\
& Optical post & TR75/M-JP & Thorlabs \\
& Pedestal post holder & PH75E/M & Thorlabs \\
& Clamping fork & CF125C/M & Thorlabs \\
\hline
Between 13 \& 14 & Cage assembly rod & ER2-P4 & Thorlabs \\
& Z-axis translation mount & SM1Z & Thorlabs \\
& FC/PC fiber adapter plate & SM1FC & Thorlabs \\
& Single mode fiber & P1-630A-FC-1 & Thorlabs \\
\hline
\end{tabular}
\end{table*}

\begin{table}
\caption{Breadboard on which the optical cage system is mounted.
One corner (75 $\times$ 75 mm$^2$ in area) must be cut out to secure the space for the objective lens.
}
\label{tab_a3}
\begin{tabular}{lll}
\hline
\multicolumn{1}{c}{Item} & \multicolumn{1}{c}{Model \#} & \multicolumn{1}{c}{Manufacturer} \\
\hline
Breadboard & MB2020/M & Thorlabs \\
Post & P200/M ($\times$3) & Thorlabs \\
Mounting post base & PB1 ($\times$3) & Thorlabs \\
\hline
\end{tabular}
\end{table}

\begin{table}
\caption{$B_0$ control.
See Table~\ref{tab_parameter} for the dimensions of the respective magnets.
In our setup, the rotary stage was later replaced by a motorized stage X-RSW60A-E03 (Zaber Technologies),
in order to automate the field tuning procedure.
On the other hand, full manual control of the magnet position is also possible by using a micrometer instead of the actuator (\#16).
Note that the actuator we currently use interferes with the two of the posts (P200/M listed in Table~\ref{tab_a3}).
The use of a smaller actuator or micrometer allows us to address all the NV axes.
As suggested in Sec.~\ref{sec_magnet}, a goniometer may be added to fine-tune the tilt of the magnet.
}
\label{tab_a4}
\begin{tabular}{dlll}
\hline
\multicolumn{1}{c}{\#} & \multicolumn{1}{c}{Item} & \multicolumn{1}{c}{Model \#} & \multicolumn{1}{c}{Manufacturer} \\
\hline
15.1 & Magnet \#1 & NdFeB N40 & Neomag \\
15.2 & Magnet \#2 & NdFeB N40 & Neomag \\
16 & Actuator & CONEX-TRA25CC & Newport \\
17 & Rotary stage & RP01/M & Thorlabs \\
\hline
\end{tabular}
\end{table}

\begin{table}
\caption{Electronics and microwave components shown in Fig.~\ref{fig3}.
}
\label{tab_a5}
\begin{tabular}{clll}
\hline
\multicolumn{1}{c}{\#} & \multicolumn{1}{c}{Item} & \multicolumn{1}{c}{Model \#} & \multicolumn{1}{c}{Manufacturer} \\
\hline
18 & AWG & AWG7102 & Tektronix \\
19 & DAQ & USB-6343 & National Instruments \\
20 & VSG & MG3700A & Anritsu \\
21 & Amplifier & ZHL-16W-43+ & MiniCircuits \\
22 & Circulator & FX00-0329-00 & Orient Microwave \\
23 & Switch & ZYSWA-2-50DR & MiniCircuits \\
\hline
\end{tabular}
\end{table}

\begin{table}
\caption{Additional instruments used in the measurements in Sec.~\ref{sec_photon}.
See Fig.~\ref{fig11} for the diagrams.
}
\label{tab_a6}
\begin{tabular}{lll}
\hline
\multicolumn{1}{c}{Item} & \multicolumn{1}{c}{Model \#} & \multicolumn{1}{c}{Manufacturer} \\
\hline
Fiber BS & TW670R5F1 & Thorlabs \\
Spectrometer & SpectraPro SP2558 & Princeton Instruments \\
CCD & PyLon 100B eXcelon & Princeton Instruments \\
SPCM & SPCM-AQRH-16-FC & Excelitas Technologies \\
Switch & ZYSWA-2-50DR & MiniCircuits \\
TCSPCM & PicoHarp 300 & PicoQuant \\
\hline
\end{tabular}
\end{table}

\clearpage
\section{Drawings\label{sec_cad}}
\begin{figure}[h]
\begin{center}
\includegraphics{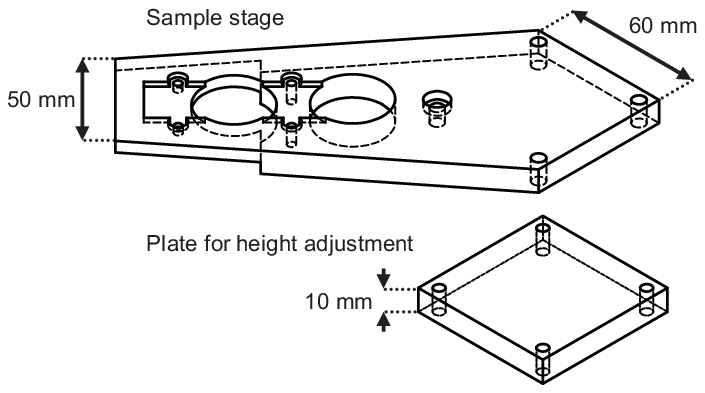}
\caption{Drawings of the sample stage and the plate.
The latter was used to adjust the height, placed between the sample stage and the linear stage.
They are fixed by three M4 $\times$ 24~mm screws and one M4 $\times$ 20~mm screw.
A microwave antenna is fixed by four M3 $\times$ 5~mm screws.
}
\label{fig13}
\end{center}
\end{figure}

\begin{figure}
\begin{center}
\includegraphics{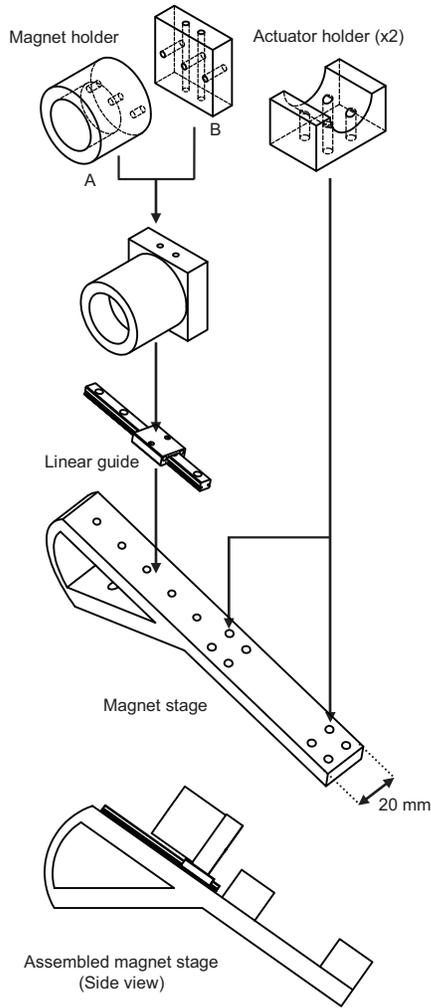}
\caption{Drawings of the parts for magnet holder/stage and the assembly.
The linear guide is commercially available from SURUGA Production Platform (Model \#: SSEB6-70).
Magnet holders A and B are fastened with three M2 $\times$ 15~mm screws;
the magnet holder and the linear guide with two M2 $\times$ 32~mm screws;
the linear guide and the magnet stage with five M2 $\times$ 10~mm screws;
the magnet stage and the actuator holders with eight M3 $\times$ 15~mm screws.
}
\label{fig14}
\end{center}
\end{figure}

Figures~\ref{fig13} and \ref{fig14} show drawings of the sample stage and the magnet holder/stage pictured in Fig.~\ref{fig1}(c).
Their CAD files, as well as those for Antennas \#1 and \#2, are available upon request.

\bibliography{nv3}

\begin{thebibliography}{103}%
\makeatletter
\providecommand \@ifxundefined [1]{%
 \@ifx{#1\undefined}
}%
\providecommand \@ifnum [1]{%
 \ifnum #1\expandafter \@firstoftwo
 \else \expandafter \@secondoftwo
 \fi
}%
\providecommand \@ifx [1]{%
 \ifx #1\expandafter \@firstoftwo
 \else \expandafter \@secondoftwo
 \fi
}%
\providecommand \natexlab [1]{#1}%
\providecommand \enquote  [1]{``#1''}%
\providecommand \bibnamefont  [1]{#1}%
\providecommand \bibfnamefont [1]{#1}%
\providecommand \citenamefont [1]{#1}%
\providecommand \href@noop [0]{\@secondoftwo}%
\providecommand \href [0]{\begingroup \@sanitize@url \@href}%
\providecommand \@href[1]{\@@startlink{#1}\@@href}%
\providecommand \@@href[1]{\endgroup#1\@@endlink}%
\providecommand \@sanitize@url [0]{\catcode `\\12\catcode `\$12\catcode
  `\&12\catcode `\#12\catcode `\^12\catcode `\_12\catcode `\%12\relax}%
\providecommand \@@startlink[1]{}%
\providecommand \@@endlink[0]{}%
\providecommand \url  [0]{\begingroup\@sanitize@url \@url }%
\providecommand \@url [1]{\endgroup\@href {#1}{\urlprefix }}%
\providecommand \urlprefix  [0]{URL }%
\providecommand \Eprint [0]{\href }%
\providecommand \doibase [0]{http://dx.doi.org/}%
\providecommand \selectlanguage [0]{\@gobble}%
\providecommand \bibinfo  [0]{\@secondoftwo}%
\providecommand \bibfield  [0]{\@secondoftwo}%
\providecommand \translation [1]{[#1]}%
\providecommand \BibitemOpen [0]{}%
\providecommand \bibitemStop [0]{}%
\providecommand \bibitemNoStop [0]{.\EOS\space}%
\providecommand \EOS [0]{\spacefactor3000\relax}%
\providecommand \BibitemShut  [1]{\csname bibitem#1\endcsname}%
\let\auto@bib@innerbib\@empty
\bibitem [{\citenamefont {Degen}, \citenamefont {Reinhard},\ and\ \citenamefont
  {Cappellaro}(2017)}]{DRC17}%
  \BibitemOpen
  \bibfield  {author} {\bibinfo {author} {\bibfnamefont {C.~L.}\ \bibnamefont
  {Degen}}, \bibinfo {author} {\bibfnamefont {F.}~\bibnamefont {Reinhard}}, \
  and\ \bibinfo {author} {\bibfnamefont {P.}~\bibnamefont {Cappellaro}},\
  }\bibfield  {title} {\enquote {\bibinfo {title} {Quantum sensing},}\
  }\href@noop {} {\bibfield  {journal} {\bibinfo  {journal} {Rev.\ Mod.\
  Phys.}\ }\textbf {\bibinfo {volume} {89}},\ \bibinfo {pages} {035002}
  (\bibinfo {year} {2017})}\BibitemShut {NoStop}%
\bibitem [{\citenamefont {Maze}\ \emph {et~al.}(2008)\citenamefont {Maze},
  \citenamefont {Stanwix}, \citenamefont {Hodges}, \citenamefont {Hong},
  \citenamefont {Taylor}, \citenamefont {Cappellaro}, \citenamefont {Jiang},
  \citenamefont {Gurudev~Dutt}, \citenamefont {Togan}, \citenamefont {Zibrov},
  \citenamefont {Yacoby}, \citenamefont {Walsworth},\ and\ \citenamefont
  {Lukin}}]{MSH+08}%
  \BibitemOpen
  \bibfield  {author} {\bibinfo {author} {\bibfnamefont {J.~R.}\ \bibnamefont
  {Maze}}, \bibinfo {author} {\bibfnamefont {P.~L.}\ \bibnamefont {Stanwix}},
  \bibinfo {author} {\bibfnamefont {J.~S.}\ \bibnamefont {Hodges}}, \bibinfo
  {author} {\bibfnamefont {S.}~\bibnamefont {Hong}}, \bibinfo {author}
  {\bibfnamefont {J.~M.}\ \bibnamefont {Taylor}}, \bibinfo {author}
  {\bibfnamefont {P.}~\bibnamefont {Cappellaro}}, \bibinfo {author}
  {\bibfnamefont {L.}~\bibnamefont {Jiang}}, \bibinfo {author} {\bibfnamefont
  {M.~V.}\ \bibnamefont {Gurudev~Dutt}}, \bibinfo {author} {\bibfnamefont
  {E.}~\bibnamefont {Togan}}, \bibinfo {author} {\bibfnamefont {A.~S.}\
  \bibnamefont {Zibrov}}, \bibinfo {author} {\bibfnamefont {A.}~\bibnamefont
  {Yacoby}}, \bibinfo {author} {\bibfnamefont {R.~L.}\ \bibnamefont
  {Walsworth}}, \ and\ \bibinfo {author} {\bibfnamefont {M.~D.}\ \bibnamefont
  {Lukin}},\ }\bibfield  {title} {\enquote {\bibinfo {title} {Nanoscale
  magnetic sensing with an individual electronic spin in diamond},}\
  }\href@noop {} {\bibfield  {journal} {\bibinfo  {journal} {Nature}\ }\textbf
  {\bibinfo {volume} {455}},\ \bibinfo {pages} {644} (\bibinfo {year}
  {2008})}\BibitemShut {NoStop}%
\bibitem [{\citenamefont {Balasubramanian}\ \emph {et~al.}(2008)\citenamefont
  {Balasubramanian}, \citenamefont {Chan}, \citenamefont {Kolesov},
  \citenamefont {Al-Hmoud}, \citenamefont {Tisler}, \citenamefont {Shin},
  \citenamefont {Kim}, \citenamefont {Wojcik}, \citenamefont {Hemmer},
  \citenamefont {Krueger}, \citenamefont {Hanke}, \citenamefont
  {Leitenstorfer}, \citenamefont {Bratschitsch}, \citenamefont {Jelezko},\ and\
  \citenamefont {Wrachtrup}}]{BCK+08}%
  \BibitemOpen
  \bibfield  {author} {\bibinfo {author} {\bibfnamefont {G.}~\bibnamefont
  {Balasubramanian}}, \bibinfo {author} {\bibfnamefont {I.~Y.}\ \bibnamefont
  {Chan}}, \bibinfo {author} {\bibfnamefont {R.}~\bibnamefont {Kolesov}},
  \bibinfo {author} {\bibfnamefont {M.}~\bibnamefont {Al-Hmoud}}, \bibinfo
  {author} {\bibfnamefont {J.}~\bibnamefont {Tisler}}, \bibinfo {author}
  {\bibfnamefont {C.}~\bibnamefont {Shin}}, \bibinfo {author} {\bibfnamefont
  {C.}~\bibnamefont {Kim}}, \bibinfo {author} {\bibfnamefont {A.}~\bibnamefont
  {Wojcik}}, \bibinfo {author} {\bibfnamefont {P.~R.}\ \bibnamefont {Hemmer}},
  \bibinfo {author} {\bibfnamefont {A.}~\bibnamefont {Krueger}}, \bibinfo
  {author} {\bibfnamefont {T.}~\bibnamefont {Hanke}}, \bibinfo {author}
  {\bibfnamefont {A.}~\bibnamefont {Leitenstorfer}}, \bibinfo {author}
  {\bibfnamefont {R.}~\bibnamefont {Bratschitsch}}, \bibinfo {author}
  {\bibfnamefont {F.}~\bibnamefont {Jelezko}}, \ and\ \bibinfo {author}
  {\bibfnamefont {J.}~\bibnamefont {Wrachtrup}},\ }\bibfield  {title} {\enquote
  {\bibinfo {title} {Nanoscale imaging magnetometry with diamond spins under
  ambient conditions},}\ }\href@noop {} {\bibfield  {journal} {\bibinfo
  {journal} {Nature}\ }\textbf {\bibinfo {volume} {455}},\ \bibinfo {pages}
  {648} (\bibinfo {year} {2008})}\BibitemShut {NoStop}%
\bibitem [{\citenamefont {Rondin}\ \emph {et~al.}(2014)\citenamefont {Rondin},
  \citenamefont {Tetienne}, \citenamefont {Hingant}, \citenamefont {Roch},
  \citenamefont {Maletinsky},\ and\ \citenamefont {Jacques}}]{RTH+14}%
  \BibitemOpen
  \bibfield  {author} {\bibinfo {author} {\bibfnamefont {L.}~\bibnamefont
  {Rondin}}, \bibinfo {author} {\bibfnamefont {J.-P.}\ \bibnamefont
  {Tetienne}}, \bibinfo {author} {\bibfnamefont {T.}~\bibnamefont {Hingant}},
  \bibinfo {author} {\bibfnamefont {J.-F.}\ \bibnamefont {Roch}}, \bibinfo
  {author} {\bibfnamefont {P.}~\bibnamefont {Maletinsky}}, \ and\ \bibinfo
  {author} {\bibfnamefont {V.}~\bibnamefont {Jacques}},\ }\bibfield  {title}
  {\enquote {\bibinfo {title} {Magnetometry with nitrogen-vacancy defects in
  diamond},}\ }\href@noop {} {\bibfield  {journal} {\bibinfo  {journal} {Rep.\
  Prog.\ Phys.}\ }\textbf {\bibinfo {volume} {77}},\ \bibinfo {pages} {056503}
  (\bibinfo {year} {2014})}\BibitemShut {NoStop}%
\bibitem [{\citenamefont {Schirhagl}\ \emph {et~al.}(2014)\citenamefont
  {Schirhagl}, \citenamefont {Chang}, \citenamefont {Loretz},\ and\
  \citenamefont {Degen}}]{SCLD14}%
  \BibitemOpen
  \bibfield  {author} {\bibinfo {author} {\bibfnamefont {R.}~\bibnamefont
  {Schirhagl}}, \bibinfo {author} {\bibfnamefont {K.}~\bibnamefont {Chang}},
  \bibinfo {author} {\bibfnamefont {M.}~\bibnamefont {Loretz}}, \ and\ \bibinfo
  {author} {\bibfnamefont {C.~L.}\ \bibnamefont {Degen}},\ }\bibfield  {title}
  {\enquote {\bibinfo {title} {Nitrogen-{V}acancy {C}enters in {D}iamond:
  {N}anoscale {S}ensors for {P}hysics and {B}iology},}\ }\href@noop {}
  {\bibfield  {journal} {\bibinfo  {journal} {Annu.\ Rev.\ Phys.\ Chem.}\
  }\textbf {\bibinfo {volume} {65}},\ \bibinfo {pages} {83} (\bibinfo {year}
  {2014})}\BibitemShut {NoStop}%
\bibitem [{\citenamefont {Casola}, \citenamefont {van~der Sar},\ and\
  \citenamefont {Yacoby}(2018)}]{CvdSY18}%
  \BibitemOpen
  \bibfield  {author} {\bibinfo {author} {\bibfnamefont {F.}~\bibnamefont
  {Casola}}, \bibinfo {author} {\bibfnamefont {T.}~\bibnamefont {van~der Sar}},
  \ and\ \bibinfo {author} {\bibfnamefont {A.}~\bibnamefont {Yacoby}},\
  }\bibfield  {title} {\enquote {\bibinfo {title} {Probing condensed matter
  physics with magnetometry based on nitrogen-vacancy centres in diamond},}\
  }\href@noop {} {\bibfield  {journal} {\bibinfo  {journal} {Nat.\ Rev.\
  Mater.}\ }\textbf {\bibinfo {volume} {3}},\ \bibinfo {pages} {17088}
  (\bibinfo {year} {2018})}\BibitemShut {NoStop}%
\bibitem [{\citenamefont {Abe}\ and\ \citenamefont {Sasaki}(2018)}]{AS18}%
  \BibitemOpen
  \bibfield  {author} {\bibinfo {author} {\bibfnamefont {E.}~\bibnamefont
  {Abe}}\ and\ \bibinfo {author} {\bibfnamefont {K.}~\bibnamefont {Sasaki}},\
  }\bibfield  {title} {\enquote {\bibinfo {title} {Tutorial: {M}agnetic
  resonance with nitrogen-vacancy centers in diamond---microwave engineering,
  materials science, and magnetometry},}\ }\href@noop {} {\bibfield  {journal}
  {\bibinfo  {journal} {J.\ Appl.\ Phys.}\ }\textbf {\bibinfo {volume} {123}},\
  \bibinfo {pages} {161101} (\bibinfo {year} {2018})}\BibitemShut {NoStop}%
\bibitem [{\citenamefont {Mamin}\ \emph {et~al.}(2013)\citenamefont {Mamin},
  \citenamefont {Kim}, \citenamefont {Sherwood}, \citenamefont {Rettner},
  \citenamefont {Ohno}, \citenamefont {Awschalom},\ and\ \citenamefont
  {Rugar}}]{MKS+13}%
  \BibitemOpen
  \bibfield  {author} {\bibinfo {author} {\bibfnamefont {H.~J.}\ \bibnamefont
  {Mamin}}, \bibinfo {author} {\bibfnamefont {M.}~\bibnamefont {Kim}}, \bibinfo
  {author} {\bibfnamefont {M.~H.}\ \bibnamefont {Sherwood}}, \bibinfo {author}
  {\bibfnamefont {C.~T.}\ \bibnamefont {Rettner}}, \bibinfo {author}
  {\bibfnamefont {K.}~\bibnamefont {Ohno}}, \bibinfo {author} {\bibfnamefont
  {D.~D.}\ \bibnamefont {Awschalom}}, \ and\ \bibinfo {author} {\bibfnamefont
  {D.}~\bibnamefont {Rugar}},\ }\bibfield  {title} {\enquote {\bibinfo {title}
  {Nanoscale {N}uclear {M}agnetic {R}esonance with a {N}itrogen-{V}acancy
  {S}pin {S}ensor},}\ }\href@noop {} {\bibfield  {journal} {\bibinfo  {journal}
  {Science}\ }\textbf {\bibinfo {volume} {339}},\ \bibinfo {pages} {557}
  (\bibinfo {year} {2013})}\BibitemShut {NoStop}%
\bibitem [{\citenamefont {Staudacher}\ \emph {et~al.}(2013)\citenamefont
  {Staudacher}, \citenamefont {Shi}, \citenamefont {Pezzagna}, \citenamefont
  {Meijer}, \citenamefont {Du}, \citenamefont {Meriles}, \citenamefont
  {Reinhard},\ and\ \citenamefont {Wrachtrup}}]{SSP+13}%
  \BibitemOpen
  \bibfield  {author} {\bibinfo {author} {\bibfnamefont {T.}~\bibnamefont
  {Staudacher}}, \bibinfo {author} {\bibfnamefont {F.}~\bibnamefont {Shi}},
  \bibinfo {author} {\bibfnamefont {S.}~\bibnamefont {Pezzagna}}, \bibinfo
  {author} {\bibfnamefont {J.}~\bibnamefont {Meijer}}, \bibinfo {author}
  {\bibfnamefont {J.}~\bibnamefont {Du}}, \bibinfo {author} {\bibfnamefont
  {C.~A.}\ \bibnamefont {Meriles}}, \bibinfo {author} {\bibfnamefont
  {F.}~\bibnamefont {Reinhard}}, \ and\ \bibinfo {author} {\bibfnamefont
  {J.}~\bibnamefont {Wrachtrup}},\ }\bibfield  {title} {\enquote {\bibinfo
  {title} {Nuclear {M}agnetic {R}esonance {S}pectroscopy on a
  (5-{N}anometer)$^3$ {S}ample {V}olume},}\ }\href@noop {} {\bibfield
  {journal} {\bibinfo  {journal} {Science}\ }\textbf {\bibinfo {volume}
  {339}},\ \bibinfo {pages} {561} (\bibinfo {year} {2013})}\BibitemShut
  {NoStop}%
\bibitem [{\citenamefont {M{\"u}ller}\ \emph {et~al.}(2014)\citenamefont
  {M{\"u}ller}, \citenamefont {Kong}, \citenamefont {Cai}, \citenamefont
  {Melentijevi{\'c}}, \citenamefont {Stacey}, \citenamefont {Markham},
  \citenamefont {Twitchen}, \citenamefont {Isoya}, \citenamefont {Pezzagna},
  \citenamefont {Meijer}, \citenamefont {Du}, \citenamefont {Plenio},
  \citenamefont {Naydenov}, \citenamefont {McGuinness},\ and\ \citenamefont
  {Jelezko}}]{MKC+14}%
  \BibitemOpen
  \bibfield  {author} {\bibinfo {author} {\bibfnamefont {C.}~\bibnamefont
  {M{\"u}ller}}, \bibinfo {author} {\bibfnamefont {X.}~\bibnamefont {Kong}},
  \bibinfo {author} {\bibfnamefont {J.-M.}\ \bibnamefont {Cai}}, \bibinfo
  {author} {\bibfnamefont {K.}~\bibnamefont {Melentijevi{\'c}}}, \bibinfo
  {author} {\bibfnamefont {A.}~\bibnamefont {Stacey}}, \bibinfo {author}
  {\bibfnamefont {M.}~\bibnamefont {Markham}}, \bibinfo {author} {\bibfnamefont
  {D.}~\bibnamefont {Twitchen}}, \bibinfo {author} {\bibfnamefont
  {J.}~\bibnamefont {Isoya}}, \bibinfo {author} {\bibfnamefont
  {S.}~\bibnamefont {Pezzagna}}, \bibinfo {author} {\bibfnamefont
  {J.}~\bibnamefont {Meijer}}, \bibinfo {author} {\bibfnamefont {J.~F.}\
  \bibnamefont {Du}}, \bibinfo {author} {\bibfnamefont {M.~B.}\ \bibnamefont
  {Plenio}}, \bibinfo {author} {\bibfnamefont {B.}~\bibnamefont {Naydenov}},
  \bibinfo {author} {\bibfnamefont {L.~P.}\ \bibnamefont {McGuinness}}, \ and\
  \bibinfo {author} {\bibfnamefont {F.}~\bibnamefont {Jelezko}},\ }\bibfield
  {title} {\enquote {\bibinfo {title} {Nuclear magnetic resonance spectroscopy
  with single spin sensitivity},}\ }\href@noop {} {\bibfield  {journal}
  {\bibinfo  {journal} {Nat.\ Commun.}\ }\textbf {\bibinfo {volume} {5}},\
  \bibinfo {pages} {4703} (\bibinfo {year} {2014})}\BibitemShut {NoStop}%
\bibitem [{\citenamefont {H{\"a}berle}\ \emph {et~al.}(2015)\citenamefont
  {H{\"a}berle}, \citenamefont {Schmid-Lorch}, \citenamefont {Reinhard},\ and\
  \citenamefont {Wrachtrup}}]{HSR+15}%
  \BibitemOpen
  \bibfield  {author} {\bibinfo {author} {\bibfnamefont {T.}~\bibnamefont
  {H{\"a}berle}}, \bibinfo {author} {\bibfnamefont {D.}~\bibnamefont
  {Schmid-Lorch}}, \bibinfo {author} {\bibfnamefont {F.}~\bibnamefont
  {Reinhard}}, \ and\ \bibinfo {author} {\bibfnamefont {J.}~\bibnamefont
  {Wrachtrup}},\ }\bibfield  {title} {\enquote {\bibinfo {title} {Nanoscale
  nuclear magnetic imaging with chemical contrast},}\ }\href@noop {} {\bibfield
   {journal} {\bibinfo  {journal} {Nat.\ Nanotechnol.}\ }\textbf {\bibinfo
  {volume} {10}},\ \bibinfo {pages} {125} (\bibinfo {year} {2015})}\BibitemShut
  {NoStop}%
\bibitem [{\citenamefont {DeVience}\ \emph {et~al.}(2015)\citenamefont
  {DeVience}, \citenamefont {Pham}, \citenamefont {Lovchinsky}, \citenamefont
  {Sushkov}, \citenamefont {Bar-Gill}, \citenamefont {Belthangady},
  \citenamefont {Casola}, \citenamefont {Corbett}, \citenamefont {Zhang},
  \citenamefont {Lukin}, \citenamefont {Park}, \citenamefont {Yacoby},\ and\
  \citenamefont {Walsworth}}]{DPL+15}%
  \BibitemOpen
  \bibfield  {author} {\bibinfo {author} {\bibfnamefont {S.~J.}\ \bibnamefont
  {DeVience}}, \bibinfo {author} {\bibfnamefont {L.~M.}\ \bibnamefont {Pham}},
  \bibinfo {author} {\bibfnamefont {I.}~\bibnamefont {Lovchinsky}}, \bibinfo
  {author} {\bibfnamefont {A.~O.}\ \bibnamefont {Sushkov}}, \bibinfo {author}
  {\bibfnamefont {N.}~\bibnamefont {Bar-Gill}}, \bibinfo {author}
  {\bibfnamefont {C.}~\bibnamefont {Belthangady}}, \bibinfo {author}
  {\bibfnamefont {F.}~\bibnamefont {Casola}}, \bibinfo {author} {\bibfnamefont
  {M.}~\bibnamefont {Corbett}}, \bibinfo {author} {\bibfnamefont
  {H.}~\bibnamefont {Zhang}}, \bibinfo {author} {\bibfnamefont
  {M.}~\bibnamefont {Lukin}}, \bibinfo {author} {\bibfnamefont
  {H.}~\bibnamefont {Park}}, \bibinfo {author} {\bibfnamefont {A.}~\bibnamefont
  {Yacoby}}, \ and\ \bibinfo {author} {\bibfnamefont {R.~L.}\ \bibnamefont
  {Walsworth}},\ }\bibfield  {title} {\enquote {\bibinfo {title} {Nanoscale
  {NMR} spectroscopy and imaging of multiple nulcear species},}\ }\href@noop {}
  {\bibfield  {journal} {\bibinfo  {journal} {Nat.\ Nanotechnol.}\ }\textbf
  {\bibinfo {volume} {10}},\ \bibinfo {pages} {129} (\bibinfo {year}
  {2015})}\BibitemShut {NoStop}%
\bibitem [{\citenamefont {Kong}\ \emph {et~al.}(2015)\citenamefont {Kong},
  \citenamefont {Stark}, \citenamefont {Du}, \citenamefont {McGuinness},\ and\
  \citenamefont {Jelezko}}]{KSD+15}%
  \BibitemOpen
  \bibfield  {author} {\bibinfo {author} {\bibfnamefont {X.}~\bibnamefont
  {Kong}}, \bibinfo {author} {\bibfnamefont {A.}~\bibnamefont {Stark}},
  \bibinfo {author} {\bibfnamefont {J.}~\bibnamefont {Du}}, \bibinfo {author}
  {\bibfnamefont {L.~P.}\ \bibnamefont {McGuinness}}, \ and\ \bibinfo {author}
  {\bibfnamefont {F.}~\bibnamefont {Jelezko}},\ }\bibfield  {title} {\enquote
  {\bibinfo {title} {Towards {C}hemical {S}tructure {R}esolution with
  {N}anoscale {N}uclear {M}agnetic {R}esonance {S}pectroscopy},}\ }\href@noop
  {} {\bibfield  {journal} {\bibinfo  {journal} {Phys.\ Rev.\ Appl.}\ }\textbf
  {\bibinfo {volume} {4}},\ \bibinfo {pages} {024004} (\bibinfo {year}
  {2015})}\BibitemShut {NoStop}%
\bibitem [{\citenamefont {Staudacher}\ \emph {et~al.}(2015)\citenamefont
  {Staudacher}, \citenamefont {Raatz}, \citenamefont {Pezzagna}, \citenamefont
  {Meijer}, \citenamefont {Reinhard}, \citenamefont {Meriles},\ and\
  \citenamefont {Wrachtrup}}]{SRP+15}%
  \BibitemOpen
  \bibfield  {author} {\bibinfo {author} {\bibfnamefont {T.}~\bibnamefont
  {Staudacher}}, \bibinfo {author} {\bibfnamefont {N.}~\bibnamefont {Raatz}},
  \bibinfo {author} {\bibfnamefont {S.}~\bibnamefont {Pezzagna}}, \bibinfo
  {author} {\bibfnamefont {J.}~\bibnamefont {Meijer}}, \bibinfo {author}
  {\bibfnamefont {F.}~\bibnamefont {Reinhard}}, \bibinfo {author}
  {\bibfnamefont {C.~A.}\ \bibnamefont {Meriles}}, \ and\ \bibinfo {author}
  {\bibfnamefont {J.}~\bibnamefont {Wrachtrup}},\ }\bibfield  {title} {\enquote
  {\bibinfo {title} {Probing molecular dynamics at the nanoscale via an
  individual paramagnetic centre},}\ }\href@noop {} {\bibfield  {journal}
  {\bibinfo  {journal} {Nat.\ Commun.}\ }\textbf {\bibinfo {volume} {6}},\
  \bibinfo {pages} {8527} (\bibinfo {year} {2015})}\BibitemShut {NoStop}%
\bibitem [{\citenamefont {Perunicic}\ \emph {et~al.}(2016)\citenamefont
  {Perunicic}, \citenamefont {Hill}, \citenamefont {Hall},\ and\ \citenamefont
  {Hollenberg}}]{PHHH16}%
  \BibitemOpen
  \bibfield  {author} {\bibinfo {author} {\bibfnamefont {V.~S.}\ \bibnamefont
  {Perunicic}}, \bibinfo {author} {\bibfnamefont {C.~D.}\ \bibnamefont {Hill}},
  \bibinfo {author} {\bibfnamefont {L.~T.}\ \bibnamefont {Hall}}, \ and\
  \bibinfo {author} {\bibfnamefont {L.~C.~L.}\ \bibnamefont {Hollenberg}},\
  }\bibfield  {title} {\enquote {\bibinfo {title} {A quantum spin-probe
  molecular microscope},}\ }\href@noop {} {\bibfield  {journal} {\bibinfo
  {journal} {Nat.\ Commun.}\ }\textbf {\bibinfo {volume} {7}},\ \bibinfo
  {pages} {12667} (\bibinfo {year} {2016})}\BibitemShut {NoStop}%
\bibitem [{\citenamefont {Kehayias}\ \emph {et~al.}(2017)\citenamefont
  {Kehayias}, \citenamefont {Jarmola}, \citenamefont {Mosavian}, \citenamefont
  {Fescenko}, \citenamefont {Benito}, \citenamefont {Laraoui}, \citenamefont
  {Smits}, \citenamefont {Bougas}, \citenamefont {Budker}, \citenamefont
  {Neumann}, \citenamefont {Brueck},\ and\ \citenamefont {Acosta}}]{KJM+17}%
  \BibitemOpen
  \bibfield  {author} {\bibinfo {author} {\bibfnamefont {P.}~\bibnamefont
  {Kehayias}}, \bibinfo {author} {\bibfnamefont {A.}~\bibnamefont {Jarmola}},
  \bibinfo {author} {\bibfnamefont {N.}~\bibnamefont {Mosavian}}, \bibinfo
  {author} {\bibfnamefont {I.}~\bibnamefont {Fescenko}}, \bibinfo {author}
  {\bibfnamefont {F.~M.}\ \bibnamefont {Benito}}, \bibinfo {author}
  {\bibfnamefont {A.}~\bibnamefont {Laraoui}}, \bibinfo {author} {\bibfnamefont
  {J.}~\bibnamefont {Smits}}, \bibinfo {author} {\bibfnamefont
  {L.}~\bibnamefont {Bougas}}, \bibinfo {author} {\bibfnamefont
  {D.}~\bibnamefont {Budker}}, \bibinfo {author} {\bibfnamefont
  {A.}~\bibnamefont {Neumann}}, \bibinfo {author} {\bibfnamefont {S.~R.~J.}\
  \bibnamefont {Brueck}}, \ and\ \bibinfo {author} {\bibfnamefont {V.~M.}\
  \bibnamefont {Acosta}},\ }\bibfield  {title} {\enquote {\bibinfo {title}
  {Solution nuclear magnetic resonance spectroscopy on a nanostructured diamond
  chip},}\ }\href@noop {} {\bibfield  {journal} {\bibinfo  {journal} {Nat.\
  Commun.}\ }\textbf {\bibinfo {volume} {8}},\ \bibinfo {pages} {188} (\bibinfo
  {year} {2017})}\BibitemShut {NoStop}%
\bibitem [{\citenamefont {Sushkov}\ \emph {et~al.}(2014)\citenamefont
  {Sushkov}, \citenamefont {Lovchinsky}, \citenamefont {Chisholm},
  \citenamefont {Walsworth}, \citenamefont {Park},\ and\ \citenamefont
  {Lukin}}]{SLC+14}%
  \BibitemOpen
  \bibfield  {author} {\bibinfo {author} {\bibfnamefont {A.~O.}\ \bibnamefont
  {Sushkov}}, \bibinfo {author} {\bibfnamefont {I.}~\bibnamefont {Lovchinsky}},
  \bibinfo {author} {\bibfnamefont {N.}~\bibnamefont {Chisholm}}, \bibinfo
  {author} {\bibfnamefont {R.~L.}\ \bibnamefont {Walsworth}}, \bibinfo {author}
  {\bibfnamefont {H.}~\bibnamefont {Park}}, \ and\ \bibinfo {author}
  {\bibfnamefont {M.~D.}\ \bibnamefont {Lukin}},\ }\bibfield  {title} {\enquote
  {\bibinfo {title} {Magnetic {R}esonance {D}etection of {I}ndividual {P}roton
  {S}pins {U}sing {Q}uantum {R}eporters},}\ }\href@noop {} {\bibfield
  {journal} {\bibinfo  {journal} {Phys.\ Rev.\ Lett.}\ }\textbf {\bibinfo
  {volume} {113}},\ \bibinfo {pages} {197601} (\bibinfo {year}
  {2014})}\BibitemShut {NoStop}%
\bibitem [{\citenamefont {Lovchinsky}\ \emph {et~al.}(2016)\citenamefont
  {Lovchinsky}, \citenamefont {Sushkov}, \citenamefont {Urbach}, \citenamefont
  {de~Leon}, \citenamefont {Choi}, \citenamefont {De~Greve}, \citenamefont
  {Evans}, \citenamefont {Gertner}, \citenamefont {Bersin}, \citenamefont
  {M{\"u}ller}, \citenamefont {McGuinness}, \citenamefont {Jelezko},
  \citenamefont {Walsworth}, \citenamefont {Park},\ and\ \citenamefont
  {Lukin}}]{LSU+16}%
  \BibitemOpen
  \bibfield  {author} {\bibinfo {author} {\bibfnamefont {I.}~\bibnamefont
  {Lovchinsky}}, \bibinfo {author} {\bibfnamefont {A.~O.}\ \bibnamefont
  {Sushkov}}, \bibinfo {author} {\bibfnamefont {E.}~\bibnamefont {Urbach}},
  \bibinfo {author} {\bibfnamefont {N.~P.}\ \bibnamefont {de~Leon}}, \bibinfo
  {author} {\bibfnamefont {S.}~\bibnamefont {Choi}}, \bibinfo {author}
  {\bibfnamefont {K.}~\bibnamefont {De~Greve}}, \bibinfo {author}
  {\bibfnamefont {R.}~\bibnamefont {Evans}}, \bibinfo {author} {\bibfnamefont
  {R.}~\bibnamefont {Gertner}}, \bibinfo {author} {\bibfnamefont
  {E.}~\bibnamefont {Bersin}}, \bibinfo {author} {\bibfnamefont
  {C.}~\bibnamefont {M{\"u}ller}}, \bibinfo {author} {\bibfnamefont
  {L.}~\bibnamefont {McGuinness}}, \bibinfo {author} {\bibfnamefont
  {F.}~\bibnamefont {Jelezko}}, \bibinfo {author} {\bibfnamefont {R.~L.}\
  \bibnamefont {Walsworth}}, \bibinfo {author} {\bibfnamefont {H.}~\bibnamefont
  {Park}}, \ and\ \bibinfo {author} {\bibfnamefont {M.~D.}\ \bibnamefont
  {Lukin}},\ }\bibfield  {title} {\enquote {\bibinfo {title} {Nuclear magnetic
  resonance detection and spectroscopy of single proteins using quantum
  logic},}\ }\href@noop {} {\bibfield  {journal} {\bibinfo  {journal}
  {Science}\ }\textbf {\bibinfo {volume} {351}},\ \bibinfo {pages} {836}
  (\bibinfo {year} {2016})}\BibitemShut {NoStop}%
\bibitem [{\citenamefont {Schmitt}\ \emph {et~al.}(2017)\citenamefont
  {Schmitt}, \citenamefont {Gefen}, \citenamefont {St{\"u}rner}, \citenamefont
  {Unden}, \citenamefont {Wolff}, \citenamefont {M{\"u}ller}, \citenamefont
  {Scheuer}, \citenamefont {Naydenov}, \citenamefont {Markham}, \citenamefont
  {Pezzagna}, \citenamefont {Meijer}, \citenamefont {Schwarz}, \citenamefont
  {Plenio}, \citenamefont {Retzker}, \citenamefont {McGuinness},\ and\
  \citenamefont {Jelezko}}]{SGS+17}%
  \BibitemOpen
  \bibfield  {author} {\bibinfo {author} {\bibfnamefont {S.}~\bibnamefont
  {Schmitt}}, \bibinfo {author} {\bibfnamefont {T.}~\bibnamefont {Gefen}},
  \bibinfo {author} {\bibfnamefont {F.~M.}\ \bibnamefont {St{\"u}rner}},
  \bibinfo {author} {\bibfnamefont {T.}~\bibnamefont {Unden}}, \bibinfo
  {author} {\bibfnamefont {G.}~\bibnamefont {Wolff}}, \bibinfo {author}
  {\bibfnamefont {C.}~\bibnamefont {M{\"u}ller}}, \bibinfo {author}
  {\bibfnamefont {J.}~\bibnamefont {Scheuer}}, \bibinfo {author} {\bibfnamefont
  {B.}~\bibnamefont {Naydenov}}, \bibinfo {author} {\bibfnamefont
  {M.}~\bibnamefont {Markham}}, \bibinfo {author} {\bibfnamefont
  {S.}~\bibnamefont {Pezzagna}}, \bibinfo {author} {\bibfnamefont
  {J.}~\bibnamefont {Meijer}}, \bibinfo {author} {\bibfnamefont
  {I.}~\bibnamefont {Schwarz}}, \bibinfo {author} {\bibfnamefont
  {M.}~\bibnamefont {Plenio}}, \bibinfo {author} {\bibfnamefont
  {A.}~\bibnamefont {Retzker}}, \bibinfo {author} {\bibfnamefont {L.~P.}\
  \bibnamefont {McGuinness}}, \ and\ \bibinfo {author} {\bibfnamefont
  {F.}~\bibnamefont {Jelezko}},\ }\bibfield  {title} {\enquote {\bibinfo
  {title} {Submillihertz magnetic spectroscopy performed with a nanoscale
  quantum sensor},}\ }\href@noop {} {\bibfield  {journal} {\bibinfo  {journal}
  {Science}\ }\textbf {\bibinfo {volume} {356}},\ \bibinfo {pages} {832}
  (\bibinfo {year} {2017})}\BibitemShut {NoStop}%
\bibitem [{\citenamefont {Boss}\ \emph {et~al.}(2017)\citenamefont {Boss},
  \citenamefont {Cujia}, \citenamefont {Zopes},\ and\ \citenamefont
  {Degen}}]{BCZD17}%
  \BibitemOpen
  \bibfield  {author} {\bibinfo {author} {\bibfnamefont {J.~M.}\ \bibnamefont
  {Boss}}, \bibinfo {author} {\bibfnamefont {K.~S.}\ \bibnamefont {Cujia}},
  \bibinfo {author} {\bibfnamefont {J.}~\bibnamefont {Zopes}}, \ and\ \bibinfo
  {author} {\bibfnamefont {C.~L.}\ \bibnamefont {Degen}},\ }\bibfield  {title}
  {\enquote {\bibinfo {title} {Quantum sensing with arbitrary frequency
  resolusion},}\ }\href@noop {} {\bibfield  {journal} {\bibinfo  {journal}
  {Science}\ }\textbf {\bibinfo {volume} {356}},\ \bibinfo {pages} {837}
  (\bibinfo {year} {2017})}\BibitemShut {NoStop}%
\bibitem [{\citenamefont {Aslam}\ \emph {et~al.}(2017)\citenamefont {Aslam},
  \citenamefont {Pfender}, \citenamefont {Neumann}, \citenamefont {Reuter},
  \citenamefont {Zappe}, \citenamefont {de~Oliveira}, \citenamefont
  {Denisenko}, \citenamefont {Sumiya}, \citenamefont {Onoda}, \citenamefont
  {Isoya},\ and\ \citenamefont {Wrachtrup}}]{APN+17}%
  \BibitemOpen
  \bibfield  {author} {\bibinfo {author} {\bibfnamefont {N.}~\bibnamefont
  {Aslam}}, \bibinfo {author} {\bibfnamefont {M.}~\bibnamefont {Pfender}},
  \bibinfo {author} {\bibfnamefont {P.}~\bibnamefont {Neumann}}, \bibinfo
  {author} {\bibfnamefont {R.}~\bibnamefont {Reuter}}, \bibinfo {author}
  {\bibfnamefont {A.}~\bibnamefont {Zappe}}, \bibinfo {author} {\bibfnamefont
  {F.~F.}\ \bibnamefont {de~Oliveira}}, \bibinfo {author} {\bibfnamefont
  {A.}~\bibnamefont {Denisenko}}, \bibinfo {author} {\bibfnamefont
  {H.}~\bibnamefont {Sumiya}}, \bibinfo {author} {\bibfnamefont
  {S.}~\bibnamefont {Onoda}}, \bibinfo {author} {\bibfnamefont
  {J.}~\bibnamefont {Isoya}}, \ and\ \bibinfo {author} {\bibfnamefont
  {J.}~\bibnamefont {Wrachtrup}},\ }\bibfield  {title} {\enquote {\bibinfo
  {title} {Nanoscale nuclear magnetic resonance with chemical resolution},}\
  }\href@noop {} {\bibfield  {journal} {\bibinfo  {journal} {Science}\ }\textbf
  {\bibinfo {volume} {357}},\ \bibinfo {pages} {67} (\bibinfo {year}
  {2017})}\BibitemShut {NoStop}%
\bibitem [{\citenamefont {Glenn}\ \emph {et~al.}(2018)\citenamefont {Glenn},
  \citenamefont {Bucher}, \citenamefont {Lee}, \citenamefont {Lukin},
  \citenamefont {Park},\ and\ \citenamefont {Walsworth}}]{GBL+18}%
  \BibitemOpen
  \bibfield  {author} {\bibinfo {author} {\bibfnamefont {D.~R.}\ \bibnamefont
  {Glenn}}, \bibinfo {author} {\bibfnamefont {D.~B.}\ \bibnamefont {Bucher}},
  \bibinfo {author} {\bibfnamefont {J.}~\bibnamefont {Lee}}, \bibinfo {author}
  {\bibfnamefont {M.~D.}\ \bibnamefont {Lukin}}, \bibinfo {author}
  {\bibfnamefont {H.}~\bibnamefont {Park}}, \ and\ \bibinfo {author}
  {\bibfnamefont {R.~L.}\ \bibnamefont {Walsworth}},\ }\bibfield  {title}
  {\enquote {\bibinfo {title} {High-resolution magnetic resonance spectroscopy
  using a solid-state spin sensor},}\ }\href@noop {} {\bibfield  {journal}
  {\bibinfo  {journal} {Nature}\ }\textbf {\bibinfo {volume} {555}},\ \bibinfo
  {pages} {351} (\bibinfo {year} {2018})}\BibitemShut {NoStop}%
\bibitem [{\citenamefont {Smits}\ \emph {et~al.}(2019)\citenamefont {Smits},
  \citenamefont {Damron}, \citenamefont {Kehayias}, \citenamefont {McDowell},
  \citenamefont {Mosavian}, \citenamefont {Fescenko}, \citenamefont {Ristoff},
  \citenamefont {Laraoui}, \citenamefont {Jarmola},\ and\ \citenamefont
  {Acosta}}]{SDK+19}%
  \BibitemOpen
  \bibfield  {author} {\bibinfo {author} {\bibfnamefont {J.}~\bibnamefont
  {Smits}}, \bibinfo {author} {\bibfnamefont {J.~T.}\ \bibnamefont {Damron}},
  \bibinfo {author} {\bibfnamefont {P.}~\bibnamefont {Kehayias}}, \bibinfo
  {author} {\bibfnamefont {A.~F.}\ \bibnamefont {McDowell}}, \bibinfo {author}
  {\bibfnamefont {N.}~\bibnamefont {Mosavian}}, \bibinfo {author}
  {\bibfnamefont {I.}~\bibnamefont {Fescenko}}, \bibinfo {author}
  {\bibfnamefont {N.}~\bibnamefont {Ristoff}}, \bibinfo {author} {\bibfnamefont
  {A.}~\bibnamefont {Laraoui}}, \bibinfo {author} {\bibfnamefont
  {A.}~\bibnamefont {Jarmola}}, \ and\ \bibinfo {author} {\bibfnamefont
  {V.~M.}\ \bibnamefont {Acosta}},\ }\bibfield  {title} {\enquote {\bibinfo
  {title} {Two-dimensional nuclear magnetic resonance spectroscopy with a
  microfluidic diamond quantum sensor},}\ }\href@noop {} {\bibfield  {journal}
  {\bibinfo  {journal} {Sci.\ Adv.}\ }\textbf {\bibinfo {volume} {5}},\
  \bibinfo {pages} {eaaw} (\bibinfo {year} {2019})}\BibitemShut {NoStop}%
\bibitem [{\citenamefont {Pfender}\ \emph {et~al.}(2019)\citenamefont
  {Pfender}, \citenamefont {Wang}, \citenamefont {Sumiya}, \citenamefont
  {Onoda}, \citenamefont {Yang}, \citenamefont {Dasari}, \citenamefont
  {Neumann}, \citenamefont {Pan}, \citenamefont {Isoya}, \citenamefont {Liu},\
  and\ \citenamefont {Wrachtrup}}]{PWS+19}%
  \BibitemOpen
  \bibfield  {author} {\bibinfo {author} {\bibfnamefont {M.}~\bibnamefont
  {Pfender}}, \bibinfo {author} {\bibfnamefont {P.}~\bibnamefont {Wang}},
  \bibinfo {author} {\bibfnamefont {H.}~\bibnamefont {Sumiya}}, \bibinfo
  {author} {\bibfnamefont {S.}~\bibnamefont {Onoda}}, \bibinfo {author}
  {\bibfnamefont {W.}~\bibnamefont {Yang}}, \bibinfo {author} {\bibfnamefont
  {D.~B.~R.}\ \bibnamefont {Dasari}}, \bibinfo {author} {\bibfnamefont
  {P.}~\bibnamefont {Neumann}}, \bibinfo {author} {\bibfnamefont {X.-Y.}\
  \bibnamefont {Pan}}, \bibinfo {author} {\bibfnamefont {J.}~\bibnamefont
  {Isoya}}, \bibinfo {author} {\bibfnamefont {R.-B.}\ \bibnamefont {Liu}}, \
  and\ \bibinfo {author} {\bibfnamefont {J.}~\bibnamefont {Wrachtrup}},\
  }\bibfield  {title} {\enquote {\bibinfo {title} {High-resolution spectroscopy
  of single nuclear spins via sequential weak measurements},}\ }\href@noop {}
  {\bibfield  {journal} {\bibinfo  {journal} {Nat.\ Commun.}\ }\textbf
  {\bibinfo {volume} {10}},\ \bibinfo {pages} {594} (\bibinfo {year}
  {2019})}\BibitemShut {NoStop}%
\bibitem [{\citenamefont {Cujia}\ \emph {et~al.}(2019)\citenamefont {Cujia},
  \citenamefont {Boss}, \citenamefont {Herb}, \citenamefont {Zopes},\ and\
  \citenamefont {Degen}}]{CBH+19}%
  \BibitemOpen
  \bibfield  {author} {\bibinfo {author} {\bibfnamefont {K.~S.}\ \bibnamefont
  {Cujia}}, \bibinfo {author} {\bibfnamefont {J.~M.}\ \bibnamefont {Boss}},
  \bibinfo {author} {\bibfnamefont {K.}~\bibnamefont {Herb}}, \bibinfo {author}
  {\bibfnamefont {J.}~\bibnamefont {Zopes}}, \ and\ \bibinfo {author}
  {\bibfnamefont {C.~L.}\ \bibnamefont {Degen}},\ }\bibfield  {title} {\enquote
  {\bibinfo {title} {Tracking the precession of single nuclear spins by weak
  measurements},}\ }\href@noop {} {\bibfield  {journal} {\bibinfo  {journal}
  {Nature}\ }\textbf {\bibinfo {volume} {571}},\ \bibinfo {pages} {230}
  (\bibinfo {year} {2019})}\BibitemShut {NoStop}%
\bibitem [{\citenamefont {Zhao}\ \emph {et~al.}(2012)\citenamefont {Zhao},
  \citenamefont {Honert}, \citenamefont {Schmid}, \citenamefont {Klas},
  \citenamefont {Isoya}, \citenamefont {Markham}, \citenamefont {Twitchen},
  \citenamefont {Jelezko}, \citenamefont {Liu}, \citenamefont {Fedder},\ and\
  \citenamefont {Wrachtrup}}]{ZHS+12}%
  \BibitemOpen
  \bibfield  {author} {\bibinfo {author} {\bibfnamefont {N.}~\bibnamefont
  {Zhao}}, \bibinfo {author} {\bibfnamefont {J.}~\bibnamefont {Honert}},
  \bibinfo {author} {\bibfnamefont {B.}~\bibnamefont {Schmid}}, \bibinfo
  {author} {\bibfnamefont {M.}~\bibnamefont {Klas}}, \bibinfo {author}
  {\bibfnamefont {J.}~\bibnamefont {Isoya}}, \bibinfo {author} {\bibfnamefont
  {M.}~\bibnamefont {Markham}}, \bibinfo {author} {\bibfnamefont
  {D.}~\bibnamefont {Twitchen}}, \bibinfo {author} {\bibfnamefont
  {F.}~\bibnamefont {Jelezko}}, \bibinfo {author} {\bibfnamefont {R.~B.}\
  \bibnamefont {Liu}}, \bibinfo {author} {\bibfnamefont {H.}~\bibnamefont
  {Fedder}}, \ and\ \bibinfo {author} {\bibfnamefont {J.}~\bibnamefont
  {Wrachtrup}},\ }\bibfield  {title} {\enquote {\bibinfo {title} {Sensing
  single remote nuclear spins},}\ }\href@noop {} {\bibfield  {journal}
  {\bibinfo  {journal} {Nat.\ Nanotechnol.}\ }\textbf {\bibinfo {volume} {7}},\
  \bibinfo {pages} {657} (\bibinfo {year} {2012})}\BibitemShut {NoStop}%
\bibitem [{\citenamefont {Zopes}\ \emph
  {et~al.}(2018{\natexlab{a}})\citenamefont {Zopes}, \citenamefont {Cujia},
  \citenamefont {Sasaki}, \citenamefont {Boss}, \citenamefont {Itoh},\ and\
  \citenamefont {Degen}}]{ZCS+18}%
  \BibitemOpen
  \bibfield  {author} {\bibinfo {author} {\bibfnamefont {J.}~\bibnamefont
  {Zopes}}, \bibinfo {author} {\bibfnamefont {K.~S.}\ \bibnamefont {Cujia}},
  \bibinfo {author} {\bibfnamefont {K.}~\bibnamefont {Sasaki}}, \bibinfo
  {author} {\bibfnamefont {J.~M.}\ \bibnamefont {Boss}}, \bibinfo {author}
  {\bibfnamefont {K.~M.}\ \bibnamefont {Itoh}}, \ and\ \bibinfo {author}
  {\bibfnamefont {C.~L.}\ \bibnamefont {Degen}},\ }\bibfield  {title} {\enquote
  {\bibinfo {title} {Three-dimensional localization spectroscopy of individual
  nuclear spins with sub-{A}ngstrom resolution},}\ }\href@noop {} {\bibfield
  {journal} {\bibinfo  {journal} {Nat.\ Commun.}\ }\textbf {\bibinfo {volume}
  {9}},\ \bibinfo {pages} {4678} (\bibinfo {year}
  {2018}{\natexlab{a}})}\BibitemShut {NoStop}%
\bibitem [{\citenamefont {Zopes}\ \emph
  {et~al.}(2018{\natexlab{b}})\citenamefont {Zopes}, \citenamefont {Herb},
  \citenamefont {Cujia},\ and\ \citenamefont {Degen}}]{ZHCD18}%
  \BibitemOpen
  \bibfield  {author} {\bibinfo {author} {\bibfnamefont {J.}~\bibnamefont
  {Zopes}}, \bibinfo {author} {\bibfnamefont {K.}~\bibnamefont {Herb}},
  \bibinfo {author} {\bibfnamefont {K.~S.}\ \bibnamefont {Cujia}}, \ and\
  \bibinfo {author} {\bibfnamefont {C.~L.}\ \bibnamefont {Degen}},\ }\bibfield
  {title} {\enquote {\bibinfo {title} {Three-{D}imensional {N}uclear {S}pin
  {P}ositioning {U}sing {C}oherent {R}adio-{F}requency {C}ontrol},}\
  }\href@noop {} {\bibfield  {journal} {\bibinfo  {journal} {Phys.\ Rev.\
  Lett.}\ }\textbf {\bibinfo {volume} {121}},\ \bibinfo {pages} {170801}
  (\bibinfo {year} {2018}{\natexlab{b}})}\BibitemShut {NoStop}%
\bibitem [{\citenamefont {Sasaki}, \citenamefont {Itoh},\ and\ \citenamefont
  {Abe}(2018)}]{SIA18}%
  \BibitemOpen
  \bibfield  {author} {\bibinfo {author} {\bibfnamefont {K.}~\bibnamefont
  {Sasaki}}, \bibinfo {author} {\bibfnamefont {K.~M.}\ \bibnamefont {Itoh}}, \
  and\ \bibinfo {author} {\bibfnamefont {E.}~\bibnamefont {Abe}},\ }\bibfield
  {title} {\enquote {\bibinfo {title} {Determination of the position of a
  single nuclear spin from free nuclear precessions detected by a solid-state
  quantum sensor},}\ }\href@noop {} {\bibfield  {journal} {\bibinfo  {journal}
  {Phys.\ Rev.\ B}\ }\textbf {\bibinfo {volume} {98}},\ \bibinfo {pages}
  {121405} (\bibinfo {year} {2018})}\BibitemShut {NoStop}%
\bibitem [{\citenamefont {Abobeih}\ \emph {et~al.}(2019)\citenamefont
  {Abobeih}, \citenamefont {Randall}, \citenamefont {Bradley}, \citenamefont
  {Bartling}, \citenamefont {Bakker}, \citenamefont {Degen}, \citenamefont
  {Markham}, \citenamefont {Twitchen},\ and\ \citenamefont
  {Taminiau}}]{ARB+19}%
  \BibitemOpen
  \bibfield  {author} {\bibinfo {author} {\bibfnamefont {M.~H.}\ \bibnamefont
  {Abobeih}}, \bibinfo {author} {\bibfnamefont {J.}~\bibnamefont {Randall}},
  \bibinfo {author} {\bibfnamefont {C.~E.}\ \bibnamefont {Bradley}}, \bibinfo
  {author} {\bibfnamefont {H.~P.}\ \bibnamefont {Bartling}}, \bibinfo {author}
  {\bibfnamefont {M.~A.}\ \bibnamefont {Bakker}}, \bibinfo {author}
  {\bibfnamefont {M.~J.}\ \bibnamefont {Degen}}, \bibinfo {author}
  {\bibfnamefont {M.}~\bibnamefont {Markham}}, \bibinfo {author} {\bibfnamefont
  {D.~J.}\ \bibnamefont {Twitchen}}, \ and\ \bibinfo {author} {\bibfnamefont
  {T.~H.}\ \bibnamefont {Taminiau}},\ }\bibfield  {title} {\enquote {\bibinfo
  {title} {Atomic-scale imaging of a 27-nuclear-spin cluster using a quantum
  sensor},}\ }\href@noop {} {\bibfield  {journal} {\bibinfo  {journal}
  {Nature}\ }\textbf {\bibinfo {volume} {576}},\ \bibinfo {pages} {411}
  (\bibinfo {year} {2019})}\BibitemShut {NoStop}%
\bibitem [{att()}]{attocube}%
  \BibitemOpen
  \href@noop {} {}\bibinfo {note} {{a}ttocube
  (https://www.attocube.com/)}\BibitemShut {NoStop}%
\bibitem [{qna()}]{qnami}%
  \BibitemOpen
  \href@noop {} {}\bibinfo {note} {Qnami (https://qnami.ch/)}\BibitemShut
  {NoStop}%
\bibitem [{qut()}]{qutools}%
  \BibitemOpen
  \href@noop {} {}\bibinfo {note} {{q}utools
  (https://www.qutools.com/)}\BibitemShut {NoStop}%
\bibitem [{qza()}]{qzabre}%
  \BibitemOpen
  \href@noop {} {}\bibinfo {note} {{QZ}abre (https://qzabre.com/)}\BibitemShut
  {NoStop}%
\bibitem [{\citenamefont {Zhang}\ \emph {et~al.}(2018)\citenamefont {Zhang},
  \citenamefont {Belvin}, \citenamefont {Li}, \citenamefont {Wang},
  \citenamefont {Wainwright}, \citenamefont {Berg},\ and\ \citenamefont
  {Bridger}}]{ZBL+18}%
  \BibitemOpen
  \bibfield  {author} {\bibinfo {author} {\bibfnamefont {H.}~\bibnamefont
  {Zhang}}, \bibinfo {author} {\bibfnamefont {C.}~\bibnamefont {Belvin}},
  \bibinfo {author} {\bibfnamefont {W.}~\bibnamefont {Li}}, \bibinfo {author}
  {\bibfnamefont {J.}~\bibnamefont {Wang}}, \bibinfo {author} {\bibfnamefont
  {J.}~\bibnamefont {Wainwright}}, \bibinfo {author} {\bibfnamefont
  {R.}~\bibnamefont {Berg}}, \ and\ \bibinfo {author} {\bibfnamefont
  {J.}~\bibnamefont {Bridger}},\ }\bibfield  {title} {\enquote {\bibinfo
  {title} {Little bits of diamond: {O}ptically detected magnetic resonance of
  nitrogen-vacancy centers},}\ }\href@noop {} {\bibfield  {journal} {\bibinfo
  {journal} {Am.\ J.\ Phys.}\ }\textbf {\bibinfo {volume} {86}},\ \bibinfo
  {pages} {225} (\bibinfo {year} {2018})}\BibitemShut {NoStop}%
\bibitem [{\citenamefont {Bucher}\ \emph {et~al.}(2019)\citenamefont {Bucher},
  \citenamefont {Aude~Craik}, \citenamefont {Backlund}, \citenamefont {Turner},
  \citenamefont {Ben-Dor}, \citenamefont {Glenn},\ and\ \citenamefont
  {Walsworth}}]{BAB+19}%
  \BibitemOpen
  \bibfield  {author} {\bibinfo {author} {\bibfnamefont {D.~B.}\ \bibnamefont
  {Bucher}}, \bibinfo {author} {\bibfnamefont {D.~P.~L.}\ \bibnamefont
  {Aude~Craik}}, \bibinfo {author} {\bibfnamefont {M.~P.}\ \bibnamefont
  {Backlund}}, \bibinfo {author} {\bibfnamefont {M.~J.}\ \bibnamefont
  {Turner}}, \bibinfo {author} {\bibfnamefont {O.}~\bibnamefont {Ben-Dor}},
  \bibinfo {author} {\bibfnamefont {D.~R.}\ \bibnamefont {Glenn}}, \ and\
  \bibinfo {author} {\bibfnamefont {R.~L.}\ \bibnamefont {Walsworth}},\
  }\bibfield  {title} {\enquote {\bibinfo {title} {Quantum diamond spectrometer
  for nanoscale {NMR} and {ESR} spectroscopy},}\ }\href@noop {} {\bibfield
  {journal} {\bibinfo  {journal} {Nat. Protoc.}\ }\textbf {\bibinfo {volume}
  {14}},\ \bibinfo {pages} {2707} (\bibinfo {year} {2019})}\BibitemShut
  {NoStop}%
\bibitem [{\citenamefont {St{\"u}ner}\ \emph {et~al.}(2019)\citenamefont
  {St{\"u}ner}, \citenamefont {Brenneis}, \citenamefont {Kassel}, \citenamefont
  {Wostradowski}, \citenamefont {R{\"o}lver}, \citenamefont {Fuchs},
  \citenamefont {Nakamura}, \citenamefont {Sumiya}, \citenamefont {Onoda},
  \citenamefont {Isoya},\ and\ \citenamefont {Jelezko}}]{SBK+19}%
  \BibitemOpen
  \bibfield  {author} {\bibinfo {author} {\bibfnamefont {F.~M.}\ \bibnamefont
  {St{\"u}ner}}, \bibinfo {author} {\bibfnamefont {A.}~\bibnamefont
  {Brenneis}}, \bibinfo {author} {\bibfnamefont {J.}~\bibnamefont {Kassel}},
  \bibinfo {author} {\bibfnamefont {U.}~\bibnamefont {Wostradowski}}, \bibinfo
  {author} {\bibfnamefont {R.}~\bibnamefont {R{\"o}lver}}, \bibinfo {author}
  {\bibfnamefont {T.}~\bibnamefont {Fuchs}}, \bibinfo {author} {\bibfnamefont
  {K.}~\bibnamefont {Nakamura}}, \bibinfo {author} {\bibfnamefont
  {H.}~\bibnamefont {Sumiya}}, \bibinfo {author} {\bibfnamefont
  {S.}~\bibnamefont {Onoda}}, \bibinfo {author} {\bibfnamefont
  {J.}~\bibnamefont {Isoya}}, \ and\ \bibinfo {author} {\bibfnamefont
  {F.}~\bibnamefont {Jelezko}},\ }\bibfield  {title} {\enquote {\bibinfo
  {title} {Compact integrated magnetometer based on nitrogen-vacancy centres in
  diamond},}\ }\href@noop {} {\bibfield  {journal} {\bibinfo  {journal} {Diam.\
  Relat.\ Mater.}\ }\textbf {\bibinfo {volume} {93}},\ \bibinfo {pages} {59}
  (\bibinfo {year} {2019})}\BibitemShut {NoStop}%
\bibitem [{\citenamefont {Webb}\ \emph {et~al.}(2019)\citenamefont {Webb},
  \citenamefont {Clement}, \citenamefont {Troise}, \citenamefont {Ahmadi},
  \citenamefont {Johansen}, \citenamefont {Huck},\ and\ \citenamefont
  {Andersen}}]{WCT+19}%
  \BibitemOpen
  \bibfield  {author} {\bibinfo {author} {\bibfnamefont {J.~L.}\ \bibnamefont
  {Webb}}, \bibinfo {author} {\bibfnamefont {J.~D.}\ \bibnamefont {Clement}},
  \bibinfo {author} {\bibfnamefont {L.}~\bibnamefont {Troise}}, \bibinfo
  {author} {\bibfnamefont {S.}~\bibnamefont {Ahmadi}}, \bibinfo {author}
  {\bibfnamefont {G.~J.}\ \bibnamefont {Johansen}}, \bibinfo {author}
  {\bibfnamefont {A.}~\bibnamefont {Huck}}, \ and\ \bibinfo {author}
  {\bibfnamefont {U.~L.}\ \bibnamefont {Andersen}},\ }\bibfield  {title}
  {\enquote {\bibinfo {title} {Nanotesla sensitivity magnetic field sensing
  using a compact diamond nitrogen-vacancy magnetometer},}\ }\href@noop {}
  {\bibfield  {journal} {\bibinfo  {journal} {Appl.\ Phys.\ Lett.}\ }\textbf
  {\bibinfo {volume} {114}},\ \bibinfo {pages} {231103} (\bibinfo {year}
  {2019})}\BibitemShut {NoStop}%
\bibitem [{\citenamefont {Kim}\ \emph {et~al.}(2019)\citenamefont {Kim},
  \citenamefont {Ibrahim}, \citenamefont {Foy}, \citenamefont {Trusheim},
  \citenamefont {Han},\ and\ \citenamefont {Englund}}]{KIF+19}%
  \BibitemOpen
  \bibfield  {author} {\bibinfo {author} {\bibfnamefont {D.}~\bibnamefont
  {Kim}}, \bibinfo {author} {\bibfnamefont {M.~I.}\ \bibnamefont {Ibrahim}},
  \bibinfo {author} {\bibfnamefont {C.}~\bibnamefont {Foy}}, \bibinfo {author}
  {\bibfnamefont {M.~E.}\ \bibnamefont {Trusheim}}, \bibinfo {author}
  {\bibfnamefont {R.}~\bibnamefont {Han}}, \ and\ \bibinfo {author}
  {\bibfnamefont {D.~R.}\ \bibnamefont {Englund}},\ }\bibfield  {title}
  {\enquote {\bibinfo {title} {A {CMOS}-integrated quantum sensor based on
  nitrogen-vacancy centres},}\ }\href@noop {} {\bibfield  {journal} {\bibinfo
  {journal} {Nat.\ Electron.}\ }\textbf {\bibinfo {volume} {2}},\ \bibinfo
  {pages} {284} (\bibinfo {year} {2019})}\BibitemShut {NoStop}%
\bibitem [{\citenamefont {Jelezko}\ and\ \citenamefont
  {Wrachtrup}(2006)}]{JW06}%
  \BibitemOpen
  \bibfield  {author} {\bibinfo {author} {\bibfnamefont {F.}~\bibnamefont
  {Jelezko}}\ and\ \bibinfo {author} {\bibfnamefont {J.}~\bibnamefont
  {Wrachtrup}},\ }\bibfield  {title} {\enquote {\bibinfo {title} {Single defect
  centres in diamond: {A} review},}\ }\href@noop {} {\bibfield  {journal}
  {\bibinfo  {journal} {phys.\ stat.\ sol.\ (a)}\ }\textbf {\bibinfo {volume}
  {203}},\ \bibinfo {pages} {3207} (\bibinfo {year} {2006})}\BibitemShut
  {NoStop}%
\bibitem [{\citenamefont {Doherty}\ \emph {et~al.}(2011)\citenamefont
  {Doherty}, \citenamefont {Manson}, \citenamefont {Delaney},\ and\
  \citenamefont {Hollenberg}}]{DMD+11}%
  \BibitemOpen
  \bibfield  {author} {\bibinfo {author} {\bibfnamefont {M.~W.}\ \bibnamefont
  {Doherty}}, \bibinfo {author} {\bibfnamefont {N.~B.}\ \bibnamefont {Manson}},
  \bibinfo {author} {\bibfnamefont {P.}~\bibnamefont {Delaney}}, \ and\
  \bibinfo {author} {\bibfnamefont {L.~C.~L.}\ \bibnamefont {Hollenberg}},\
  }\bibfield  {title} {\enquote {\bibinfo {title} {The negatively charged
  nitrogen-vacancy centre in diamond: the electronic solution},}\ }\href@noop
  {} {\bibfield  {journal} {\bibinfo  {journal} {New J.\ Phys.}\ }\textbf
  {\bibinfo {volume} {13}},\ \bibinfo {pages} {025019} (\bibinfo {year}
  {2011})}\BibitemShut {NoStop}%
\bibitem [{\citenamefont {Maze}\ \emph {et~al.}(2011)\citenamefont {Maze},
  \citenamefont {Gali}, \citenamefont {Togan}, \citenamefont {Chu},
  \citenamefont {Trifonov}, \citenamefont {Kaxiras},\ and\ \citenamefont
  {Lukin}}]{MGT+11}%
  \BibitemOpen
  \bibfield  {author} {\bibinfo {author} {\bibfnamefont {J.~R.}\ \bibnamefont
  {Maze}}, \bibinfo {author} {\bibfnamefont {A.}~\bibnamefont {Gali}}, \bibinfo
  {author} {\bibfnamefont {E.}~\bibnamefont {Togan}}, \bibinfo {author}
  {\bibfnamefont {Y.}~\bibnamefont {Chu}}, \bibinfo {author} {\bibfnamefont
  {A.}~\bibnamefont {Trifonov}}, \bibinfo {author} {\bibfnamefont
  {E.}~\bibnamefont {Kaxiras}}, \ and\ \bibinfo {author} {\bibfnamefont
  {M.~D.}\ \bibnamefont {Lukin}},\ }\bibfield  {title} {\enquote {\bibinfo
  {title} {Properties of nitrogen-vacancy centers in diamond: the group
  theoretic approach},}\ }\href@noop {} {\bibfield  {journal} {\bibinfo
  {journal} {New J.\ Phys.}\ }\textbf {\bibinfo {volume} {13}},\ \bibinfo
  {pages} {025015} (\bibinfo {year} {2011})}\BibitemShut {NoStop}%
\bibitem [{\citenamefont {Doherty}\ \emph {et~al.}(2013)\citenamefont
  {Doherty}, \citenamefont {Manson}, \citenamefont {Delaney}, \citenamefont
  {Jelezko}, \citenamefont {Wrachtrup},\ and\ \citenamefont
  {Hollenberg}}]{DMD+13}%
  \BibitemOpen
  \bibfield  {author} {\bibinfo {author} {\bibfnamefont {M.~W.}\ \bibnamefont
  {Doherty}}, \bibinfo {author} {\bibfnamefont {N.~B.}\ \bibnamefont {Manson}},
  \bibinfo {author} {\bibfnamefont {P.}~\bibnamefont {Delaney}}, \bibinfo
  {author} {\bibfnamefont {F.}~\bibnamefont {Jelezko}}, \bibinfo {author}
  {\bibfnamefont {J.}~\bibnamefont {Wrachtrup}}, \ and\ \bibinfo {author}
  {\bibfnamefont {L.~C.~L.}\ \bibnamefont {Hollenberg}},\ }\bibfield  {title}
  {\enquote {\bibinfo {title} {The nitrogen-vacancy colour centre in
  diamond},}\ }\href@noop {} {\bibfield  {journal} {\bibinfo  {journal} {Phys.\
  Rep.}\ }\textbf {\bibinfo {volume} {528}},\ \bibinfo {pages} {1} (\bibinfo
  {year} {2013})}\BibitemShut {NoStop}%
\bibitem [{\citenamefont {Ohashi}\ \emph {et~al.}(2013)\citenamefont {Ohashi},
  \citenamefont {Rosskopf}, \citenamefont {Watanabe}, \citenamefont {Loretz},
  \citenamefont {Tao}, \citenamefont {Hauert}, \citenamefont {Tomizawa},
  \citenamefont {Ishikawa}, \citenamefont {Ishi-Hayase}, \citenamefont
  {Shikata}, \citenamefont {Degen},\ and\ \citenamefont {Itoh}}]{ORW+13}%
  \BibitemOpen
  \bibfield  {author} {\bibinfo {author} {\bibfnamefont {K.}~\bibnamefont
  {Ohashi}}, \bibinfo {author} {\bibfnamefont {T.}~\bibnamefont {Rosskopf}},
  \bibinfo {author} {\bibfnamefont {H.}~\bibnamefont {Watanabe}}, \bibinfo
  {author} {\bibfnamefont {M.}~\bibnamefont {Loretz}}, \bibinfo {author}
  {\bibfnamefont {Y.}~\bibnamefont {Tao}}, \bibinfo {author} {\bibfnamefont
  {R.}~\bibnamefont {Hauert}}, \bibinfo {author} {\bibfnamefont
  {S.}~\bibnamefont {Tomizawa}}, \bibinfo {author} {\bibfnamefont
  {T.}~\bibnamefont {Ishikawa}}, \bibinfo {author} {\bibfnamefont
  {J.}~\bibnamefont {Ishi-Hayase}}, \bibinfo {author} {\bibfnamefont
  {S.}~\bibnamefont {Shikata}}, \bibinfo {author} {\bibfnamefont {C.~L.}\
  \bibnamefont {Degen}}, \ and\ \bibinfo {author} {\bibfnamefont {K.~M.}\
  \bibnamefont {Itoh}},\ }\bibfield  {title} {\enquote {\bibinfo {title}
  {Negatively {C}harged {N}itrogen-{V}acancy {C}enters in a 5 nm {T}hin
  $^{12}${C} {D}iamond {F}ilm},}\ }\href@noop {} {\bibfield  {journal}
  {\bibinfo  {journal} {Nano Lett.}\ }\textbf {\bibinfo {volume} {13}},\
  \bibinfo {pages} {4733} (\bibinfo {year} {2013})}\BibitemShut {NoStop}%
\bibitem [{\citenamefont {Meijer}\ \emph {et~al.}(2005)\citenamefont {Meijer},
  \citenamefont {Burchard}, \citenamefont {Domhan}, \citenamefont {Wittmann},
  \citenamefont {Gaebel}, \citenamefont {Popa}, \citenamefont {Jelezko},\ and\
  \citenamefont {Wrachtrup}}]{MBD+05}%
  \BibitemOpen
  \bibfield  {author} {\bibinfo {author} {\bibfnamefont {J.}~\bibnamefont
  {Meijer}}, \bibinfo {author} {\bibfnamefont {B.}~\bibnamefont {Burchard}},
  \bibinfo {author} {\bibfnamefont {M.}~\bibnamefont {Domhan}}, \bibinfo
  {author} {\bibfnamefont {C.}~\bibnamefont {Wittmann}}, \bibinfo {author}
  {\bibfnamefont {T.}~\bibnamefont {Gaebel}}, \bibinfo {author} {\bibfnamefont
  {I.}~\bibnamefont {Popa}}, \bibinfo {author} {\bibfnamefont {F.}~\bibnamefont
  {Jelezko}}, \ and\ \bibinfo {author} {\bibfnamefont {J.}~\bibnamefont
  {Wrachtrup}},\ }\bibfield  {title} {\enquote {\bibinfo {title} {Generation of
  single color centers by focused nitrogen implantation},}\ }\href@noop {}
  {\bibfield  {journal} {\bibinfo  {journal} {Appl.\ Phys.\ Lett}\ }\textbf
  {\bibinfo {volume} {87}},\ \bibinfo {pages} {261909} (\bibinfo {year}
  {2005})}\BibitemShut {NoStop}%
\bibitem [{\citenamefont {Ofori-Okai}\ \emph {et~al.}(2012)\citenamefont
  {Ofori-Okai}, \citenamefont {Pezzagna}, \citenamefont {Chang}, \citenamefont
  {Loretz}, \citenamefont {Schirhagl}, \citenamefont {Tao}, \citenamefont
  {Moores}, \citenamefont {Groot-Berning}, \citenamefont {Meijer},\ and\
  \citenamefont {Degen}}]{OPC+12}%
  \BibitemOpen
  \bibfield  {author} {\bibinfo {author} {\bibfnamefont {B.~K.}\ \bibnamefont
  {Ofori-Okai}}, \bibinfo {author} {\bibfnamefont {S.}~\bibnamefont
  {Pezzagna}}, \bibinfo {author} {\bibfnamefont {K.}~\bibnamefont {Chang}},
  \bibinfo {author} {\bibfnamefont {M.}~\bibnamefont {Loretz}}, \bibinfo
  {author} {\bibfnamefont {R.}~\bibnamefont {Schirhagl}}, \bibinfo {author}
  {\bibfnamefont {Y.}~\bibnamefont {Tao}}, \bibinfo {author} {\bibfnamefont
  {B.~A.}\ \bibnamefont {Moores}}, \bibinfo {author} {\bibfnamefont
  {K.}~\bibnamefont {Groot-Berning}}, \bibinfo {author} {\bibfnamefont
  {J.}~\bibnamefont {Meijer}}, \ and\ \bibinfo {author} {\bibfnamefont {C.~L.}\
  \bibnamefont {Degen}},\ }\bibfield  {title} {\enquote {\bibinfo {title} {Spin
  properties of very shallow nitrogen vacancy defects in diamond},}\
  }\href@noop {} {\bibfield  {journal} {\bibinfo  {journal} {Phys.\ Rev.\ B}\
  }\textbf {\bibinfo {volume} {86}},\ \bibinfo {pages} {081406} (\bibinfo
  {year} {2012})}\BibitemShut {NoStop}%
\bibitem [{\citenamefont {Ohno}\ \emph {et~al.}(2012)\citenamefont {Ohno},
  \citenamefont {Heremans}, \citenamefont {Bassett}, \citenamefont {Myers},
  \citenamefont {Toyli}, \citenamefont {Bleszynski~Jayich}, \citenamefont
  {Palmstr{\/o}m},\ and\ \citenamefont {Awschalom}}]{OHB+12}%
  \BibitemOpen
  \bibfield  {author} {\bibinfo {author} {\bibfnamefont {K.}~\bibnamefont
  {Ohno}}, \bibinfo {author} {\bibfnamefont {F.~J.}\ \bibnamefont {Heremans}},
  \bibinfo {author} {\bibfnamefont {L.~C.}\ \bibnamefont {Bassett}}, \bibinfo
  {author} {\bibfnamefont {B.~A.}\ \bibnamefont {Myers}}, \bibinfo {author}
  {\bibfnamefont {D.~M.}\ \bibnamefont {Toyli}}, \bibinfo {author}
  {\bibfnamefont {A.~C.}\ \bibnamefont {Bleszynski~Jayich}}, \bibinfo {author}
  {\bibfnamefont {C.~J.}\ \bibnamefont {Palmstr{\/o}m}}, \ and\ \bibinfo
  {author} {\bibfnamefont {D.~D.}\ \bibnamefont {Awschalom}},\ }\bibfield
  {title} {\enquote {\bibinfo {title} {Engineering shallow spins in diamond
  with nitrogen delta-doping},}\ }\href@noop {} {\bibfield  {journal} {\bibinfo
   {journal} {Appl.\ Phys.\ Lett.}\ }\textbf {\bibinfo {volume} {101}},\
  \bibinfo {pages} {082413} (\bibinfo {year} {2012})}\BibitemShut {NoStop}%
\bibitem [{\citenamefont {Ohno}\ \emph {et~al.}(2014)\citenamefont {Ohno},
  \citenamefont {Heremans}, \citenamefont {de~las Casas}, \citenamefont
  {Myers}, \citenamefont {Alem{\'a}n}, \citenamefont {Bleszynski~Jayich},\ and\
  \citenamefont {Awschalom}}]{OHdlC+14}%
  \BibitemOpen
  \bibfield  {author} {\bibinfo {author} {\bibfnamefont {K.}~\bibnamefont
  {Ohno}}, \bibinfo {author} {\bibfnamefont {F.~J.}\ \bibnamefont {Heremans}},
  \bibinfo {author} {\bibfnamefont {C.~F.}\ \bibnamefont {de~las Casas}},
  \bibinfo {author} {\bibfnamefont {B.~A.}\ \bibnamefont {Myers}}, \bibinfo
  {author} {\bibfnamefont {B.~J.}\ \bibnamefont {Alem{\'a}n}}, \bibinfo
  {author} {\bibfnamefont {A.~C.}\ \bibnamefont {Bleszynski~Jayich}}, \ and\
  \bibinfo {author} {\bibfnamefont {D.~D.}\ \bibnamefont {Awschalom}},\
  }\bibfield  {title} {\enquote {\bibinfo {title} {Three-dimensional
  localization of spins in diamond using $^{12}${C} implantation},}\
  }\href@noop {} {\bibfield  {journal} {\bibinfo  {journal} {Appl.\ Phys.\
  Lett.}\ }\textbf {\bibinfo {volume} {105}},\ \bibinfo {pages} {052406}
  (\bibinfo {year} {2014})}\BibitemShut {NoStop}%
\bibitem [{\citenamefont {Huang}\ \emph {et~al.}(2013)\citenamefont {Huang},
  \citenamefont {Li}, \citenamefont {Santori}, \citenamefont {Acosta},
  \citenamefont {Faraon}, \citenamefont {Ishikawa}, \citenamefont {Wu},
  \citenamefont {Winston}, \citenamefont {Williams},\ and\ \citenamefont
  {Beausoleil}}]{HLS+13}%
  \BibitemOpen
  \bibfield  {author} {\bibinfo {author} {\bibfnamefont {Z.}~\bibnamefont
  {Huang}}, \bibinfo {author} {\bibfnamefont {W.-D.}\ \bibnamefont {Li}},
  \bibinfo {author} {\bibfnamefont {C.}~\bibnamefont {Santori}}, \bibinfo
  {author} {\bibfnamefont {V.~M.}\ \bibnamefont {Acosta}}, \bibinfo {author}
  {\bibfnamefont {A.}~\bibnamefont {Faraon}}, \bibinfo {author} {\bibfnamefont
  {T.}~\bibnamefont {Ishikawa}}, \bibinfo {author} {\bibfnamefont
  {W.}~\bibnamefont {Wu}}, \bibinfo {author} {\bibfnamefont {D.}~\bibnamefont
  {Winston}}, \bibinfo {author} {\bibfnamefont {R.~S.}\ \bibnamefont
  {Williams}}, \ and\ \bibinfo {author} {\bibfnamefont {R.~G.}\ \bibnamefont
  {Beausoleil}},\ }\bibfield  {title} {\enquote {\bibinfo {title} {Diamond
  nitrogen-vacancy centers created by scanning focused helium ion beam and
  annealing},}\ }\href@noop {} {\bibfield  {journal} {\bibinfo  {journal}
  {Appl.\ Phys.\ Lett.}\ }\textbf {\bibinfo {volume} {103}},\ \bibinfo {pages}
  {081906} (\bibinfo {year} {2013})}\BibitemShut {NoStop}%
\bibitem [{\citenamefont {Wrachtrup}\ \emph {et~al.}(2013)\citenamefont
  {Wrachtrup}, \citenamefont {Jelezko}, \citenamefont {Grotz},\ and\
  \citenamefont {McGuinness}}]{WJGM13}%
  \BibitemOpen
  \bibfield  {author} {\bibinfo {author} {\bibfnamefont {J.}~\bibnamefont
  {Wrachtrup}}, \bibinfo {author} {\bibfnamefont {F.}~\bibnamefont {Jelezko}},
  \bibinfo {author} {\bibfnamefont {B.}~\bibnamefont {Grotz}}, \ and\ \bibinfo
  {author} {\bibfnamefont {L.}~\bibnamefont {McGuinness}},\ }\bibfield  {title}
  {\enquote {\bibinfo {title} {Nitrogen-vacancy centers close to surfaces},}\
  }\href@noop {} {\bibfield  {journal} {\bibinfo  {journal} {MRS Bull.}\
  }\textbf {\bibinfo {volume} {38}},\ \bibinfo {pages} {149} (\bibinfo {year}
  {2013})}\BibitemShut {NoStop}%
\bibitem [{\citenamefont {Rittweger}\ \emph {et~al.}(2009)\citenamefont
  {Rittweger}, \citenamefont {Han}, \citenamefont {Irvine}, \citenamefont
  {Eggeling},\ and\ \citenamefont {Hell}}]{RHI+09}%
  \BibitemOpen
  \bibfield  {author} {\bibinfo {author} {\bibfnamefont {E.}~\bibnamefont
  {Rittweger}}, \bibinfo {author} {\bibfnamefont {K.~Y.}\ \bibnamefont {Han}},
  \bibinfo {author} {\bibfnamefont {S.~E.}\ \bibnamefont {Irvine}}, \bibinfo
  {author} {\bibfnamefont {C.}~\bibnamefont {Eggeling}}, \ and\ \bibinfo
  {author} {\bibfnamefont {S.~W.}\ \bibnamefont {Hell}},\ }\bibfield  {title}
  {\enquote {\bibinfo {title} {{STED} microscopy reveals crystal colour centres
  with nanometric resolution},}\ }\href@noop {} {\bibfield  {journal} {\bibinfo
   {journal} {Nat.\ Photon.}\ }\textbf {\bibinfo {volume} {3}},\ \bibinfo
  {pages} {144} (\bibinfo {year} {2009})}\BibitemShut {NoStop}%
\bibitem [{\citenamefont {Maurer}\ \emph {et~al.}(2010)\citenamefont {Maurer},
  \citenamefont {Maze}, \citenamefont {Stanwix}, \citenamefont {Jiang},
  \citenamefont {Gorshkov}, \citenamefont {Zibrov}, \citenamefont {Harke},
  \citenamefont {Hodges}, \citenamefont {Zibrov}, \citenamefont {Yacoby},
  \citenamefont {Twitchen}, \citenamefont {Hell}, \citenamefont {Walsworth}, ,\
  and\ \citenamefont {Lukin}}]{MMS+10}%
  \BibitemOpen
  \bibfield  {author} {\bibinfo {author} {\bibfnamefont {P.~C.}\ \bibnamefont
  {Maurer}}, \bibinfo {author} {\bibfnamefont {J.~R.}\ \bibnamefont {Maze}},
  \bibinfo {author} {\bibfnamefont {P.~L.}\ \bibnamefont {Stanwix}}, \bibinfo
  {author} {\bibfnamefont {L.}~\bibnamefont {Jiang}}, \bibinfo {author}
  {\bibfnamefont {A.~V.}\ \bibnamefont {Gorshkov}}, \bibinfo {author}
  {\bibfnamefont {A.~A.}\ \bibnamefont {Zibrov}}, \bibinfo {author}
  {\bibfnamefont {B.}~\bibnamefont {Harke}}, \bibinfo {author} {\bibfnamefont
  {J.~S.}\ \bibnamefont {Hodges}}, \bibinfo {author} {\bibfnamefont {A.~S.}\
  \bibnamefont {Zibrov}}, \bibinfo {author} {\bibfnamefont {A.}~\bibnamefont
  {Yacoby}}, \bibinfo {author} {\bibfnamefont {D.}~\bibnamefont {Twitchen}},
  \bibinfo {author} {\bibfnamefont {S.~W.}\ \bibnamefont {Hell}}, \bibinfo
  {author} {\bibfnamefont {R.~L.}\ \bibnamefont {Walsworth}}, , \ and\ \bibinfo
  {author} {\bibfnamefont {M.~D.}\ \bibnamefont {Lukin}},\ }\bibfield  {title}
  {\enquote {\bibinfo {title} {Far-field optical imaging and manipulation of
  individual spins with nanoscale resolution},}\ }\href@noop {} {\bibfield
  {journal} {\bibinfo  {journal} {Nat.\ Phys.}\ }\textbf {\bibinfo {volume}
  {6}},\ \bibinfo {pages} {912} (\bibinfo {year} {2010})}\BibitemShut {NoStop}%
\bibitem [{\citenamefont {Pfender}\ \emph {et~al.}(2014)\citenamefont
  {Pfender}, \citenamefont {Aslam}, \citenamefont {Waldherr}, \citenamefont
  {Neumann},\ and\ \citenamefont {Wrachtrup}}]{PAW+14}%
  \BibitemOpen
  \bibfield  {author} {\bibinfo {author} {\bibfnamefont {M.}~\bibnamefont
  {Pfender}}, \bibinfo {author} {\bibfnamefont {N.}~\bibnamefont {Aslam}},
  \bibinfo {author} {\bibfnamefont {G.}~\bibnamefont {Waldherr}}, \bibinfo
  {author} {\bibfnamefont {P.}~\bibnamefont {Neumann}}, \ and\ \bibinfo
  {author} {\bibfnamefont {J.}~\bibnamefont {Wrachtrup}},\ }\bibfield  {title}
  {\enquote {\bibinfo {title} {Single-spin stochastic optical reconstruction
  microscopy},}\ }\href@noop {} {\bibfield  {journal} {\bibinfo  {journal}
  {Proc.\ Natl.\ Acad.\ Sci.\ USA}\ }\textbf {\bibinfo {volume} {111}},\
  \bibinfo {pages} {14669} (\bibinfo {year} {2014})}\BibitemShut {NoStop}%
\bibitem [{\citenamefont {Jaskula}\ \emph {et~al.}(2017)\citenamefont
  {Jaskula}, \citenamefont {Bauch}, \citenamefont {Arroyo-Camejo},
  \citenamefont {Lukin}, \citenamefont {Hell}, \citenamefont {Trifonov},\ and\
  \citenamefont {Walsworth}}]{JBA+17}%
  \BibitemOpen
  \bibfield  {author} {\bibinfo {author} {\bibfnamefont {J.-C.}\ \bibnamefont
  {Jaskula}}, \bibinfo {author} {\bibfnamefont {E.}~\bibnamefont {Bauch}},
  \bibinfo {author} {\bibfnamefont {S.}~\bibnamefont {Arroyo-Camejo}}, \bibinfo
  {author} {\bibfnamefont {M.~D.}\ \bibnamefont {Lukin}}, \bibinfo {author}
  {\bibfnamefont {S.~W.}\ \bibnamefont {Hell}}, \bibinfo {author}
  {\bibfnamefont {A.~S.}\ \bibnamefont {Trifonov}}, \ and\ \bibinfo {author}
  {\bibfnamefont {R.~L.}\ \bibnamefont {Walsworth}},\ }\bibfield  {title}
  {\enquote {\bibinfo {title} {Superresolution optical magnetic imaging and
  spectroscopy using individual electronic spins in diamond},}\ }\href@noop {}
  {\bibfield  {journal} {\bibinfo  {journal} {Opt.\ Express}\ }\textbf
  {\bibinfo {volume} {25}},\ \bibinfo {pages} {11048} (\bibinfo {year}
  {2017})}\BibitemShut {NoStop}%
\bibitem [{\citenamefont {Hell}\ \emph {et~al.}(1993)\citenamefont {Hell},
  \citenamefont {Reiner}, \citenamefont {Cremer},\ and\ \citenamefont
  {Stelzer}}]{HRCS93}%
  \BibitemOpen
  \bibfield  {author} {\bibinfo {author} {\bibfnamefont {S.}~\bibnamefont
  {Hell}}, \bibinfo {author} {\bibfnamefont {G.}~\bibnamefont {Reiner}},
  \bibinfo {author} {\bibfnamefont {C.}~\bibnamefont {Cremer}}, \ and\ \bibinfo
  {author} {\bibfnamefont {E.~H.~K.}\ \bibnamefont {Stelzer}},\ }\bibfield
  {title} {\enquote {\bibinfo {title} {Aberrations in confocal fluorescence
  microscopy induced by mismatches in refractive index},}\ }\href@noop {}
  {\bibfield  {journal} {\bibinfo  {journal} {J.\ Microsc.}\ }\textbf {\bibinfo
  {volume} {169}},\ \bibinfo {pages} {391} (\bibinfo {year}
  {1993})}\BibitemShut {NoStop}%
\bibitem [{\citenamefont {Sage}\ \emph {et~al.}(2012)\citenamefont {Sage},
  \citenamefont {Pham}, \citenamefont {Bar-Gill}, \citenamefont {Belthangady},
  \citenamefont {Lukin}, \citenamefont {Yacoby},\ and\ \citenamefont
  {Walsworth}}]{LPB+12}%
  \BibitemOpen
  \bibfield  {author} {\bibinfo {author} {\bibfnamefont {D.~L.}\ \bibnamefont
  {Sage}}, \bibinfo {author} {\bibfnamefont {L.~M.}\ \bibnamefont {Pham}},
  \bibinfo {author} {\bibfnamefont {N.}~\bibnamefont {Bar-Gill}}, \bibinfo
  {author} {\bibfnamefont {C.}~\bibnamefont {Belthangady}}, \bibinfo {author}
  {\bibfnamefont {M.~D.}\ \bibnamefont {Lukin}}, \bibinfo {author}
  {\bibfnamefont {A.}~\bibnamefont {Yacoby}}, \ and\ \bibinfo {author}
  {\bibfnamefont {R.~L.}\ \bibnamefont {Walsworth}},\ }\bibfield  {title}
  {\enquote {\bibinfo {title} {Efficient photon detection from color centers in
  a diamond optical waveguide},}\ }\href@noop {} {\bibfield  {journal}
  {\bibinfo  {journal} {Phys.\ Rev.\ B}\ }\textbf {\bibinfo {volume} {85}},\
  \bibinfo {pages} {121202} (\bibinfo {year} {2012})}\BibitemShut {NoStop}%
\bibitem [{\citenamefont {Epstein}\ \emph {et~al.}(2005)\citenamefont
  {Epstein}, \citenamefont {Mendoza}, \citenamefont {Kato},\ and\ \citenamefont
  {Awschalom}}]{EMKA05}%
  \BibitemOpen
  \bibfield  {author} {\bibinfo {author} {\bibfnamefont {R.~J.}\ \bibnamefont
  {Epstein}}, \bibinfo {author} {\bibfnamefont {F.~M.}\ \bibnamefont
  {Mendoza}}, \bibinfo {author} {\bibfnamefont {Y.~K.}\ \bibnamefont {Kato}}, \
  and\ \bibinfo {author} {\bibfnamefont {D.~D.}\ \bibnamefont {Awschalom}},\
  }\bibfield  {title} {\enquote {\bibinfo {title} {Anisotropic interactions of
  a single spin and dark-spin spectroscopy in diamond},}\ }\href@noop {}
  {\bibfield  {journal} {\bibinfo  {journal} {Nat.\ Phys.}\ }\textbf {\bibinfo
  {volume} {1}},\ \bibinfo {pages} {94} (\bibinfo {year} {2005})}\BibitemShut
  {NoStop}%
\bibitem [{\citenamefont {Zheng}\ \emph {et~al.}(2017)\citenamefont {Zheng},
  \citenamefont {Liapis}, \citenamefont {Chen}, \citenamefont {Black},\ and\
  \citenamefont {Englund}}]{ZLC+13}%
  \BibitemOpen
  \bibfield  {author} {\bibinfo {author} {\bibfnamefont {J.}~\bibnamefont
  {Zheng}}, \bibinfo {author} {\bibfnamefont {A.~C.}\ \bibnamefont {Liapis}},
  \bibinfo {author} {\bibfnamefont {E.~H.}\ \bibnamefont {Chen}}, \bibinfo
  {author} {\bibfnamefont {C.~T.}\ \bibnamefont {Black}}, \ and\ \bibinfo
  {author} {\bibfnamefont {D.}~\bibnamefont {Englund}},\ }\bibfield  {title}
  {\enquote {\bibinfo {title} {Chirped circular dielectric gratings for
  near-unity collection efficiency from quantum emitters in bulk diamond},}\
  }\href@noop {} {\bibfield  {journal} {\bibinfo  {journal} {Opt.\ Express}\
  }\textbf {\bibinfo {volume} {25}},\ \bibinfo {pages} {32420} (\bibinfo {year}
  {2017})}\BibitemShut {NoStop}%
\bibitem [{\citenamefont {Atat{\"u}re}\ \emph {et~al.}(2018)\citenamefont
  {Atat{\"u}re}, \citenamefont {Englund}, \citenamefont {Vamivakas},
  \citenamefont {Lee},\ and\ \citenamefont {Wrachtrup}}]{AEV+18}%
  \BibitemOpen
  \bibfield  {author} {\bibinfo {author} {\bibfnamefont {M.}~\bibnamefont
  {Atat{\"u}re}}, \bibinfo {author} {\bibfnamefont {D.}~\bibnamefont
  {Englund}}, \bibinfo {author} {\bibfnamefont {N.}~\bibnamefont {Vamivakas}},
  \bibinfo {author} {\bibfnamefont {S.-Y.}\ \bibnamefont {Lee}}, \ and\
  \bibinfo {author} {\bibfnamefont {J.}~\bibnamefont {Wrachtrup}},\ }\bibfield
  {title} {\enquote {\bibinfo {title} {Material platforms for spin-based
  photonic quantum technologies},}\ }\href@noop {} {\bibfield  {journal}
  {\bibinfo  {journal} {Nat.\ Rev.\ Mater.}\ }\textbf {\bibinfo {volume} {3}},\
  \bibinfo {pages} {38} (\bibinfo {year} {2018})}\BibitemShut {NoStop}%
\bibitem [{\citenamefont {Awschalom}\ \emph {et~al.}(2018)\citenamefont
  {Awschalom}, \citenamefont {Hanson}, \citenamefont {Wrachtrup},\ and\
  \citenamefont {Zhou}}]{AHWZ18}%
  \BibitemOpen
  \bibfield  {author} {\bibinfo {author} {\bibfnamefont {D.~D.}\ \bibnamefont
  {Awschalom}}, \bibinfo {author} {\bibfnamefont {R.}~\bibnamefont {Hanson}},
  \bibinfo {author} {\bibfnamefont {J.}~\bibnamefont {Wrachtrup}}, \ and\
  \bibinfo {author} {\bibfnamefont {B.~B.}\ \bibnamefont {Zhou}},\ }\bibfield
  {title} {\enquote {\bibinfo {title} {Quantum technologies with optically
  interfaced solid-state spins},}\ }\href@noop {} {\bibfield  {journal}
  {\bibinfo  {journal} {Nat.\ Photon.}\ }\textbf {\bibinfo {volume} {12}},\
  \bibinfo {pages} {516} (\bibinfo {year} {2018})}\BibitemShut {NoStop}%
\bibitem [{\citenamefont {Hopper}, \citenamefont {Shulevitz},\ and\
  \citenamefont {Bassett}(2018)}]{HSB18}%
  \BibitemOpen
  \bibfield  {author} {\bibinfo {author} {\bibfnamefont {D.~A.}\ \bibnamefont
  {Hopper}}, \bibinfo {author} {\bibfnamefont {H.~J.}\ \bibnamefont
  {Shulevitz}}, \ and\ \bibinfo {author} {\bibfnamefont {L.~C.}\ \bibnamefont
  {Bassett}},\ }\bibfield  {title} {\enquote {\bibinfo {title} {Spin {R}eadout
  {T}echniques of the {N}itrogen-{V}acancy {C}enter in {D}iamond},}\
  }\href@noop {} {\bibfield  {journal} {\bibinfo  {journal} {Micromachines}\
  }\textbf {\bibinfo {volume} {9}},\ \bibinfo {pages} {437} (\bibinfo {year}
  {2018})}\BibitemShut {NoStop}%
\bibitem [{\citenamefont {Sasaki}\ \emph {et~al.}(2016)\citenamefont {Sasaki},
  \citenamefont {Monnai}, \citenamefont {Saijo}, \citenamefont {Fujita},
  \citenamefont {Watanabe}, \citenamefont {Ishi-Hayase}, \citenamefont {Itoh},\
  and\ \citenamefont {Abe}}]{SMS+16}%
  \BibitemOpen
  \bibfield  {author} {\bibinfo {author} {\bibfnamefont {K.}~\bibnamefont
  {Sasaki}}, \bibinfo {author} {\bibfnamefont {Y.}~\bibnamefont {Monnai}},
  \bibinfo {author} {\bibfnamefont {S.}~\bibnamefont {Saijo}}, \bibinfo
  {author} {\bibfnamefont {R.}~\bibnamefont {Fujita}}, \bibinfo {author}
  {\bibfnamefont {H.}~\bibnamefont {Watanabe}}, \bibinfo {author}
  {\bibfnamefont {J.}~\bibnamefont {Ishi-Hayase}}, \bibinfo {author}
  {\bibfnamefont {K.~M.}\ \bibnamefont {Itoh}}, \ and\ \bibinfo {author}
  {\bibfnamefont {E.}~\bibnamefont {Abe}},\ }\bibfield  {title} {\enquote
  {\bibinfo {title} {Broadband, large-area microwave antenna for optically
  detected magnetic resonance of nitrogen-vacancy centers in diamond},}\
  }\href@noop {} {\bibfield  {journal} {\bibinfo  {journal} {Rev.\ Sci.\
  Instrum.}\ }\textbf {\bibinfo {volume} {87}},\ \bibinfo {pages} {053904}
  (\bibinfo {year} {2016})}\BibitemShut {NoStop}%
\bibitem [{\citenamefont {Tetienne}\ \emph {et~al.}(2012)\citenamefont
  {Tetienne}, \citenamefont {Rondin}, \citenamefont {Spinicelli}, \citenamefont
  {Chipaux}, \citenamefont {Debuisschert}, \citenamefont {Roch},\ and\
  \citenamefont {Jacques}}]{TRS+12}%
  \BibitemOpen
  \bibfield  {author} {\bibinfo {author} {\bibfnamefont {J.-P.}\ \bibnamefont
  {Tetienne}}, \bibinfo {author} {\bibfnamefont {L.}~\bibnamefont {Rondin}},
  \bibinfo {author} {\bibfnamefont {P.}~\bibnamefont {Spinicelli}}, \bibinfo
  {author} {\bibfnamefont {M.}~\bibnamefont {Chipaux}}, \bibinfo {author}
  {\bibfnamefont {T.}~\bibnamefont {Debuisschert}}, \bibinfo {author}
  {\bibfnamefont {J.-F.}\ \bibnamefont {Roch}}, \ and\ \bibinfo {author}
  {\bibfnamefont {V.}~\bibnamefont {Jacques}},\ }\bibfield  {title} {\enquote
  {\bibinfo {title} {Magnetic-field-dependent photodynamics of single {NV}
  defects in diamond: an application to qualitative all-optical magnetic
  imaging},}\ }\href@noop {} {\bibfield  {journal} {\bibinfo  {journal} {New
  J.\ Phys.}\ }\textbf {\bibinfo {volume} {14}},\ \bibinfo {pages} {103033}
  (\bibinfo {year} {2012})}\BibitemShut {NoStop}%
\bibitem [{\citenamefont {Herrmann}\ \emph {et~al.}(2016)\citenamefont
  {Herrmann}, \citenamefont {Appleton}, \citenamefont {Sasaki}, \citenamefont
  {Monnai}, \citenamefont {Teraji}, \citenamefont {Itoh},\ and\ \citenamefont
  {Abe}}]{HAS+16}%
  \BibitemOpen
  \bibfield  {author} {\bibinfo {author} {\bibfnamefont {J.}~\bibnamefont
  {Herrmann}}, \bibinfo {author} {\bibfnamefont {M.~A.}\ \bibnamefont
  {Appleton}}, \bibinfo {author} {\bibfnamefont {K.}~\bibnamefont {Sasaki}},
  \bibinfo {author} {\bibfnamefont {Y.}~\bibnamefont {Monnai}}, \bibinfo
  {author} {\bibfnamefont {T.}~\bibnamefont {Teraji}}, \bibinfo {author}
  {\bibfnamefont {K.~M.}\ \bibnamefont {Itoh}}, \ and\ \bibinfo {author}
  {\bibfnamefont {E.}~\bibnamefont {Abe}},\ }\bibfield  {title} {\enquote
  {\bibinfo {title} {Polarization- and frequency-tunable microwave circuit for
  selective excitation of nitrogen-vacancy spins in diamond},}\ }\href@noop {}
  {\bibfield  {journal} {\bibinfo  {journal} {Appl.\ Phys.\ Lett.}\ }\textbf
  {\bibinfo {volume} {109}},\ \bibinfo {pages} {183111} (\bibinfo {year}
  {2016})}\BibitemShut {NoStop}%
\bibitem [{\citenamefont {Masuyama}\ \emph {et~al.}(2018)\citenamefont
  {Masuyama}, \citenamefont {Mizuno}, \citenamefont {Ozawa}, \citenamefont
  {Ishiwata}, \citenamefont {Hatano}, \citenamefont {Ohshima}, \citenamefont
  {Iwasaki},\ and\ \citenamefont {Hatano}}]{MMO+18}%
  \BibitemOpen
  \bibfield  {author} {\bibinfo {author} {\bibfnamefont {Y.}~\bibnamefont
  {Masuyama}}, \bibinfo {author} {\bibfnamefont {K.}~\bibnamefont {Mizuno}},
  \bibinfo {author} {\bibfnamefont {H.}~\bibnamefont {Ozawa}}, \bibinfo
  {author} {\bibfnamefont {H.}~\bibnamefont {Ishiwata}}, \bibinfo {author}
  {\bibfnamefont {Y.}~\bibnamefont {Hatano}}, \bibinfo {author} {\bibfnamefont
  {T.}~\bibnamefont {Ohshima}}, \bibinfo {author} {\bibfnamefont
  {T.}~\bibnamefont {Iwasaki}}, \ and\ \bibinfo {author} {\bibfnamefont
  {M.}~\bibnamefont {Hatano}},\ }\bibfield  {title} {\enquote {\bibinfo {title}
  {Extending coherence time of macro-scale diamond magnetometer by dynamical
  decoupling with coplanar waveguide resonator},}\ }\href@noop {} {\bibfield
  {journal} {\bibinfo  {journal} {Rev.\ Sci.\ Instrum.}\ }\textbf {\bibinfo
  {volume} {89}},\ \bibinfo {pages} {125007} (\bibinfo {year}
  {2018})}\BibitemShut {NoStop}%
\bibitem [{sof()}]{software}%
  \BibitemOpen
  \href@noop {} {}\bibinfo {note} {Our MATLAB software is available upon
  request.}\BibitemShut {Stop}%
\bibitem [{\citenamefont {Vandersypen}\ and\ \citenamefont
  {Chuang}(2004)}]{VC04}%
  \BibitemOpen
  \bibfield  {author} {\bibinfo {author} {\bibfnamefont {L.~M.~K.}\
  \bibnamefont {Vandersypen}}\ and\ \bibinfo {author} {\bibfnamefont {I.~L.}\
  \bibnamefont {Chuang}},\ }\bibfield  {title} {\enquote {\bibinfo {title}
  {{NMR} techniques for quantum control and computation},}\ }\href@noop {}
  {\bibfield  {journal} {\bibinfo  {journal} {Rev.\ Mod.\ Phys.}\ }\textbf
  {\bibinfo {volume} {76}},\ \bibinfo {pages} {1037} (\bibinfo {year}
  {2004})}\BibitemShut {NoStop}%
\bibitem [{\citenamefont {Ramsey}(1950)}]{R50}%
  \BibitemOpen
  \bibfield  {author} {\bibinfo {author} {\bibfnamefont {N.~F.}\ \bibnamefont
  {Ramsey}},\ }\bibfield  {title} {\enquote {\bibinfo {title} {A {M}olecular
  {B}eam {R}esonance {M}ethod with {S}eparated {O}scillating {F}ields},}\
  }\href@noop {} {\bibfield  {journal} {\bibinfo  {journal} {Phys.\ Rev.}\
  }\textbf {\bibinfo {volume} {780}},\ \bibinfo {pages} {695} (\bibinfo {year}
  {1950})}\BibitemShut {NoStop}%
\bibitem [{\citenamefont {Hahn}(1950)}]{E50}%
  \BibitemOpen
  \bibfield  {author} {\bibinfo {author} {\bibfnamefont {E.~L.}\ \bibnamefont
  {Hahn}},\ }\bibfield  {title} {\enquote {\bibinfo {title} {Spin {E}choes},}\
  }\href@noop {} {\bibfield  {journal} {\bibinfo  {journal} {Phys.\ Rev.}\
  }\textbf {\bibinfo {volume} {80}},\ \bibinfo {pages} {580} (\bibinfo {year}
  {1950})}\BibitemShut {NoStop}%
\bibitem [{\citenamefont {de~Lange}\ \emph {et~al.}(2010)\citenamefont
  {de~Lange}, \citenamefont {Wang}, \citenamefont {Riste}, \citenamefont
  {Dobrovitski},\ and\ \citenamefont {Hanson}}]{dLWR+10}%
  \BibitemOpen
  \bibfield  {author} {\bibinfo {author} {\bibfnamefont {G.}~\bibnamefont
  {de~Lange}}, \bibinfo {author} {\bibfnamefont {Z.~H.}\ \bibnamefont {Wang}},
  \bibinfo {author} {\bibfnamefont {D.}~\bibnamefont {Riste}}, \bibinfo
  {author} {\bibfnamefont {V.~V.}\ \bibnamefont {Dobrovitski}}, \ and\ \bibinfo
  {author} {\bibfnamefont {R.}~\bibnamefont {Hanson}},\ }\bibfield  {title}
  {\enquote {\bibinfo {title} {Universal {D}ynamical {D}ecoupling of a {S}ingle
  {S}olid-{S}tate {S}pin from a {S}pin {B}ath},}\ }\href@noop {} {\bibfield
  {journal} {\bibinfo  {journal} {Science}\ }\textbf {\bibinfo {volume}
  {330}},\ \bibinfo {pages} {60} (\bibinfo {year} {2010})}\BibitemShut
  {NoStop}%
\bibitem [{\citenamefont {Bar-Gill}\ \emph {et~al.}(2012)\citenamefont
  {Bar-Gill}, \citenamefont {Pham}, \citenamefont {Belthangady}, \citenamefont
  {Le~Sage}, \citenamefont {Cappellaro}, \citenamefont {Maze},\ and\
  \citenamefont {Lukin}}]{BPB+12}%
  \BibitemOpen
  \bibfield  {author} {\bibinfo {author} {\bibfnamefont {N.}~\bibnamefont
  {Bar-Gill}}, \bibinfo {author} {\bibfnamefont {L.~M.}\ \bibnamefont {Pham}},
  \bibinfo {author} {\bibfnamefont {C.}~\bibnamefont {Belthangady}}, \bibinfo
  {author} {\bibfnamefont {D.}~\bibnamefont {Le~Sage}}, \bibinfo {author}
  {\bibfnamefont {P.}~\bibnamefont {Cappellaro}}, \bibinfo {author}
  {\bibfnamefont {J.~R.}\ \bibnamefont {Maze}}, \ and\ \bibinfo {author}
  {\bibfnamefont {M.~D.}\ \bibnamefont {Lukin}},\ }\bibfield  {title} {\enquote
  {\bibinfo {title} {Suppression of spin-bath dynamics for improved coherence
  of multi-spin-qubit systems},}\ }\href@noop {} {\bibfield  {journal}
  {\bibinfo  {journal} {Nat.\ Commun.}\ }\textbf {\bibinfo {volume} {3}},\
  \bibinfo {pages} {858} (\bibinfo {year} {2012})}\BibitemShut {NoStop}%
\bibitem [{\citenamefont {Bar-Gill}\ \emph {et~al.}(2013)\citenamefont
  {Bar-Gill}, \citenamefont {Pham}, \citenamefont {Jarmola}, \citenamefont
  {Budker},\ and\ \citenamefont {Walsworth}}]{BPJ+13}%
  \BibitemOpen
  \bibfield  {author} {\bibinfo {author} {\bibfnamefont {N.}~\bibnamefont
  {Bar-Gill}}, \bibinfo {author} {\bibfnamefont {L.~M.}\ \bibnamefont {Pham}},
  \bibinfo {author} {\bibfnamefont {A.}~\bibnamefont {Jarmola}}, \bibinfo
  {author} {\bibfnamefont {D.}~\bibnamefont {Budker}}, \ and\ \bibinfo {author}
  {\bibfnamefont {R.~L.}\ \bibnamefont {Walsworth}},\ }\bibfield  {title}
  {\enquote {\bibinfo {title} {Solid-state electronic spin coherence time
  approaching one second},}\ }\href@noop {} {\bibfield  {journal} {\bibinfo
  {journal} {Nat.\ Commun.}\ }\textbf {\bibinfo {volume} {4}},\ \bibinfo
  {pages} {1743} (\bibinfo {year} {2013})}\BibitemShut {NoStop}%
\bibitem [{\citenamefont {Abobeih}\ \emph {et~al.}(2018)\citenamefont
  {Abobeih}, \citenamefont {Cramer}, \citenamefont {Bakker}, \citenamefont
  {Kalb}, \citenamefont {Markham}, \citenamefont {Twitchen},\ and\
  \citenamefont {Taminiau}}]{ACB+18}%
  \BibitemOpen
  \bibfield  {author} {\bibinfo {author} {\bibfnamefont {M.~H.}\ \bibnamefont
  {Abobeih}}, \bibinfo {author} {\bibfnamefont {J.}~\bibnamefont {Cramer}},
  \bibinfo {author} {\bibfnamefont {M.~A.}\ \bibnamefont {Bakker}}, \bibinfo
  {author} {\bibfnamefont {N.}~\bibnamefont {Kalb}}, \bibinfo {author}
  {\bibfnamefont {M.}~\bibnamefont {Markham}}, \bibinfo {author} {\bibfnamefont
  {D.~J.}\ \bibnamefont {Twitchen}}, \ and\ \bibinfo {author} {\bibfnamefont
  {T.~H.}\ \bibnamefont {Taminiau}},\ }\bibfield  {title} {\enquote {\bibinfo
  {title} {One-second coherence for a single electron spin coupled to a
  multi-qubit nuclear-spin environment},}\ }\href@noop {} {\bibfield  {journal}
  {\bibinfo  {journal} {Nat.\ Commun.}\ }\textbf {\bibinfo {volume} {9}},\
  \bibinfo {pages} {2552} (\bibinfo {year} {2018})}\BibitemShut {NoStop}%
\bibitem [{\citenamefont {Carr}\ and\ \citenamefont {Purcell}(1954)}]{CP54}%
  \BibitemOpen
  \bibfield  {author} {\bibinfo {author} {\bibfnamefont {H.~Y.}\ \bibnamefont
  {Carr}}\ and\ \bibinfo {author} {\bibfnamefont {E.~M.}\ \bibnamefont
  {Purcell}},\ }\bibfield  {title} {\enquote {\bibinfo {title} {Effect of
  diffusion on free precession in nuclear magnetic resonance experiments},}\
  }\href@noop {} {\bibfield  {journal} {\bibinfo  {journal} {Phys.\ Rev.}\
  }\textbf {\bibinfo {volume} {94}},\ \bibinfo {pages} {630} (\bibinfo {year}
  {1954})}\BibitemShut {NoStop}%
\bibitem [{\citenamefont {Meiboom}\ and\ \citenamefont {Gill}(1958)}]{MG58}%
  \BibitemOpen
  \bibfield  {author} {\bibinfo {author} {\bibfnamefont {S.}~\bibnamefont
  {Meiboom}}\ and\ \bibinfo {author} {\bibfnamefont {D.}~\bibnamefont {Gill}},\
  }\bibfield  {title} {\enquote {\bibinfo {title} {Modified {S}pin-{E}cho
  {M}ethod for {M}easuring {N}uclear {R}elaxation {T}imes},}\ }\href@noop {}
  {\bibfield  {journal} {\bibinfo  {journal} {Rev.\ Sci.\ Instrum.}\ }\textbf
  {\bibinfo {volume} {29}},\ \bibinfo {pages} {688} (\bibinfo {year}
  {1958})}\BibitemShut {NoStop}%
\bibitem [{\citenamefont {Gullion}, \citenamefont {Baker},\ and\ \citenamefont
  {Conradi}(1990)}]{GBC90}%
  \BibitemOpen
  \bibfield  {author} {\bibinfo {author} {\bibfnamefont {T.}~\bibnamefont
  {Gullion}}, \bibinfo {author} {\bibfnamefont {D.~B.}\ \bibnamefont {Baker}},
  \ and\ \bibinfo {author} {\bibfnamefont {M.~S.}\ \bibnamefont {Conradi}},\
  }\bibfield  {title} {\enquote {\bibinfo {title} {New, compensated
  {C}arr-{P}urcell sequences},}\ }\href@noop {} {\bibfield  {journal} {\bibinfo
   {journal} {J.\ Mag.\ Res.}\ }\textbf {\bibinfo {volume} {89}},\ \bibinfo
  {pages} {479} (\bibinfo {year} {1990})}\BibitemShut {NoStop}%
\bibitem [{\citenamefont {Laraoui}\ \emph {et~al.}(2013)\citenamefont
  {Laraoui}, \citenamefont {Dolde}, \citenamefont {Burk}, \citenamefont
  {Reinhard}, \citenamefont {Wrachtrup},\ and\ \citenamefont
  {Meriles}}]{LDB+13}%
  \BibitemOpen
  \bibfield  {author} {\bibinfo {author} {\bibfnamefont {A.}~\bibnamefont
  {Laraoui}}, \bibinfo {author} {\bibfnamefont {F.}~\bibnamefont {Dolde}},
  \bibinfo {author} {\bibfnamefont {C.}~\bibnamefont {Burk}}, \bibinfo {author}
  {\bibfnamefont {F.}~\bibnamefont {Reinhard}}, \bibinfo {author}
  {\bibfnamefont {J.}~\bibnamefont {Wrachtrup}}, \ and\ \bibinfo {author}
  {\bibfnamefont {C.~A.}\ \bibnamefont {Meriles}},\ }\bibfield  {title}
  {\enquote {\bibinfo {title} {High-resolution correlation spectroscopy of
  $^{13}${C} spins near a nitrogen vacancy centre in diamond},}\ }\href@noop {}
  {\bibfield  {journal} {\bibinfo  {journal} {Nat.\ Commun.}\ }\textbf
  {\bibinfo {volume} {4}},\ \bibinfo {pages} {1651} (\bibinfo {year}
  {2013})}\BibitemShut {NoStop}%
\bibitem [{\citenamefont {Boss}\ \emph {et~al.}(2016)\citenamefont {Boss},
  \citenamefont {Chang}, \citenamefont {Armijo}, \citenamefont {Cujia},
  \citenamefont {Rosskopf}, \citenamefont {Maze},\ and\ \citenamefont
  {Degen}}]{BCA+16}%
  \BibitemOpen
  \bibfield  {author} {\bibinfo {author} {\bibfnamefont {J.~M.}\ \bibnamefont
  {Boss}}, \bibinfo {author} {\bibfnamefont {K.}~\bibnamefont {Chang}},
  \bibinfo {author} {\bibfnamefont {J.}~\bibnamefont {Armijo}}, \bibinfo
  {author} {\bibfnamefont {K.}~\bibnamefont {Cujia}}, \bibinfo {author}
  {\bibfnamefont {T.}~\bibnamefont {Rosskopf}}, \bibinfo {author}
  {\bibfnamefont {J.~R.}\ \bibnamefont {Maze}}, \ and\ \bibinfo {author}
  {\bibfnamefont {C.~L.}\ \bibnamefont {Degen}},\ }\bibfield  {title} {\enquote
  {\bibinfo {title} {One- and {T}wo-{D}imensional {N}uclear {M}agnetic
  {R}esonance {S}pectroscopy with a {D}iamond {Q}uantum {S}ensor},}\
  }\href@noop {} {\bibfield  {journal} {\bibinfo  {journal} {Phys.\ Rev.\
  Lett.}\ }\textbf {\bibinfo {volume} {116}},\ \bibinfo {pages} {197601}
  (\bibinfo {year} {2016})}\BibitemShut {NoStop}%
\bibitem [{\citenamefont {Zopes}\ \emph {et~al.}(2017)\citenamefont {Zopes},
  \citenamefont {Sasaki}, \citenamefont {Cujia}, \citenamefont {Boss},
  \citenamefont {Chang}, \citenamefont {Segawa}, \citenamefont {Itoh},\ and\
  \citenamefont {Degen}}]{ZSC+17}%
  \BibitemOpen
  \bibfield  {author} {\bibinfo {author} {\bibfnamefont {J.}~\bibnamefont
  {Zopes}}, \bibinfo {author} {\bibfnamefont {K.}~\bibnamefont {Sasaki}},
  \bibinfo {author} {\bibfnamefont {K.~S.}\ \bibnamefont {Cujia}}, \bibinfo
  {author} {\bibfnamefont {J.~M.}\ \bibnamefont {Boss}}, \bibinfo {author}
  {\bibfnamefont {K.}~\bibnamefont {Chang}}, \bibinfo {author} {\bibfnamefont
  {T.~F.}\ \bibnamefont {Segawa}}, \bibinfo {author} {\bibfnamefont {K.~M.}\
  \bibnamefont {Itoh}}, \ and\ \bibinfo {author} {\bibfnamefont {C.~L.}\
  \bibnamefont {Degen}},\ }\bibfield  {title} {\enquote {\bibinfo {title}
  {High-{R}esolution {Q}uantum {S}ensing with {S}haped {C}ontrol {P}ulses},}\
  }\href@noop {} {\bibfield  {journal} {\bibinfo  {journal} {Phys.\ Rev.\
  Lett.}\ }\textbf {\bibinfo {volume} {119}},\ \bibinfo {pages} {260501}
  (\bibinfo {year} {2017})}\BibitemShut {NoStop}%
\bibitem [{\citenamefont {Ito}\ \emph {et~al.}(2017)\citenamefont {Ito},
  \citenamefont {Saito}, \citenamefont {Sasaki}, \citenamefont {Watanabe},
  \citenamefont {Teraji}, \citenamefont {Itoh},\ and\ \citenamefont
  {Abe}}]{ISS+17}%
  \BibitemOpen
  \bibfield  {author} {\bibinfo {author} {\bibfnamefont {K.}~\bibnamefont
  {Ito}}, \bibinfo {author} {\bibfnamefont {H.}~\bibnamefont {Saito}}, \bibinfo
  {author} {\bibfnamefont {K.}~\bibnamefont {Sasaki}}, \bibinfo {author}
  {\bibfnamefont {H.}~\bibnamefont {Watanabe}}, \bibinfo {author}
  {\bibfnamefont {T.}~\bibnamefont {Teraji}}, \bibinfo {author} {\bibfnamefont
  {K.~M.}\ \bibnamefont {Itoh}}, \ and\ \bibinfo {author} {\bibfnamefont
  {E.}~\bibnamefont {Abe}},\ }\bibfield  {title} {\enquote {\bibinfo {title}
  {Nitrogen-vacancy centers created by {N}$^+$ ion implantation through
  screening {S}i{O}$_2$ layers on diamond},}\ }\href@noop {} {\bibfield
  {journal} {\bibinfo  {journal} {Appl.\ Phys.\ Lett.}\ }\textbf {\bibinfo
  {volume} {110}},\ \bibinfo {pages} {213105} (\bibinfo {year}
  {2017})}\BibitemShut {NoStop}%
\bibitem [{\citenamefont {Kupce}\ and\ \citenamefont {Freeman}(1995)}]{KF95}%
  \BibitemOpen
  \bibfield  {author} {\bibinfo {author} {\bibfnamefont {E.}~\bibnamefont
  {Kupce}}\ and\ \bibinfo {author} {\bibfnamefont {R.}~\bibnamefont
  {Freeman}},\ }\bibfield  {title} {\enquote {\bibinfo {title} {Adiabatic
  {P}ulses for {W}ideband {I}nversion and {B}roadband {D}ecoupling},}\
  }\href@noop {} {\bibfield  {journal} {\bibinfo  {journal} {J.\ Mag.\ Res.,
  Ser. A}\ }\textbf {\bibinfo {volume} {115}},\ \bibinfo {pages} {273}
  (\bibinfo {year} {1995})}\BibitemShut {NoStop}%
\bibitem [{\citenamefont {Childress}\ \emph {et~al.}(2006)\citenamefont
  {Childress}, \citenamefont {Gurudev~Dutt}, \citenamefont {Taylor},
  \citenamefont {Zibrov}, \citenamefont {Jelezko}, \citenamefont {Wrachtrup},
  \citenamefont {Hemmer},\ and\ \citenamefont {Lukin}}]{CDT+06}%
  \BibitemOpen
  \bibfield  {author} {\bibinfo {author} {\bibfnamefont {L.}~\bibnamefont
  {Childress}}, \bibinfo {author} {\bibfnamefont {M.~V.}\ \bibnamefont
  {Gurudev~Dutt}}, \bibinfo {author} {\bibfnamefont {J.~M.}\ \bibnamefont
  {Taylor}}, \bibinfo {author} {\bibfnamefont {A.~S.}\ \bibnamefont {Zibrov}},
  \bibinfo {author} {\bibfnamefont {F.}~\bibnamefont {Jelezko}}, \bibinfo
  {author} {\bibfnamefont {J.}~\bibnamefont {Wrachtrup}}, \bibinfo {author}
  {\bibfnamefont {P.~R.}\ \bibnamefont {Hemmer}}, \ and\ \bibinfo {author}
  {\bibfnamefont {M.~D.}\ \bibnamefont {Lukin}},\ }\bibfield  {title} {\enquote
  {\bibinfo {title} {Coherent {D}ynamics of {C}oupled {E}lectron and {N}uclear
  {S}pin {Q}ubits in {D}iamond},}\ }\href@noop {} {\bibfield  {journal}
  {\bibinfo  {journal} {Science}\ }\textbf {\bibinfo {volume} {314}},\ \bibinfo
  {pages} {281} (\bibinfo {year} {2006})}\BibitemShut {NoStop}%
\bibitem [{\citenamefont {Maze}, \citenamefont {Taylor},\ and\ \citenamefont
  {Lukin}(2008)}]{MTL08}%
  \BibitemOpen
  \bibfield  {author} {\bibinfo {author} {\bibfnamefont {J.~R.}\ \bibnamefont
  {Maze}}, \bibinfo {author} {\bibfnamefont {J.~M.}\ \bibnamefont {Taylor}}, \
  and\ \bibinfo {author} {\bibfnamefont {M.~D.}\ \bibnamefont {Lukin}},\
  }\bibfield  {title} {\enquote {\bibinfo {title} {Electron spin decoherence of
  single nitrogen-vacancy defects in diamond},}\ }\href@noop {} {\bibfield
  {journal} {\bibinfo  {journal} {Phys.\ Rev.\ B}\ }\textbf {\bibinfo {volume}
  {78}},\ \bibinfo {pages} {094303} (\bibinfo {year} {2008})}\BibitemShut
  {NoStop}%
\bibitem [{\citenamefont {Zhao}\ \emph {et~al.}(2011)\citenamefont {Zhao},
  \citenamefont {Hu}, \citenamefont {Ho}, \citenamefont {Wan},\ and\
  \citenamefont {Liu}}]{ZHH+11}%
  \BibitemOpen
  \bibfield  {author} {\bibinfo {author} {\bibfnamefont {N.}~\bibnamefont
  {Zhao}}, \bibinfo {author} {\bibfnamefont {J.-L.}\ \bibnamefont {Hu}},
  \bibinfo {author} {\bibfnamefont {S.-W.}\ \bibnamefont {Ho}}, \bibinfo
  {author} {\bibfnamefont {J.~T.~K.}\ \bibnamefont {Wan}}, \ and\ \bibinfo
  {author} {\bibfnamefont {R.~B.}\ \bibnamefont {Liu}},\ }\bibfield  {title}
  {\enquote {\bibinfo {title} {Atomic-scale magnetometry of distant nuclear
  spin cluster via nitrogen-vacancy spin in diamond},}\ }\href@noop {}
  {\bibfield  {journal} {\bibinfo  {journal} {Nat.\ Nanotechnol.}\ }\textbf
  {\bibinfo {volume} {6}},\ \bibinfo {pages} {242} (\bibinfo {year}
  {2011})}\BibitemShut {NoStop}%
\bibitem [{\citenamefont {Taminiau}\ \emph {et~al.}(2012)\citenamefont
  {Taminiau}, \citenamefont {Wagenaar}, \citenamefont {van~der Sar},
  \citenamefont {Jelezko}, \citenamefont {Dobrovitski},\ and\ \citenamefont
  {Hanson}}]{TWS+12}%
  \BibitemOpen
  \bibfield  {author} {\bibinfo {author} {\bibfnamefont {T.~H.}\ \bibnamefont
  {Taminiau}}, \bibinfo {author} {\bibfnamefont {J.~J.~T.}\ \bibnamefont
  {Wagenaar}}, \bibinfo {author} {\bibfnamefont {T.}~\bibnamefont {van~der
  Sar}}, \bibinfo {author} {\bibfnamefont {F.}~\bibnamefont {Jelezko}},
  \bibinfo {author} {\bibfnamefont {V.~V.}\ \bibnamefont {Dobrovitski}}, \ and\
  \bibinfo {author} {\bibfnamefont {R.}~\bibnamefont {Hanson}},\ }\bibfield
  {title} {\enquote {\bibinfo {title} {Detection and {C}ontrol of {I}ndividual
  {N}uclear {S}pin {U}sing a {W}eakly {C}oupled {E}lectron {S}pin},}\
  }\href@noop {} {\bibfield  {journal} {\bibinfo  {journal} {Phys.\ Rev.\
  Lett.}\ }\textbf {\bibinfo {volume} {109}},\ \bibinfo {pages} {137602}
  (\bibinfo {year} {2012})}\BibitemShut {NoStop}%
\bibitem [{\citenamefont {Kolkowitz}\ \emph {et~al.}(2012)\citenamefont
  {Kolkowitz}, \citenamefont {Unterreithmeier}, \citenamefont {Bennett},\ and\
  \citenamefont {Lukin}}]{KUBL12}%
  \BibitemOpen
  \bibfield  {author} {\bibinfo {author} {\bibfnamefont {S.}~\bibnamefont
  {Kolkowitz}}, \bibinfo {author} {\bibfnamefont {Q.~P.}\ \bibnamefont
  {Unterreithmeier}}, \bibinfo {author} {\bibfnamefont {S.~D.}\ \bibnamefont
  {Bennett}}, \ and\ \bibinfo {author} {\bibfnamefont {M.~D.}\ \bibnamefont
  {Lukin}},\ }\bibfield  {title} {\enquote {\bibinfo {title} {Sensing {D}istant
  {N}uclear {S}pins with a {S}ingle {E}lectron {S}pin},}\ }\href@noop {}
  {\bibfield  {journal} {\bibinfo  {journal} {Phys.\ Rev.\ Lett.}\ }\textbf
  {\bibinfo {volume} {109}},\ \bibinfo {pages} {137601} (\bibinfo {year}
  {2012})}\BibitemShut {NoStop}%
\bibitem [{\citenamefont {Loretz}\ \emph {et~al.}(2015)\citenamefont {Loretz},
  \citenamefont {Boss}, \citenamefont {Rosskopf}, \citenamefont {Mamin},
  \citenamefont {Rugar},\ and\ \citenamefont {Degen}}]{LBR+15}%
  \BibitemOpen
  \bibfield  {author} {\bibinfo {author} {\bibfnamefont {M.}~\bibnamefont
  {Loretz}}, \bibinfo {author} {\bibfnamefont {J.~M.}\ \bibnamefont {Boss}},
  \bibinfo {author} {\bibfnamefont {T.}~\bibnamefont {Rosskopf}}, \bibinfo
  {author} {\bibfnamefont {H.~J.}\ \bibnamefont {Mamin}}, \bibinfo {author}
  {\bibfnamefont {D.}~\bibnamefont {Rugar}}, \ and\ \bibinfo {author}
  {\bibfnamefont {C.~L.}\ \bibnamefont {Degen}},\ }\bibfield  {title} {\enquote
  {\bibinfo {title} {Spurious {H}armonic {R}esponse of {M}ultipulse {Q}uantum
  {S}ensing {S}equences},}\ }\href@noop {} {\bibfield  {journal} {\bibinfo
  {journal} {Phys.\ Rev.\ X}\ }\textbf {\bibinfo {volume} {5}},\ \bibinfo
  {pages} {021009} (\bibinfo {year} {2015})}\BibitemShut {NoStop}%
\bibitem [{\citenamefont {Sangtawesin}\ \emph {et~al.}(2014)\citenamefont
  {Sangtawesin}, \citenamefont {Brundage}, \citenamefont {Atkins},\ and\
  \citenamefont {Petta}}]{SBAP14}%
  \BibitemOpen
  \bibfield  {author} {\bibinfo {author} {\bibfnamefont {S.}~\bibnamefont
  {Sangtawesin}}, \bibinfo {author} {\bibfnamefont {T.~O.}\ \bibnamefont
  {Brundage}}, \bibinfo {author} {\bibfnamefont {Z.~J.}\ \bibnamefont
  {Atkins}}, \ and\ \bibinfo {author} {\bibfnamefont {J.~R.}\ \bibnamefont
  {Petta}},\ }\bibfield  {title} {\enquote {\bibinfo {title} {Highly tunable
  formation of nitrogen-vacancy centers via ion implantation},}\ }\href@noop {}
  {\bibfield  {journal} {\bibinfo  {journal} {Appl.\ Phys.\ Lett.}\ }\textbf
  {\bibinfo {volume} {105}},\ \bibinfo {pages} {063107} (\bibinfo {year}
  {2014})}\BibitemShut {NoStop}%
\bibitem [{\citenamefont {de~Oliveira}\ \emph {et~al.}(2017)\citenamefont
  {de~Oliveira}, \citenamefont {Antonov}, \citenamefont {Wang}, \citenamefont
  {Neumann}, \citenamefont {Momenzadeh}, \citenamefont {H{\"a}u{\ss}ermann},
  \citenamefont {Pasquarelli}, \citenamefont {Denisenko},\ and\ \citenamefont
  {Wrachtrup}}]{dOAW+17}%
  \BibitemOpen
  \bibfield  {author} {\bibinfo {author} {\bibfnamefont {F.~F.}\ \bibnamefont
  {de~Oliveira}}, \bibinfo {author} {\bibfnamefont {D.}~\bibnamefont
  {Antonov}}, \bibinfo {author} {\bibfnamefont {Y.}~\bibnamefont {Wang}},
  \bibinfo {author} {\bibfnamefont {P.}~\bibnamefont {Neumann}}, \bibinfo
  {author} {\bibfnamefont {S.~A.}\ \bibnamefont {Momenzadeh}}, \bibinfo
  {author} {\bibfnamefont {T.}~\bibnamefont {H{\"a}u{\ss}ermann}}, \bibinfo
  {author} {\bibfnamefont {A.}~\bibnamefont {Pasquarelli}}, \bibinfo {author}
  {\bibfnamefont {A.}~\bibnamefont {Denisenko}}, \ and\ \bibinfo {author}
  {\bibfnamefont {J.}~\bibnamefont {Wrachtrup}},\ }\bibfield  {title} {\enquote
  {\bibinfo {title} {Tailoring spin defects in diamond by lattice charging},}\
  }\href@noop {} {\bibfield  {journal} {\bibinfo  {journal} {Nat.\ Commun.}\
  }\textbf {\bibinfo {volume} {8}},\ \bibinfo {pages} {15409} (\bibinfo {year}
  {2017})}\BibitemShut {NoStop}%
\bibitem [{\citenamefont {Sangtawesin}\ \emph {et~al.}(2019)\citenamefont
  {Sangtawesin}, \citenamefont {Dwyer}, \citenamefont {Srinivasan},
  \citenamefont {Allred}, \citenamefont {Rodgers}, \citenamefont {De~Greve},
  \citenamefont {Stacey}, \citenamefont {Dontschuk}, \citenamefont {O'Donnell},
  \citenamefont {Hu}, \citenamefont {Evans}, \citenamefont {Jaye},
  \citenamefont {Fischer}, \citenamefont {Markham}, \citenamefont {Twitchen},
  \citenamefont {Park}, \citenamefont {Lukin},\ and\ \citenamefont
  {de~Leon}}]{SDS+19}%
  \BibitemOpen
  \bibfield  {author} {\bibinfo {author} {\bibfnamefont {S.}~\bibnamefont
  {Sangtawesin}}, \bibinfo {author} {\bibfnamefont {B.~L.}\ \bibnamefont
  {Dwyer}}, \bibinfo {author} {\bibfnamefont {S.}~\bibnamefont {Srinivasan}},
  \bibinfo {author} {\bibfnamefont {J.~J.}\ \bibnamefont {Allred}}, \bibinfo
  {author} {\bibfnamefont {L.~V.~H.}\ \bibnamefont {Rodgers}}, \bibinfo
  {author} {\bibfnamefont {K.}~\bibnamefont {De~Greve}}, \bibinfo {author}
  {\bibfnamefont {A.}~\bibnamefont {Stacey}}, \bibinfo {author} {\bibfnamefont
  {N.}~\bibnamefont {Dontschuk}}, \bibinfo {author} {\bibfnamefont {K.~M.}\
  \bibnamefont {O'Donnell}}, \bibinfo {author} {\bibfnamefont {D.}~\bibnamefont
  {Hu}}, \bibinfo {author} {\bibfnamefont {D.~A.}\ \bibnamefont {Evans}},
  \bibinfo {author} {\bibfnamefont {C.}~\bibnamefont {Jaye}}, \bibinfo {author}
  {\bibfnamefont {D.~A.}\ \bibnamefont {Fischer}}, \bibinfo {author}
  {\bibfnamefont {M.~L.}\ \bibnamefont {Markham}}, \bibinfo {author}
  {\bibfnamefont {D.~J.}\ \bibnamefont {Twitchen}}, \bibinfo {author}
  {\bibfnamefont {H.}~\bibnamefont {Park}}, \bibinfo {author} {\bibfnamefont
  {M.~D.}\ \bibnamefont {Lukin}}, \ and\ \bibinfo {author} {\bibfnamefont
  {N.~P.}\ \bibnamefont {de~Leon}},\ }\bibfield  {title} {\enquote {\bibinfo
  {title} {Origins of diamond surface noise probed by correlating single spin
  measurements with surface spectroscopy},}\ }\href@noop {} {\bibfield
  {journal} {\bibinfo  {journal} {Phys.\ Rev.\ X}\ }\textbf {\bibinfo {volume}
  {9}},\ \bibinfo {pages} {031052} (\bibinfo {year} {2019})}\BibitemShut
  {NoStop}%
\bibitem [{\citenamefont {Pham}\ \emph {et~al.}(2016)\citenamefont {Pham},
  \citenamefont {DeVience}, \citenamefont {Casola}, \citenamefont {Lovchinsky},
  \citenamefont {Sushkov}, \citenamefont {Bersin}, \citenamefont {Lee},
  \citenamefont {Urbach}, \citenamefont {Cappellaro}, \citenamefont {Park},
  \citenamefont {Yacoby}, \citenamefont {Lukin},\ and\ \citenamefont
  {Walsworth}}]{PDC+16}%
  \BibitemOpen
  \bibfield  {author} {\bibinfo {author} {\bibfnamefont {L.~M.}\ \bibnamefont
  {Pham}}, \bibinfo {author} {\bibfnamefont {S.~J.}\ \bibnamefont {DeVience}},
  \bibinfo {author} {\bibfnamefont {F.}~\bibnamefont {Casola}}, \bibinfo
  {author} {\bibfnamefont {I.}~\bibnamefont {Lovchinsky}}, \bibinfo {author}
  {\bibfnamefont {A.~O.}\ \bibnamefont {Sushkov}}, \bibinfo {author}
  {\bibfnamefont {E.}~\bibnamefont {Bersin}}, \bibinfo {author} {\bibfnamefont
  {J.}~\bibnamefont {Lee}}, \bibinfo {author} {\bibfnamefont {E.}~\bibnamefont
  {Urbach}}, \bibinfo {author} {\bibfnamefont {P.}~\bibnamefont {Cappellaro}},
  \bibinfo {author} {\bibfnamefont {H.}~\bibnamefont {Park}}, \bibinfo {author}
  {\bibfnamefont {A.}~\bibnamefont {Yacoby}}, \bibinfo {author} {\bibfnamefont
  {M.}~\bibnamefont {Lukin}}, \ and\ \bibinfo {author} {\bibfnamefont {R.~L.}\
  \bibnamefont {Walsworth}},\ }\bibfield  {title} {\enquote {\bibinfo {title}
  {{NMR} technique for determining the depth of shallow nitrogen-vacancy
  centers in diamond},}\ }\href@noop {} {\bibfield  {journal} {\bibinfo
  {journal} {Phys.\ Rev.\ B}\ }\textbf {\bibinfo {volume} {93}},\ \bibinfo
  {pages} {045425} (\bibinfo {year} {2016})}\BibitemShut {NoStop}%
\bibitem [{\citenamefont {Loretz}\ \emph {et~al.}(2014)\citenamefont {Loretz},
  \citenamefont {Pezzagna}, \citenamefont {Meijer},\ and\ \citenamefont
  {Degen}}]{LPMD14}%
  \BibitemOpen
  \bibfield  {author} {\bibinfo {author} {\bibfnamefont {M.}~\bibnamefont
  {Loretz}}, \bibinfo {author} {\bibfnamefont {S.}~\bibnamefont {Pezzagna}},
  \bibinfo {author} {\bibfnamefont {J.}~\bibnamefont {Meijer}}, \ and\ \bibinfo
  {author} {\bibfnamefont {C.~L.}\ \bibnamefont {Degen}},\ }\bibfield  {title}
  {\enquote {\bibinfo {title} {Nanoscale nuclear magnetic resonance with a
  1.9-nm-deep nitrogen-vacancy sensor},}\ }\href@noop {} {\bibfield  {journal}
  {\bibinfo  {journal} {Appl.\ Phys.\ Lett.}\ }\textbf {\bibinfo {volume}
  {104}},\ \bibinfo {pages} {033102} (\bibinfo {year} {2014})}\BibitemShut
  {NoStop}%
\bibitem [{\citenamefont {Lee}, \citenamefont {Okuno},\ and\ \citenamefont
  {Cavagnero}(2014)}]{LOC14}%
  \BibitemOpen
  \bibfield  {author} {\bibinfo {author} {\bibfnamefont {J.~H.}\ \bibnamefont
  {Lee}}, \bibinfo {author} {\bibfnamefont {Y.}~\bibnamefont {Okuno}}, \ and\
  \bibinfo {author} {\bibfnamefont {S.}~\bibnamefont {Cavagnero}},\ }\bibfield
  {title} {\enquote {\bibinfo {title} {Sensitivity {E}nhancement in {S}olution
  {NMR}: {E}merging {I}deas and {N}ew {F}rontiers},}\ }\href@noop {} {\bibfield
   {journal} {\bibinfo  {journal} {J.\ Magn.\ Reson.}\ }\textbf {\bibinfo
  {volume} {241}},\ \bibinfo {pages} {18} (\bibinfo {year} {2014})}\BibitemShut
  {NoStop}%
\bibitem [{\citenamefont {Ardenkjaer-Larsen}\ \emph {et~al.}(2015)\citenamefont
  {Ardenkjaer-Larsen}, \citenamefont {Boebinger}, \citenamefont {Comment},
  \citenamefont {Duckett}, \citenamefont {Edison}, \citenamefont {Engelke},
  \citenamefont {Griesinger}, \citenamefont {Griffin}, \citenamefont {Hilty},
  \citenamefont {Maeda}, \citenamefont {Parigi}, \citenamefont {Prisner},
  \citenamefont {Ravera}, \citenamefont {van Bentum}, \citenamefont {Vega},
  \citenamefont {Webb}, \citenamefont {Luchinat}, \citenamefont {Schwalbe},\
  and\ \citenamefont {Frydman}}]{ABC+15}%
  \BibitemOpen
  \bibfield  {author} {\bibinfo {author} {\bibfnamefont {J.-H.}\ \bibnamefont
  {Ardenkjaer-Larsen}}, \bibinfo {author} {\bibfnamefont {G.~S.}\ \bibnamefont
  {Boebinger}}, \bibinfo {author} {\bibfnamefont {A.}~\bibnamefont {Comment}},
  \bibinfo {author} {\bibfnamefont {S.}~\bibnamefont {Duckett}}, \bibinfo
  {author} {\bibfnamefont {A.~S.}\ \bibnamefont {Edison}}, \bibinfo {author}
  {\bibfnamefont {F.}~\bibnamefont {Engelke}}, \bibinfo {author} {\bibfnamefont
  {C.}~\bibnamefont {Griesinger}}, \bibinfo {author} {\bibfnamefont {R.~G.}\
  \bibnamefont {Griffin}}, \bibinfo {author} {\bibfnamefont {C.}~\bibnamefont
  {Hilty}}, \bibinfo {author} {\bibfnamefont {H.}~\bibnamefont {Maeda}},
  \bibinfo {author} {\bibfnamefont {G.}~\bibnamefont {Parigi}}, \bibinfo
  {author} {\bibfnamefont {T.}~\bibnamefont {Prisner}}, \bibinfo {author}
  {\bibfnamefont {E.}~\bibnamefont {Ravera}}, \bibinfo {author} {\bibfnamefont
  {J.}~\bibnamefont {van Bentum}}, \bibinfo {author} {\bibfnamefont
  {S.}~\bibnamefont {Vega}}, \bibinfo {author} {\bibfnamefont {A.}~\bibnamefont
  {Webb}}, \bibinfo {author} {\bibfnamefont {C.}~\bibnamefont {Luchinat}},
  \bibinfo {author} {\bibfnamefont {H.}~\bibnamefont {Schwalbe}}, \ and\
  \bibinfo {author} {\bibfnamefont {L.}~\bibnamefont {Frydman}},\ }\bibfield
  {title} {\enquote {\bibinfo {title} {Facing and {O}vercoming {S}ensitivity
  {C}hallenges in {B}iomolecular {NMR} {S}pectroscopy},}\ }\href@noop {}
  {\bibfield  {journal} {\bibinfo  {journal} {Angew.\ Chem.\ Int.\ Ed.}\
  }\textbf {\bibinfo {volume} {54}},\ \bibinfo {pages} {9162} (\bibinfo {year}
  {2015})}\BibitemShut {NoStop}%
\bibitem [{\citenamefont {Davies}\ \emph {et~al.}(1992)\citenamefont {Davies},
  \citenamefont {Lawson}, \citenamefont {Collins}, \citenamefont {Mainwood},\
  and\ \citenamefont {Sharp}}]{DLC+92}%
  \BibitemOpen
  \bibfield  {author} {\bibinfo {author} {\bibfnamefont {G.}~\bibnamefont
  {Davies}}, \bibinfo {author} {\bibfnamefont {S.~C.}\ \bibnamefont {Lawson}},
  \bibinfo {author} {\bibfnamefont {A.~T.}\ \bibnamefont {Collins}}, \bibinfo
  {author} {\bibfnamefont {A.}~\bibnamefont {Mainwood}}, \ and\ \bibinfo
  {author} {\bibfnamefont {S.~J.}\ \bibnamefont {Sharp}},\ }\bibfield  {title}
  {\enquote {\bibinfo {title} {Vacancy-related centers in diamond},}\
  }\href@noop {} {\bibfield  {journal} {\bibinfo  {journal} {Phys.\ Rev.\ B}\
  }\textbf {\bibinfo {volume} {46}},\ \bibinfo {pages} {13157} (\bibinfo {year}
  {1992})}\BibitemShut {NoStop}%
\bibitem [{\citenamefont {Solin}\ and\ \citenamefont {Ramdas}(1970)}]{SR70}%
  \BibitemOpen
  \bibfield  {author} {\bibinfo {author} {\bibfnamefont {S.~A.}\ \bibnamefont
  {Solin}}\ and\ \bibinfo {author} {\bibfnamefont {A.~K.}\ \bibnamefont
  {Ramdas}},\ }\bibfield  {title} {\enquote {\bibinfo {title} {Raman {S}pectrum
  of {D}iamond},}\ }\href@noop {} {\bibfield  {journal} {\bibinfo  {journal}
  {Phys.\ Rev.\ B}\ }\textbf {\bibinfo {volume} {1}},\ \bibinfo {pages} {1687}
  (\bibinfo {year} {1970})}\BibitemShut {NoStop}%
\bibitem [{\citenamefont {Kurtsiefer}\ \emph {et~al.}(2000)\citenamefont
  {Kurtsiefer}, \citenamefont {Mayer}, \citenamefont {Zarda},\ and\
  \citenamefont {Weinfurter}}]{KMZW00}%
  \BibitemOpen
  \bibfield  {author} {\bibinfo {author} {\bibfnamefont {C.}~\bibnamefont
  {Kurtsiefer}}, \bibinfo {author} {\bibfnamefont {S.}~\bibnamefont {Mayer}},
  \bibinfo {author} {\bibfnamefont {P.}~\bibnamefont {Zarda}}, \ and\ \bibinfo
  {author} {\bibfnamefont {H.}~\bibnamefont {Weinfurter}},\ }\bibfield  {title}
  {\enquote {\bibinfo {title} {Stable {S}olid-{S}tate {S}ource of {S}ingle
  {P}hotons},}\ }\href@noop {} {\bibfield  {journal} {\bibinfo  {journal}
  {Phys.\ Rev.\ Lett.}\ }\textbf {\bibinfo {volume} {85}},\ \bibinfo {pages}
  {290} (\bibinfo {year} {2000})}\BibitemShut {NoStop}%
\bibitem [{\citenamefont {Brouri}\ \emph {et~al.}(2000)\citenamefont {Brouri},
  \citenamefont {Beveratos}, \citenamefont {Poizat},\ and\ \citenamefont
  {Grangier}}]{BBPG00}%
  \BibitemOpen
  \bibfield  {author} {\bibinfo {author} {\bibfnamefont {R.}~\bibnamefont
  {Brouri}}, \bibinfo {author} {\bibfnamefont {A.}~\bibnamefont {Beveratos}},
  \bibinfo {author} {\bibfnamefont {J.-P.}\ \bibnamefont {Poizat}}, \ and\
  \bibinfo {author} {\bibfnamefont {P.}~\bibnamefont {Grangier}},\ }\bibfield
  {title} {\enquote {\bibinfo {title} {Photon antibunching in the fluorescence
  of individual color centers in diamond},}\ }\href@noop {} {\bibfield
  {journal} {\bibinfo  {journal} {Opt.\ Lett.}\ }\textbf {\bibinfo {volume}
  {25}},\ \bibinfo {pages} {1294} (\bibinfo {year} {2000})}\BibitemShut
  {NoStop}%
\bibitem [{\citenamefont {Berthel}\ \emph {et~al.}(2015)\citenamefont
  {Berthel}, \citenamefont {Mollet}, \citenamefont {Dantelle}, \citenamefont
  {Gacoin}, \citenamefont {Huant},\ and\ \citenamefont {Drezet}}]{BMD+15}%
  \BibitemOpen
  \bibfield  {author} {\bibinfo {author} {\bibfnamefont {M.}~\bibnamefont
  {Berthel}}, \bibinfo {author} {\bibfnamefont {O.}~\bibnamefont {Mollet}},
  \bibinfo {author} {\bibfnamefont {G.}~\bibnamefont {Dantelle}}, \bibinfo
  {author} {\bibfnamefont {T.}~\bibnamefont {Gacoin}}, \bibinfo {author}
  {\bibfnamefont {S.}~\bibnamefont {Huant}}, \ and\ \bibinfo {author}
  {\bibfnamefont {A.}~\bibnamefont {Drezet}},\ }\bibfield  {title} {\enquote
  {\bibinfo {title} {Photophysics of single nitrogen-vacancy centers in diamond
  nanocrystals},}\ }\href@noop {} {\bibfield  {journal} {\bibinfo  {journal}
  {Phys.\ Rev.\ B}\ }\textbf {\bibinfo {volume} {91}},\ \bibinfo {pages}
  {035308} (\bibinfo {year} {2015})}\BibitemShut {NoStop}%
\bibitem [{\citenamefont {Wehner}, \citenamefont {Elkouss},\ and\ \citenamefont
  {Hanson}(2018)}]{WEH18}%
  \BibitemOpen
  \bibfield  {author} {\bibinfo {author} {\bibfnamefont {S.}~\bibnamefont
  {Wehner}}, \bibinfo {author} {\bibfnamefont {D.}~\bibnamefont {Elkouss}}, \
  and\ \bibinfo {author} {\bibfnamefont {R.}~\bibnamefont {Hanson}},\
  }\bibfield  {title} {\enquote {\bibinfo {title} {Quantum internet: {A} vision
  for the road ahead},}\ }\href@noop {} {\bibfield  {journal} {\bibinfo
  {journal} {Science}\ }\textbf {\bibinfo {volume} {362}},\ \bibinfo {pages}
  {eaam9288} (\bibinfo {year} {2018})}\BibitemShut {NoStop}%
\bibitem [{\citenamefont {Holt}(2007)}]{H07}%
  \BibitemOpen
  \bibfield  {author} {\bibinfo {author} {\bibfnamefont {K.~B.}\ \bibnamefont
  {Holt}},\ }\bibfield  {title} {\enquote {\bibinfo {title} {Diamond at the
  nanoscale: {A}pplications of diamond nanoparticles from cellular biomarkers
  to quantum computing},}\ }\href@noop {} {\bibfield  {journal} {\bibinfo
  {journal} {Phil. Trans. Roy. Soc. A}\ }\textbf {\bibinfo {volume} {365}},\
  \bibinfo {pages} {2845} (\bibinfo {year} {2007})}\BibitemShut {NoStop}%
\bibitem [{\citenamefont {Mochalin}\ \emph {et~al.}(2012)\citenamefont
  {Mochalin}, \citenamefont {Shenderova}, \citenamefont {Ho},\ and\
  \citenamefont {Gogotsi}}]{MSHG12}%
  \BibitemOpen
  \bibfield  {author} {\bibinfo {author} {\bibfnamefont {V.~N.}\ \bibnamefont
  {Mochalin}}, \bibinfo {author} {\bibfnamefont {O.}~\bibnamefont
  {Shenderova}}, \bibinfo {author} {\bibfnamefont {D.}~\bibnamefont {Ho}}, \
  and\ \bibinfo {author} {\bibfnamefont {Y.}~\bibnamefont {Gogotsi}},\
  }\bibfield  {title} {\enquote {\bibinfo {title} {The properties and
  applications of nanodiamonds},}\ }\href@noop {} {\bibfield  {journal}
  {\bibinfo  {journal} {Nat.\ Nanotechnol.}\ }\textbf {\bibinfo {volume} {7}},\
  \bibinfo {pages} {11} (\bibinfo {year} {2012})}\BibitemShut {NoStop}%
\bibitem [{\citenamefont {Wu}\ \emph {et~al.}(2016)\citenamefont {Wu},
  \citenamefont {Jelezko}, \citenamefont {Plenio},\ and\ \citenamefont
  {Weil}}]{WJPW16}%
  \BibitemOpen
  \bibfield  {author} {\bibinfo {author} {\bibfnamefont {Y.}~\bibnamefont
  {Wu}}, \bibinfo {author} {\bibfnamefont {F.}~\bibnamefont {Jelezko}},
  \bibinfo {author} {\bibfnamefont {M.~B.}\ \bibnamefont {Plenio}}, \ and\
  \bibinfo {author} {\bibfnamefont {T.}~\bibnamefont {Weil}},\ }\bibfield
  {title} {\enquote {\bibinfo {title} {Diamond {Q}uantum {D}evices in
  {B}iology},}\ }\href@noop {} {\bibfield  {journal} {\bibinfo  {journal}
  {Angew.\ Chem.\ Int.\ Ed.}\ }\textbf {\bibinfo {volume} {55}},\ \bibinfo
  {pages} {6586} (\bibinfo {year} {2016})}\BibitemShut {NoStop}%
\end{thebibliography}%
\end{document}